\documentclass[twocolumn,aps,prx,amsmath,amssymb,amsfonts,reprint,floatfix,superscriptaddress]{revtex4-2}
\usepackage{microtype} 
\usepackage{graphicx}
\usepackage{enumerate}
\usepackage{bm}
\usepackage{amsmath}
\usepackage{slashed}
\usepackage{color}
\usepackage{upgreek}
\usepackage[normalem]{ulem}
\usepackage{tabularx} 

\allowdisplaybreaks 

\usepackage{xspace} 

\usepackage{multirow} 

\newcommand{\zeroplus}{0^{\scriptscriptstyle{+}}}

\newcommand{\zomega}{z}
\newcommand{\omegaplus}{z}
\newcommand{\omegaR}{{\omega^{\scriptscriptstyle{+}}}} 
\newcommand{\izeroplus}{{\mi 0^{\scriptscriptstyle{+}}}} 
\newcommand{\omegaRinExponent}{{\omega^{+}}} 

\renewcommand{\vec}[1]{\bm{#1}}
\newcommand{\bk}{{\mathbf{k}}}
\newcommand{\bK}{{\mathbf{K}}}

\newcommand{\mi}{\mathrm{i}}
\usepackage{bbm}  

\newcommand{\smallsigma}{{\scriptscriptstyle{\Sigma}}}

\newcommand{\fs}{{\scriptscriptstyle{\textrm{FS}}}}
\newcommand{\ls}{{\scriptscriptstyle{\textrm{LS}}}}

\newcommand{\Luttinger}{{\scriptscriptstyle{\textrm{L}}}}
\newcommand{\fl}{{\scriptscriptstyle{\textrm{FL}}}}
\newcommand{\flstar}{{\scriptscriptstyle{\textrm{FL}^{\ast}}}}
\newcommand{\nfl}{{\scriptscriptstyle{\textrm{NFL}}}}

\newcommand{\bz}{{\scriptscriptstyle{\textrm{BZ}}}}

\newcommand{\fc}{{f\!c}}

\newcommand{\TFL}{T_\fl}
\newcommand{\TFLstar}{T_\flstar}
\newcommand{\TNF}{T_\nfl}
\newcommand{\FLstar}{FL${}^\ast$}
\newcommand{\nHall}{n_\mathrm{H}}
\newcommand{\RHall}{R_\mathrm{H}}
\newcommand{\Vc}{V_{\mathrm{c}}}
\newcommand{\epsilonf}{\epsilon_f}  
\newcommand{\epsilonck}{\epsilon_{c\bk}} 
\newcommand{\epsilonfk}{\epsilon_{f\bk}} 
\newcommand{\epsilonxk}{\epsilon_{x\bk}} 

\newcommand{\YRS}{{YbRh${}_2$Si${}_2$}}
\newcommand{\CRI}{{CeRhIn${}_5$}}
\newcommand{\CCI}{{CeCoIn${}_5$}}
\newcommand{\CCA}{{CeCu${}_{6-x}$Au${}_x$}}
\newcommand{\YCS}{{YbCo${}_2$Si${}_2$}}

\newcommand{\HG}{h-gap\xspace}
\newcommand{\HP}{h-pole\xspace}
\newcommand{\SP}{fs-pole\xspace}
\newcommand{\KP}{Kondo peak\xspace}

\newcommand{\braket}[1]{\langle#1\rangle}

\newcommand{\mc}[1]{\ensuremath{\mathcal{#1}}}
\newcommand{\mr}[1]{\ensuremath{\mathrm{#1}}}

\newcommand{\tr}{\ensuremath{\mathrm{Tr}}}
\newcommand{\re}{\ensuremath{\mathrm{Re}}}
\newcommand{\im}{\ensuremath{\mathrm{Im}}}

\newcommand{\ua}{\ensuremath{{\uparrow}}}
\newcommand{\da}{\ensuremath{{\downarrow}}}

\newcommand{\circled}[1]{\raisebox{.5pt}{\textcircled{\raisebox{-.9pt} {#1}}}}

\newcommand{\Eq}[1]{Eq.~\eqref{#1}}
\newcommand{\Eqs}[1]{Eqs.~\eqref{#1}}

\newcommand{\Fig}[1]{Fig.~\ref{#1}}

\newcommand{\Sec}[1]{Sec.~\ref{#1}}

\newcommand{\pdag}{{\vphantom{dagger}}}  
\newcommand{\pminus}{{\hphantom{-}}}  

\newcommand{\FigAlog}{\ref{fig:A_vs_V_T1e-11_Cluster_log}}

\makeatletter
\let\origsection\section
\renewcommand\section{\@ifstar{\starsection}{\nostarsection}}

\newcommand\nostarsection[1]
{\sectionprelude\origsection{#1}\sectionpostlude}

\newcommand\starsection[1]
{\sectionprelude\origsection*{#1}\sectionpostlude}

\let\origsubsection\subsection
\renewcommand\subsection{\@ifstar{\starsubsection}{\nostarsubsection}}

\newcommand\nostarsubsection[1]
{\sectionprelude\origsubsection{#1}\sectionpostlude}

\newcommand\starsubsection[1]
{\sectionprelude\origsubsection*{#1}\sectionpostlude}

\newcolumntype{Y}{>{\centering\arraybackslash}X}

\newcommand\sectionprelude{%
  \vspace{15pt}
}

\newcommand\sectionpostlude{%
  \vspace{0pt}
}
\makeatother

\RequirePackage[
  hyperindex,colorlinks,bookmarksnumbered,
  plainpages=true,pdfstartview=FitH]{hyperref}
\hypersetup{linkcolor=blue,urlcolor=blue,citecolor=blue} 
\usepackage{hyperref}
\usepackage[all]{hypcap}

\begin{document} 

\title{
Emergent Properties of the Periodic Anderson Model: a High-Resolution, Real-Frequency Study of Heavy-Fermion Quantum Criticality
}
\author{Andreas Gleis}
\email{andreas.gleis@lmu.de}
\affiliation{Arnold Sommerfeld Center for Theoretical Physics, 
Center for NanoScience,\looseness=-1\,  and 
Munich Center for \\ Quantum Science and Technology,\looseness=-2\, 
Ludwig-Maximilians-Universit\"at M\"unchen, 80333 Munich, Germany}
\author{Seung-Sup B.~Lee}
\email{sslee@snu.ac.kr}
\affiliation{Arnold Sommerfeld Center for Theoretical Physics, 
Center for NanoScience,\looseness=-1\,  and 
Munich Center for \\ Quantum Science and Technology,\looseness=-2\, 
Ludwig-Maximilians-Universit\"at M\"unchen, 80333 Munich, Germany}
\affiliation{Department of Physics and Astronomy, Seoul National University, Seoul 08826, Korea}
\affiliation{Center for Theoretical Physics, Seoul National University, Seoul 08826, Korea}
\author{Gabriel Kotliar}
\email{kotliar@physics.rutgers.edu}
\affiliation{Condensed Matter Physics and Materials Science Department,\looseness=-1\,  
Brookhaven National Laboratory, Upton, NY 11973, USA}
\affiliation{Department of Physics and Astronomy, Rutgers University, Piscataway, NJ 08854, USA}
\author{Jan von Delft}
\email{vondelft@lmu.de}
\affiliation{Arnold Sommerfeld Center for Theoretical Physics, 
Center for NanoScience,\looseness=-1\,  and 
Munich Center for \\ Quantum Science and Technology,\looseness=-2\, 
Ludwig-Maximilians-Universit\"at M\"unchen, 80333 Munich, Germany}

\date{\today}

\begin{abstract}
We study paramagnetic quantum criticality in the periodic Anderson model 
(PAM) using  cellular dynamical mean-field theory~(CDMFT), with the numerical renormalization group~(NRG) as a cluster impurity solver.
The PAM describes itinerant $c$ electrons hybridizing with a lattice of  localized $f$ electrons. 
At zero temperature, it exhibits a much-studied quantum phase transition 
from a Kondo phase to an RKKY phase when the hybridization is decreased through a so-called Kondo breakdown quantum critical point (KB--QCP). 
There, Kondo screening of $f$ spins by $c$ electrons breaks down,
so that $f$ excitations change their character from somewhat itinerant to 
mainly localized, while $c$ excitations remain itinerant. 
Building on \href{http://link.aps.org/doi/10.1103/PhysRevLett.101.256404}{Phys.\ Rev.\ Lett.\ 101, 256404 (2008)}, which interpreted the 
KB transition as an orbital selective Mott transition, 
we here elucidate its nature in great detail by performing a high-resolution, real-frequency study of various dynamical quantities (susceptibilities, self-energies, spectral functions). 
NRG allows us to study the quantum critical regime governed by the QCP and located between two temperature scales, $\TFL < \TNF$.
In this regime we find
fingerprints of non-Fermi-liquid (NFL) behavior in several dynamical susceptibilities.
Surprisingly, CDMFT self-consistency is essential to stabilize the QCP and the NFL regime.
The Fermi-liquid (FL) scale $\TFL$ decreases towards and vanishes at the KB--QCP; at temperatures below $\TFL$, FL behavior emerges. At $T=0$, we find the following properties. The KB transition is 
continuous.
The $f$ quasiparticle weight decreases continuously as the transition is approached from either side, vanishing only at the KB--QCP.
Therefore, the quasiparticle weight of the $f$-band is nonzero not only in the Kondo phase but also in the RKKY phase; hence, the 
FL quasiparticles comprise $c$ \textit{and} $f$ electrons in both phases.
The Fermi surface (FS) volumes in the two phases differ, implying a FS reconstruction at the KB--QCP. Whereas the large-FS Kondo phase has a two-band structure as expected, the small-FS RKKY phase unexpectedly has a three-band structure. 
We provide a detailed analysis of quasiparticle properties of both the Kondo and, for the first time, also the RKKY phase and uncover their differences. The FS reconstruction is accompanied by the appearance of a Luttinger surface on which the $f$ self-energy diverges. The volumes of the Luttinger and Fermi surfaces are related to the charge density by a generalized Luttinger sum rule. 
We interpret the small FS volume and the emergent Luttinger surface as evidence
for $f$-electron fractionalization in the RKKY phase.
Finally, we compute the temperature dependence of the Hall coefficient and the specific heat, finding good qualitative agreement with experiments.
\end{abstract}

\maketitle
\section{Introduction}
%

%
For more than twenty years, quantum criticality in heavy-fermion (HF) systems has remained a subject of ongoing experimental and theoretical research~\cite{Loehneysen2007,Stewart2001,Kirchner2020}. In this paper,
we study several open theoretical questions within a canonical model for HF systems, the periodic Anderson model (PAM) in three dimensions. Our new insights are derived from real-frequency results with unprecedented energy resolution at arbitrarily low temperatures. To set the scene, we begin with a
survey of the state of the field, focusing in particular on  aspects relevant for the subsequent discussion of our own results.
Readers well familiar with HF physics may prefer to skip directly to section~\ref{sec:overview}, which offers an outline of our own work and results.

\subsection{Heavy fermion compounds and phenomena}

HF compounds are a class of strongly correlated systems. They contain partially filled, localized $f$ orbitals featuring strong local Coulomb repulsion. These localized orbitals hybridize with weakly interacting itinerant conduction bands~($c$ bands)~\cite{Coleman2007}. Particularly interesting is the appearance of a so-called Kondo breakdown~(KB) quantum critical point~(QCP)~\cite{Coleman2001,Coleman2002}, which will be subject of this work. 
The most prominent HF compounds featuring a KB--QCP derive their $f$ orbitals from Yb or Ce. Examples are {\YRS}, {\CCA} or the so called Ce-115 family, including {\CCI} or {\CRI}. In the following, we will first introduce HF materials in general and then focus on experimental and theoretical aspects of the KB--QCP. 
In HF systems, the hybridization between $c$ and $f$ electrons in combination with the strong local repulsion of $f$ electrons generates Kondo correlations~\cite{Coleman2007}. The strong repulsion effectively leads to the formation of local moments in the $f$ orbitals. These experience an effective antiferromagnetic interaction with the $c$ electrons due to hybridization. This promotes singlet formation between $c$ and $f$ electrons, similar to the Kondo singlet formation in the single impurity Kondo or Anderson models~\cite{Hewson1993a}. At 
temperatures below some scale $\TFL$, these Kondo correlations ultimately lead to the formation of a Fermi liquid~(FL) with quasiparticles~(QP) composed of both $c$ and $f$ electrons. Due to the local nature of the $f$ electrons, these QPs usually have a large effective mass, hence the name heavy fermions. 
If Kondo correlations are strong, the $f$ electrons effectively become mobile and contribute to the density of mobile charge carriers.
This especially affects the Fermi surface~(FS) volume~\cite{Luttinger1960,Oshikawa2000,Dzyaloshinskii2003,Curtin2018,Nishikawa2018,Heath2020} and the Hall number $\nHall$~\cite{Giuliani2005}, which are both proportional to the charge density in a FL. 
Kondo correlations compete with Rudermann-Kittel-Kasuya-Yosida~(RKKY)~\cite{Ruderman1954,Kasuya1956,Yosida1957,Nejati2017} correlations. The RKKY interaction is an effective exchange interaction between $f$ electrons mediated by the $c$ electrons. If the $c$ band is close to half-filling, this interaction is antiferromagnetic and promotes $f$-$f$ singlet formation. This competes with the aforementioned $c$-$f$ singlet formation~\cite{Doniach1977}.
It is believed that quantum criticality in HF systems is largely driven by this competition between RKKY and Kondo correlations~\cite{Doniach1977,Coleman2007,Nejati2017}. Many HF materials can be tuned through a QCP by varying, e.g., magnetic fields, pressure, or doping~\cite{Loehneysen2007,Stewart2001}. At these QCPs, a transition from the Kondo correlated heavy FL to some other, often magnetically ordered phase occurs. In some HF materials, this quantum phase transition may be understood in terms of a spin density wave~(SDW) instability of the heavy FL, 
i.e.\ a magnetic transition in an itinerant electron system,
described by Hertz-Millis-Moriya theory~\cite{Hertz1976,Moriya1985,Millis1993,Sachdev2011}. The antiferromagnetic ordering occurring at this SDW--QCP leads to a doubling of the unit cell, but QPs remain intact across the transition.
In particular, the charge density involved in charge transport does not change abruptly across the SDW--QCP. It is therefore expected that the Hall coefficient, which is sensitive to the carrier density, likewise does not abruptly change at such a QCP~\cite{Coleman2001,Coleman2002}. Further, in $d=3$ spatial dimensions, such a QCP is essentially described by a $\phi^4$ theory above its upper critical dimension. The long wavelength order parameter fluctuations are therefore Gaussian. Due to that, $\omega/T$ scaling of dynamical susceptibilities, a clear sign of an interacting fixed point~\cite{Sachdev2011}, is not expected at a SDW--QCP.
Interestingly however, there is a large class of HF materials which show QCPs not compatible with the spin density wave scenario~\cite{Loehneysen2007}. Examples include {\YRS}~\cite{Trovarelli2000,Paschen2004,Gegenwart2008,Custers2003}, {\CCA}~\cite{Loehneysen1996}, {\CRI}~\cite{Shishido2005,Jiao2015} and {\CCI}~\cite{Maksimovic2022}. In these materials, experimental observations point towards a sudden localization of the $f$ electrons as the QCP is crossed from the Kondo correlation dominated heavy FL phase. In contrast to the SDW scenario, QP seem to be destroyed at this QCP~\cite{Custers2003,Coleman2001,Coleman2002}. It thus seems that the Kondo correlations between $f$ and $c$ electrons suddenly break down at the QCP, hence the name KB--QCP.
\subsection{Experimental phenomena at the KB--QCP \label{sec:overview_experiment}}
In the following, we briefly summarize some experimental results indicative of the sudden breakdown of Kondo correlations and a corresponding sudden localization of the $f$ electrons. We first focus on results close to $T=0$ that indicate that the $f$ electrons localize. After that, we discuss some remarkable dynamical and finite temperature properties of the KB--QCP. We focus on universal phenomena and omit material-specific aspects.
\textit{Fermi liquid behavior at low $T$.---}
In most HF systems, FL behavior is observed at temperatures below some FL scale $\TFL$, on either side of the KB--QCP (on the RKKY side, the FL is often antiferromagnetically ordered). Below $\TFL$, $\sim T^2$ behavior of the resistivity and $\sim T$ behavior of the specific heat is usually observed~\cite{Gegenwart2002,Custers2003,Gegenwart2008,Stockert2011}. 
\textit{FS reconstruction at $T=0$.---}
A smoking gun signal of a KB--QCP is a sudden reconstruction of the FS at $T=0$. 
In a FL, the volume of the FS is connected to the density of charge carriers by Luttingers theorem~\cite{Luttinger1960,Oshikawa2000,Seki2017,Hazra2021}. Thus, a sudden change in the FS volume is also a sign for partial localization of charge carriers. A sudden change in the carrier density has been observed in terms of a sudden jump of the Hall number $\nHall \sim 1/\RHall$ (where $\RHall$ is the Hall coefficient) in many HF materials, including \YRS~\cite{Paschen2004,Friedemann2010}, \CCA~\cite{Loehneysen1996} and \CCI~\cite{Maksimovic2022}.
Further evidence for a FS reconstruction is due to de Haas--van Alphen (dHvA) frequency measurements, with sudden jumps of dHvA frequencies observed in \CRI~\cite{Shishido2005,Jiao2015} and \CCI~\cite{Settai2001,Shishido2002,Maksimovic2022}. 
More direct access to the FS 
is provided by angle resolved photoemission~(ARPES) measurements~\cite{Kirchner2020}, which 
have by now been performed on several HF compounds like \CCI~\cite{Koitzsch2009,Koitzsch2008,Chen2017,Jang2020}, \CRI~\cite{Chen2018},
\YRS~\cite{Danzenbaecher2011,Kummer2015,Paschen2016} or \YCS~\cite{Guettler2014}. 
Close to criticality,
ARPES data on HF compounds is to date not quite
conclusive yet since 
low-temperature scans across the KB--QCP (often tuned by magnetic field or pressure) are challenging.
\textit{Possible absence of magnetic ordering.---}
The KB--QCP is not necessarily accompanied by magnetic ordering~\cite{Friedemann2009,Zhao2019,Maksimovic2022}. In \CCI, antiferromagnetic order only occurs well away from the KB--QCP inside the RKKY dominated phase where the $f$ electrons are localized~\cite{Maksimovic2022}. Further, while for pure {\YRS} antiferromagnetic ordering sets in at the KB--QCP, this may be changed by chemical pressure~\cite{Friedemann2009}. In this way, the jump of $\nHall$ can be tuned to occur either deep in the antiferromagnetic phase or deep in the paramagnetic phase. This fact suggests that the KB--QCP is not tied to magnetic ordering~\cite{Friedemann2009,Yamamoto2010}.
\textit{Continuous suppression of the FL scale to zero.---}
The above mentioned sudden FS reconstruction suggests that the KB--QCP marks a transition between two FL phases with different densities of mobile carriers. Observations on various different materials suggests that this transition is continuous: The FL scale $\TFL$ decreases continuously to zero at the KB--QCP~\cite{Custers2003,Stockert2011} and the QP mass at the QCP diverges in many compounds~\cite{Gegenwart2002,Shishido2005,Loehneysen1996} from both sides of the transition.
\textit{Onset scale for $c$-$f$ hybridization.---}
Besides the FL scale $\TFL$, another important scale in HF compounds is the scale 
 below which $c$-$f$ hybridization begins to build up. 
We denote this scale by $\TNF$, for reasons explained later. (It is often
also denoted $T_0$.) This scale is visible for instance in scanning tunneling spectroscopy~(STS) experiments~\cite{Kirchner2020} or in optical conductivity measurements in terms of a 
distinct gap in the STS or optical spectra, called hybridization gap. $\TNF$ is then the temperature below which hybridization gap formation sets in. 
This scale has been determined in many different HF compounds via STS, for instance in \CCI~\cite{Aynajian2012}, \CRI~\cite{Haze2019} and \YRS~\cite{Ernst2011,Seiro2018}, or via optical conductivity measurements, e.g. in {\YRS}~\cite{Kimura2004,Kimura2006}, {\CRI}~\cite{Mena2005}, {\CCI}~\cite{Singley2002,Mena2005} and {\CCA}~\cite{Marabelli1990}. These experiments unambiguously show that $\TNF$ is virtually unaffected by the distance
to the KB--QCP or whether the $f$ electrons are (de-)localized at $T=0$.
\textit{Strange metal behavior.---}
Close to the KB--QCP, there is a vast scale separation between $\TNF$ and the FL scale, giving rise to an intermediate quantum critical region with NFL behavior.
In this NFL region, a linear in temperature resistivity is measured universally for all of the above mentioned materials~\cite{Trovarelli2000,Prochaska2020,Loehneysen1996a,Paglione2003,Maksimovic2022,Muramatsu2001}. Further, \YRS~\cite{Trovarelli2000,Custers2003}, \CCA~\cite{Loehneysen1996,Loehneysen1996a} and \CCI~\cite{Bianchi2003} feature a $\sim T\ln(T)$ dependence of the specific heat. Both observations are in stark contrast to the $\sim T^2$ dependence of the resistivity and $\sim T$ dependence of the specific heat expected from a FL~\cite{Giuliani2005}. 
Recent shot-noise measurements on {\YRS} nanowires further indicate the absence of QP in the strange metal region~\cite{Chen2022}.
Further, dynamical susceptibilities exhibit $\omega/T$ scaling~\cite{Varma1989} at the KB--QCP. This was initially observed for the dynamical magnetic susceptibilites in UCu${}_{5-x}$Pd${}_{x}$~\cite{Aronson1995}, {\CCA}~\cite{Schroeder2000}  and CeCu${}_{6-x}$Ag${}_{x}$~\cite{Poudel2019} and very recently also for the optical conductivites of both {\YRS}~\cite{Prochaska2020} and {\CCA}~\cite{Yang2020}. Note that $\omega/T$ scaling is a clear sign for a non-Gaussian QCP~\cite{Sachdev2011}, i.e.\ the critical fixed point is an interacting one. Particularly interesting, too, are the recent observations of $\omega/T$ scaling for the optical conductivity, as it shows that the critical behavior is not limited to the magnetic degrees of freedom only, but also includes the charge degrees of freedom.
To summarize, the following phenomena seem to be almost universal for the KB--QCP:
(i) a sudden jump of $\nHall$ as the KB--QCP is crossed at $T=0$;
(ii) a sudden reconstruction of the FS as the KB--QCP is crossed at $T=0$; 
(iii) a diverging QP mass as the KB--QCP is approached from either side at $T=0$; 
(iv) a $\ln(T)$ dependence of $\gamma = C/T$ at finite temperatures above the KB--QCP; 
(v) a linear-in-$T$ dependence of the resistivity at finite temperatures above the KB--QCP; 
(vi) $\omega/T$ scaling of dynamical susceptibilities at finite temperatures above the QCP.
All of these phenomena are not compatible with a magnetic transition in an itinerant electron system.
To the best of our knowledge, a full understanding of the KB-QCP has not yet been achieved. 
\subsection{Theory of the KB--QCP: basics and challenges}
Below, we introduce the basic models which have been proposed to describe the essentials of HF physics, including the KB--QCP.
We further review some basic intuitive, qualitative notions associated with the physics of these models.
Then, we give a qualitative overview of the challenges faced when attempting to describe the KB--QCP.
Concrete approaches for tackling those challenges are reviewed in the next subsection.
The universal physics of HF systems is believed to be described by the periodic Anderson model~(PAM),
\begin{flalign}
H_{\mathrm{PAM}} &=  \sum_{i\sigma} \epsilonf 
f^\dagger_{i\sigma} f^\pdag_{i\sigma}  
+  \sum_{i}  U
 f^\dagger_{i\uparrow} f^\pdag_{i\uparrow} f^\dagger_{i\downarrow} f^\pdag_{i\downarrow}
 & \nonumber \\
\label{eq:PAM-Hamiltonian}
& \quad  +  \sum_{i\sigma} V 
\bigl(c^\dagger_{i\sigma} f^\pdag_{i\sigma} + \mathrm{h.c.} \bigr)
 + \sum_{\bk\sigma} \epsilonck 
 c^\dagger_{\bk\sigma} c^\pdag_{\bk\sigma} \,, & 
\end{flalign}
which we consider here on a 3-dimensional cubic lattice.
Here, $f^\pdag_{i\sigma}$ and $c^\pdag_{i\sigma}$ annihilate an
$f$ or $c$ electron with spin $\sigma \in\{\ua,\da\}$ at site $i$,
respectively, $c_{\bk \sigma}$ is the discrete Fourier transform of $c^\pdag_{i\sigma}$, while $\epsilonf = \epsilonf^0 - \mu$ and
$\epsilonck =  \epsilonck^0 -\mu$, 
with $\epsilonck^0 = -2t \sum_{a = x,y,z} \cos (k_a)$,  denote the local $f$ energy and $c$-band dispersion relative to the chemical potential $\mu$, respectively.
The $f$ electrons experience a strong local 
repulsion $U$, and hybridize with the $c$ electrons with hybridization strength $V$. 
At $U=0$, the PAM features a two-band structure, with a band gap determined by the hybridization strength $V$ 
(therefore also often called hybridization gap). The hybridization thereby shifts the FS such that
both $f$ and $c$ electrons are accounted for and QP in the vicinity of the FS are hybrid $c$-$f$ objects.
The low-energy physics of the Kondo correlated FL phase can 
be thought of as a renormalized version of the $U=0$ case. The interaction does not destroy the 
low-energy hybridization between $c$ and $f$ electrons, but merely renormalizes it.
When approaching the KB--QCP from the Kondo correlated phase, the interaction 
renormalizes the hybridization to ever smaller values. The point where the hybridization 
renormalizes to zero and $c$ and $f$ electrons decouple at low energies marks the KB--QCP~\cite{Si2014}.
In the RKKY correlated phase, $c$ and $f$ electrons have been argued to remain decoupled, so that the FS is that of 
the free $c$ electrons, with QP of purely $c$ electron character~\cite{Si2014,Kirchner2020}.
Surprisingly, we find a somewhat different scenario. Indeed, we show in the present work that even in the RKKY correlated phase
QP close to the FS are $c$-$f$ hybridized, see Sec.~\ref{sec:QP_properties}. 
\textit{Description of the strange metal.---}
Arguably the most challenging aspect of the KB--QCP is the strange metal behavior at finite temperatures above the QCP.
There are by now various routes to microscopically realize NFL behavior, see Ref.~\onlinecite{Chowdhury2022} for an 
extensive recent review. 
Rigorous results on NFL physics can for instance be obtained from Sachdev-Ye-Kitaev~(SYK) models~\cite{Chowdhury2022} 
or from impurity models featuring quantum phase transitions~\cite{Vojta2006}, e.g. multi-channel 
Kondo impurities~\cite{Tsvelick1984,Tsvelick1985,Emery1992,Coleman1995} or multi-impurity models~\cite{Gan1995,Sakai1990,Sakai1992,Fabrizio2003,DeLeo2004,Silva1996,Affleck1992,Affleck1995,Jones2007,Wojcik2023,Wojcik2022}.
Despite considerable recent progress~\cite{Else2021a,Else2021b,Patel2023}, it is to date not fully clarified to what extent known routes to NFL physics
connect to the strange metal behavior observed experimentally in HF materials. 
\textit{Description of the Fermi surface reconstruction.---}
Another challenging issue is to explain how the FS can change its size in the first place.
The volume of the FS is fixed to be proportional to the particle number by the Luttinger sum rule~\cite{Luttinger1960,Oshikawa2000}, which involves the combined particle number of the $c$ and $f$ electrons~\cite{Oshikawa2000}. While the FS volume matches the Luttinger sum rule prediction in the Kondo correlated phase, this is not the case in the RKKY correlated phase where the $f$ electrons seem to be missing from the FS volume. A theoretical description of the KB--QCP also needs to correctly describe both the Kondo and RKKY correlated phases, which is far from straight forward especially in the latter case.
Nevertheless, this aspect of the KB--QCP is better understood and intuitively more accessible than the strange metal physics.
\subsection{Theory of the KB--QCP: approaches}
The KB--QCP has been subject to many theoretical studies in the past, using both analytical and numerical approaches. 
Below, we briefly list what has been achieved so far and point out the main issues of the corresponding approaches. 
\textit{Numerically exact methods.---}
Significant progress on physical phenomena can be made based on exact solutions obtained with controlled numerical methods.
The main advantage is that such an approach is highly unbiased: the bare PAM or the closely related Kondo lattice model~(KLM)
is solved exactly, potentially in some simplified geometry and usually in some constrained parameter regime. Results have so far been obtained
with Quantum Monte Carlo (QMC) methods~\cite{Assaad1999,Assaad2004,Capponi2001,Berg2019,Toldin2019,Danu2020,Danu2021,Danu2022,Frank2022,Liu2022}
and the Density Matrix Renormalization Group (DMRG)~\cite{Gleis2022}, some of which have reported evidence of a KB--QCP~\cite{Danu2021,Danu2022,Gleis2022}.
Even though reports dynamical or transport properties is scarce, numerically exact studies can provide valuable benchmarks for less controlled approaches.
\textit{Slave-particle theories.---} 
Considerable conceptual progress on KB physics has been achieved using slave-particle approaches~\cite{Coleman1984,Burdin2002,Senthil2003,Senthil2004,Pepin2005,Vojta2010}. These approaches decompose the degrees of freedom of the PAM or the KLM in terms of additional fermionic or bosonic degrees of freedom (often called partons) which are subject to gauge constraints, to ensure the mapping is exact~\cite{Coleman1984,Pepin2007,Burdin2002,Senthil2003,Senthil2004,Vojta2010,Aldape2022}.
While the parton decomposition does not render the models solvable, it allows for more flexibility when constructing approximate solutions.
For instance, a certain effective low-energy form of the Hamiltonian and some effective dynamics of the gauge fields (which are static after the initial exact mapping) are usually assumed. The effective theory can then be solved by means of approximate methods, for instance by taking certain large $N$ limits and/or resorting to static mean-field theory.
One of the early successes of slave-particle approaches is the prediction of a RKKY phase in which $f$-electrons are localized and do not contribute to the FS in terms of an orbital selective Mott phase~\cite{Senthil2003,Senthil2004,Vojta2010}. The missing FS volume in the RKKY phase
was linked to emergent topological excitations of fractionalized spins~\cite{Senthil2003,Senthil2004,Bonderson2016}, thus coining the term fractionalized FL (\FLstar).
It was further established that a continuous transition between Kondo and RKKY correlated phases can exist~\cite{Senthil2004}, including a FS reconstruction 
accompanied by a sudden jump in the Hall coefficient~\cite{Coleman2005a}. Recently, by considering spatially disordered interactions~\cite{Aldape2022,Patel2023}, it has been been possible to account for a strange-metal-like $\sim T\ln T$ resistivity using a slave particle approach (though to our knowledge, a $\ln T$ correction to the $\sim T$ resistivity has not been reported in \YRS~\cite{Nguyen2021}, which shows the most extensive strange metal regimes of all known HF compounds).
\textit{Dynamical mean-field theory.---}
Dynamical mean-field theory~(DMFT)~\cite{Georges1996,Kotliar2006}  and its extensions~\cite{Kotliar2001,Maier2005,Rohringer2018}
have been successfully used in many studies on HF systems~\cite{Si2001,Sun2003,Medici2005,Sun2005,Schaefer2019,Hu2020,Osolin2017,Tanaskovic2011,DeLeo2008a,DeLeo2008}
and have lead to valuable new insights.
DMFT methods treat lattice models by mapping them on self-consistent impurity models. 
The most prominent approach, which has lead to many insights, is the extended DMFT~(EDMFT) approach to KLM~\cite{Si1999,Smith2000,Si2001,Si2003,Sun2003,Zhu2003,Glossop2007,Cai2020,Hu2021a,Hu2021b,Kandala2022,Hu2022a,Hu2022b}.
EDMFT maps the KLM on a self-consistent Bose-Fermi-Kondo~(BFK) impurity model and is able to capture a KB--QCP due to the local
competition between Kondo screening and magnetic fluctuations.
One of the main successes of EDMFT is the explanation of $\omega/T$ scaling of the dynamical spin
structure factor in {\CCA}~\cite{Schroeder2000} at the KB--QCP.
However, to the best of our knowledge, predictions of other thermodynamic and transport properties, like the linear-in-$T$ resistivity or the $T\ln T$ dependence of the specific heat, are lacking to date.
It is therefore still unclear whether the EDMFT approach correctly describes the experimentally observed strange metal behavior.
We expect though that these gaps in the literature will be filled in future studies.
A downside of the EDMFT approach is that full self-consistency leads to a first-order phase transition~\cite{Sun2003,Sun2005a} at $T>0$.
A continuous transition can be recovered by
insisting on a featureless fermionic density of states~(DOS)~\cite{Si2005}, at the cost of giving up self-consistency of the fermionic degrees of
freedom, as is routinely done in KB--QCP studies using EDMFT~\cite{Zhu2003,Glossop2007,Hu2021a}. 
This downside of EDMFT has lead to the proposal of using 2-site cellular DMFT~(CDMFT)~\cite{Kotliar2001} to study Kondo breakdown physics~\cite{Sun2005}.
Using exact diagonalization~(ED) as an impurity solver, it was shown that a 2-site CDMFT treatment of the PAM [Eq.~\eqref{eq:PAM-Hamiltonian}] can describe the KB--QCP as an orbital selective Mott transition~(OSMT) at $T=0$~\cite{DeLeo2008a,DeLeo2008}, where the $f$ electrons localize while the $c$ electrons remain itinerant. Similar studies with QMC impurity solvers~\cite{Martin2008,Martin2010,Tanaskovic2011} were however not able to find signs of a KB--QCP in the temperature range studied. Since ED suffers from limited frequency resolution while QMC has trouble reaching low temperatures, it is to date not clear to what extent CDMFT can describe KB physics. The ED study was further not able to establish conclusively whether the transition is first or second order.
%

%
\begin{figure*}[hbt!]
\includegraphics[width=\textwidth]{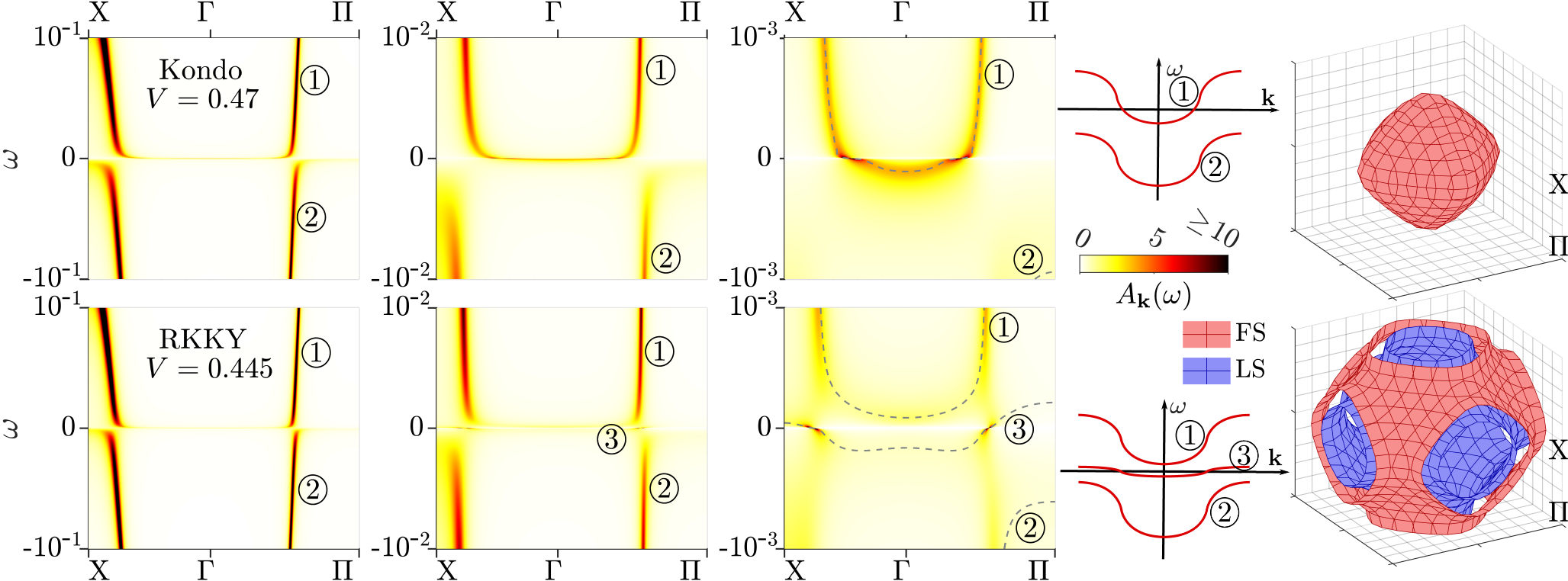}
\caption{%
Three left columns:  $\bk$ dependent spectral function 
$A_\bk(\omega)$ of the PAM at $T=0$, zooming in on frequencies 
$|\omega| \le 10^{-1}$, $10^{-2}$ and $10^{-3}$. 
 $V$ is chosen close to the KB--QCP, in either the Kondo phase (upper row)
 or the RKKY phase (lower row). At relatively high frequencies (first column), the spectral functions of both phases seem to have a similar structure, involving two bands,
 labeled \circled{1} and \circled{2}.  
 However, zooming in to lower frequencies (second and third columns) we find a striking difference: 
in the RKKY phase, a narrow third band emerges at low frequencies, labeled \circled{3}, as indicated by the dashed lines in the third column and in the schematic sketch of the band structure in the fourth column.
This difference also leads to different Fermi surfaces, shown on the far right as red surfaces; the blue surface shows the Luttinger surface in the RKKY phase.  
For more details, see Sec.~\ref{sec:Ak_vs_w}.
}
\label{fig:Ak_lin_FS}
\end{figure*}
%

%
\subsection{Overview of our main results \label{sec:overview}}
In this work, we revisit the CDMFT approach of Refs.~\onlinecite{DeLeo2008a,DeLeo2008}, now using the Numerical Renormalization Group~(NRG)~\cite{Wilson1975,Bulla2008} as an impurity solver. 
The NRG is numerically exact, produces spectral data directly on the real frequency axis and is able to access arbitrarily low temperatures and frequencies. NRG therefore eliminates the limitations of both ED and QMC for studying quantum critical phenomena. In particular, using NRG, we are able to settle the question of whether a 2-site CDMFT approximation of the PAM on a simple cubic lattice is capable of describing a KB--QCP. Furthermore, leveraging the high resolution of NRG, we find several new features of the RKKY phase which were not accessible to lower resolution methods. Most important, NRG can explore the quantum critical regime governed by the QCP. We stick to the parameters used in Refs.~\onlinecite{DeLeo2008a,DeLeo2008} and vary the $c$-$f$ hybridization strength $V$ and temperature $T$ [c.f. Eq.~\eqref{eq:PAM-Hamiltonian} and Sec.~\ref{sec:ModelMethods}]. Similar in spirit as Ref.~\onlinecite{DeLeo2008a},
we focus on purely paramagnetic solutions by artificially preventing the breaking of spin rotation symmetry.
This is motivated by experimental observations which suggest that the KB--QCP and magnetic ordering are 
distinct phenomena~\cite{Friedemann2009,Maksimovic2022,Coleman2010}. Here, we decide to focus on the paramagnetic KB--QCP
and refrain from the additional complications introduced by possible magnetic ordering.
The interplay between KB physics and symmetry
breaking will be considered in detail in future work.
The main goals of our work are to (i) establish that 2-site CDMFT is able to describes a continuous KB--QCP; (ii) establish that the QCP is governed by a NFL critical fixed point and characterize its properties; (iii) make progress on our understanding of the fate of $c$-$f$ hybridization in the vicinity of the QCP; and (iv) explore to what extent CDMFT is able to qualitatively capture the experimental phenomena described in Sec.~\ref{sec:overview_experiment}. In the process, we reveal several new aspects of the CDMFT solution. The remainder of this subsection is intended as a summary of our main results and a guide to where to find them in our paper.
(i) \textit{The KB--QCP is a continuous OSMT.---}
Using NRG, we clearly establish that 2-site CDMFT describes a continuous KB--QCP.
First and foremost, this is shown in Sec.~\ref{sec:phasediagram} and Fig.~\ref{fig:Escale}, where we present the phase diagram obtained with CDMFT-NRG.
Here, we establish the presence of two energy scales: the FL scale $\TFL$, below which we find FL behavior; and a NFL scale $\TNF (\geq \TFL)$,
which marks the onset of $c$-$f$ hybridization [c.f. Figs.~\ref{fig:A_vs_T_V046_Cluster_log} and~\ref{fig:A_vs_T_V0455_Cluster_log}] and below which we find strange-metal-like NFL behavior in the vicinity of the QCP. We find that as $V$ approaches a critical hybridization strength $V_c$ from either side, $\TFL$ continuously decreases to zero while $\TNF$ remains non-zero throughout. We identify $V_c$ as the location of a KB--QCP. While the FS volume in the Kondo correlated phase at $V>V_c$ counts both the $c$ and the $f$ electrons, it only counts the $c$ electrons in the RKKY correlated phase at $V<V_c$ [c.f. Sec.~\ref{sec:Luttinger}]. 
The KB--QCP thereby marks a continuous transition between two FL phases, which differ in their FS volumes [c.f. Figs.~\ref{fig:A_vs_V} and~\ref{fig:LT_PAM} and their corresponding sections]. The FS reconstruction, which occurs at the KB--QCP, is accompanied by the appearance of a \textit{dispersive} pole in the $f$-electron self-energy. This also implies the appearance of a Luttinger surface~\cite{Dzyaloshinskii2003}, the locus of points in the Brillouin zone at which the $f$ self-energy 
pole lies at $\omega=0$ [c.f. Secs.~\ref{sec:SingleParticleCluster} and~\ref{sec:Ak_vs_w}]. Ref.~\onlinecite{Fabrizio2023} recently suggested that Luttinger surfaces may define spinon Fermi surfaces. 
The appearance of a Luttinger surface therefore suggests that the $f$-electron is fractionalized (i.e. spinon degrees of freedom emerge at low energies as stable spin-$1/2$ excitations) in the RKKY phase.
In the parlance of Refs.~\onlinecite{Senthil2003,Senthil2004}, this suggests that the RKKY phase is a fractionalized FL (\FLstar). We will explore this more concretely in future work. Following Refs.~\onlinecite{DeLeo2008a,DeLeo2008}, \textit{we therefore identify the KB--QCP as a continuous OSMT, in which the $f$ electrons partially localize} while the $c$ electrons do not.
(ii) \textit{NFL physics at intermediate $T$ close to the QCP.---}
In the vicinity of the QCP, there is a scale separation between $\TFL$ and $\TNF$, giving rise to an intermediate NFL region extending down to $T=0$ at $V_c$ [c.f. Fig.~\ref{fig:Escale}].
Our evidence that this intermediate region is a NFL region is based on NRG finite size spectra [Fig.~\ref{fig:FiniteSize}], dynamical correlation functions [Fig.~\ref{fig:Chi_QCP}] and a $\sim T\ln T$ of the specific heat [Fig.~\ref{fig:Sent_SFC_PAM}]. In a companion paper~\cite{Gleis2023a}, we will present 
a detailed analysis of the optical conductivity, showing $\omega/T$ scaling,
and the temperature dependence of the resistivity, showing linear-in-$T$ behavior in the NFL regime.
Moreover, our KB--QCP for the PAM shows several similarities with QCPs found for the 2-impurity and 2-channel Kondo models [c.f. Sec.~\ref{sec:other_models}]. 
Very surprisingly, we find a stable NFL fixed point even though the effective 2-impurity model lacks the symmetries necessary to 
stabilize a NFL fixed point
\textit{without} self-consistency~\cite{Jones1988,Jones1989,Affleck1992,Affleck1995,Jones2007,Mitchell2012a,Mitchell2012,Sakai1990,Sakai1992,Fabrizio2003,DeLeo2004}.
We find that the CDMFT self-consistency conditions are essential for the stability of the NFL fixed point [c.f. Sec.~\ref{sec:two-stage screening} and App.~\ref{app:Hybfun}].
(iii) \textit{Fate of $c$-$f$ hybridization across the KB--QCP.---}
One of our most surprising findings is that low-energy $c$-$f$ hybridization is not destroyed as the KB--QCP is crossed from the Kondo ($V>V_c$) to the RKKY phase ($V<V_c$). We elaborate this in detail in Sections~\ref{sec:single-particle-preliminaries}, \ref{sec:SingleParticleCluster}, \ref{sec:Ak_vs_w} and \ref{sec:Luttinger}. Indeed, we find that the QP weights for both the $c$ and $f$ electrons are non-zero in \textit{both} $T=0$ phases adjacent to the KB--QCP, 
vanishing only \textit{at} at the KB--QCP [Fig.~\ref{fig:QP_para}]. This is one of our most surprising results and in stark contrast to previous work. 
It implies that, contrary to widespread belief~\cite{Kirchner2020}, the difference between the Kondo and RKKY phase is not due to non-zero versus zero $f$-electron QP weight. Instead, it is caused by a sign change of the effective $f$-level position close to the center of the Brillouin zone [Fig.~\ref{fig:QP_para} and its discussion; Sec.~\ref{sec:Luttinger}]. We connect this sign change to the aforementioned emergence of a dispersive self-energy pole [Fig.~\ref{fig:SECinv_vs_V_T1e-11_Cluster_log} and its discussion]. It thus reflects 
the orbital selective Mott nature of the RKKY phase.

The non-zero $f$-electron QP weight and the dispersive self-energy pole leads to the emergence of a \textit{third} band in the RKKY phase
[c.f. Figs.~\ref{fig:A_vs_V} and \ref{fig:SEk_vs_V} and their discussion]. 
The emergence of a third band in a model constructed from only two bands is our most striking and unexpected result. Its emergence 
is previewed in Fig.~\ref{fig:Ak_lin_FS}, showing the
total ($c$ and $f$) spectral function $A_\bk(\omega)$: at high frequency ($|\omega| \lesssim 10^{-1}$, measured in terms of the bare $c$-electron half-band width) its structure remains qualitatively unaltered as the KB--QCP is crossed---it seems as if in both cases, there is a two-band structure characteristic for HF systems.
However, as one zooms in further to lower frequencies, it becomes clear that the low frequency physics is entirely different: a third band emerges in the RKKY phase and the FS is shifted relative to that in the Kondo phase. 
The emergence of the third band is intimately tied to the emergence of a Luttinger surface [c.f. Sec.~\ref{sec:Ak_vs_w} and~\ref{sec:Luttinger}].
It was concluded in Ref.~\onlinecite{Fabrizio2023}  that Luttinger surfaces may define spinon Fermi surfaces. From that perspective, 
the third band can be viewed as a direct manifestation of the fractionalization of the $f$-electrons in the RKKY phase: their spinon degrees of freedom become independent, long-lived excitation, giving rise to the third band. 
A more concrete investigation will be the subject of future work.

(iv) \textit{Relation to experiment.---}
We repeatedly make contact to experimental observations in our manuscript.
In table~\ref{tab:exp_CDMFT}, we provide a list of experimental observations which are qualitatively reproduced by our CDMFT-NRG approach.
We include references to the relevant experimental publications and reviews (without claim of completeness) and pointers to where our corresponding CDMFT results appear in this paper.
%

\begin{table}[bt!]
\begin{tabularx}{\linewidth}{lcc}
\hline
 phenomenon & experiment & PAM, CDMFT \\ 
 \hline  
phase diagram &  \cite{Custers2003,Stockert2011,Kirchner2020} & Fig.~\ref{fig:Escale} \\[1.5mm]
sudden FS reconstruction  & \cite{Shishido2005,Jiao2015,Settai2001,Shishido2002,Maksimovic2022} & Figs.~\ref{fig:A_vs_V},~\ref{fig:LT_PAM} \\[1.5mm]  
jump of Hall coefficient  & \cite{Paschen2004,Friedemann2010,Maksimovic2022} & Fig.~\ref{fig:Hall_PAM}, App~\ref{app:THall_PAM} \\[1.5mm] 
\parbox[c]{3cm}{\raggedright control parameter
dependence of $T_{\rm Hall}$}
& \cite{Paschen2004} & Fig.~\ref{fig:Escale}  \\[3mm]
divergent QP mass  & \cite{Loehneysen1996,Gegenwart2002,Custers2003,Shishido2005} & Fig.~\ref{fig:QP_para},~\ref{fig:Sent_SFC_PAM} \\[1.5mm] 
$\TFL \to 0$ at KB--QCP  & \cite{Custers2003,Stockert2011,Kirchner2020} & Fig.~\ref{fig:Escale} \\[1.5mm] 
hybridization gap forms at &
\multirow[l]{2}{*}{
\cite{Kirchner2020, Aynajian2012,Haze2019,Ernst2011,Seiro2018,Kimura2004,Kimura2006,Mena2005,Singley2002,Mena2005,Marabelli1990}
}
&
\multirow[l]{2}{*}{
Figs.~\ref{fig:Escale},~\ref{fig:A_vs_T_V046_Cluster_log},~\ref{fig:A_vs_T_V0455_Cluster_log}
} 
\\
$\TNF$; $\TNF \neq 0$ at KB--QCP  & & 
\\[1.5mm]  
NFL: ${\sim T\ln T}$ specific heat & \cite{Trovarelli2000,Custers2003,Loehneysen1996,Bianchi2003} & Fig.~\ref{fig:Sent_SFC_PAM} \\[1.5mm]
NFL:
linear-in-$T$ resistivity & \cite{Trovarelli2000,Prochaska2020,Loehneysen1996a,Paglione2003,Maksimovic2022,Muramatsu2001} & \cite{Gleis2023a} \\[1.5mm]
NFL:
$\omega/T$ scaling & \cite{Aronson1995,Schroeder2000,Poudel2019,Prochaska2020,Yang2020}& \cite{Gleis2023a} \\[1mm]
\hline
\end{tabularx}
\caption{\label{tab:exp_CDMFT} Left: Experimental phenomena 
associated with heavy-fermion behavior that can be 
recovered qualitatively from the periodic Anderson model, treated using 2-site CDMFT+NRG. Middle: References and reviews (without claiming completeness) which have inferred these phenomena from experimental data. 
Right: Figures in this work or a follow-up paper 
\cite{Gleis2023a} exhibiting these phenomena.
}
\end{table}
To conclude our overview, we summarize the structure of the paper:
After reviewing CDMFT and NRG in Sec.~\ref{sec:ModelMethods}, we present and discuss the phase diagram in Sec.~\ref{sec:phasediagram}.
By detailed discussion of real-frequency dynamical susceptibilities and NRG finite size spectra, we demonstrate in Sec.~\ref{sec:two-stage screening} that $\TFL$ vanishes at the QCP and gives rise to NFL behavior at intermediate temperatures
below $\TNF$.
After reviewing expectations on single particle properties in HF systems in Sec.~\ref{sec:single-particle-preliminaries}, a detailed discussion of single-particle properties of the self-consistent 2IAM follows in Sec.~\ref{sec:SingleParticleCluster}. 
Using NRG, we show unambiguously that the $f$ electron QP weight is finite in \textit{both} the Kondo \textit{and} RKKY correlated phases.
In Sec.~\ref{sec:Ak_vs_w}, we discuss how the single-particle properties of the self-consistent impurity model translate to lattice properties.
There, we show that the FS indeed reconstructs across the KB--QCP. 
In Sec.~\ref{sec:Luttinger}, we discuss the details of this FS reconstruction in the context of Luttinger's theorem and present
our results for the Hall coefficient. 
Section~\ref{sec:Sommerfeld} shows results on the specific heat.
Finally, in Section~\ref{sec:other_models}, we discuss the similarities and differences between the KB--QCP in the PAM studied with 2-site CDMFT and the impurity QCPs in the two-channel and two-impurity Kondo models.
Section~\ref{sec:ConclusionOutlook} presents our conclusions and an outlook. Several appendices discuss technical details of our methods.

%

\section{Model and Methods \label{sec:ModelMethods}}%

Although the CDMFT treatment of the PAM is well-established \cite{DeLeo2008,DeLeo2008a,Tanaskovic2011}, we describe it in some detail, to introduce notation and terminology that will be used extensively in subsequent sections. 

Before starting, a general remark on notation. Matsubara
propagators analytically continued into the complex plane 
will be denoted $G(z)$, with $z \in \mathbbm{C}$. 
The corresponding retarded propagators are $G(\omegaR)$, 
with $\omegaplus = \omega + \mi \zeroplus$, $\omega\in \mathbbm{R}$. 
Ditto for self-energies.

\subsection{Periodic Anderson model}
\label{eq:PAM-definition}

We consider the PAM on a three-dimensional cubic lattice, where each lattice site hosts a non-interacting conduction $c$ orbital and an interacting localized $f$ orbital.
The Hamiltonian, $H_\mathrm{PAM}$, is given by \Eq{eq:PAM-Hamiltonian}. In this work, we
set $t = 1/6$ so that the $c$-electron half-bandwidth
is $1$, and use the latter as unit of energy. Following the choices 
of De Leo, Civelli and Kotliar \cite{DeLeo2008a,DeLeo2008}, we set 
the chemical potential to $\mu=0.2$, the $f$-level energy to 
$\epsilonf^{0}=-5.5$ and 
the $f$-level Coulomb repulsion to $U=10$, so that the system is electron-doped. When exploring the phase diagram  in Sec.~\ref{sec:phasediagram}, we will vary the $c$-$f$ hybridization $V$ and temperature $T$. 
In the momentum representation, the lattice propagators can be expressed as
\begin{align}
G_\bk(\zomega) &= 
\begin{pmatrix}
\omegaplus  - \epsilonf 
- \Sigma_{f\bk}(\zomega) & -V \\
-V & \omegaplus  - \epsilonck 
\end{pmatrix}^{-1} \nonumber \\
& = 
\begin{pmatrix}
G_{f\bk}(\zomega) & G_{\fc \bk}(\zomega) \\
G_{\fc \bk}(\zomega) & G_{c\bk}(\zomega)
\end{pmatrix} \, ,
\label{eq:Gmatrix}
\end{align}
The matrix elements in the second line of \Eq{eq:Gmatrix}, defined by computing the matrix inverse stated in the first, are given by
\begin{subequations}
\label{eq:Glatt}
\begin{align}
\label{eq:Glatt-Gf}
G_{f\bk}(\zomega) &= \bigl[\omegaplus - \epsilonf 
-  \Delta_{f\bk}(\zomega) -\Sigma_{f\bk}(\zomega)\bigr]^{-1} , 
\hspace{-1cm} & 
\\
\label{eq:Glatt-Gc}
G_{c\bk}(\zomega) &= \bigl[\omegaplus  - \epsilonck
-\Sigma_{c\bk}(\zomega)\bigr]^{-1} , 
\\
\label{eq:Glatt-Gfc}
G_{\fc \bk}(\zomega) &= \Sigma_{c\bk}(\zomega) \, G_{c\bk}(\zomega)/V = \Delta_{f\bk}(\zomega)\, G_{f\bk}(\zomega)/V \, , 
\\
\label{eq:Glatt-Deltaf}
\Delta_{f\bk}(\zomega) & = V^2
\bigl[\omegaplus  -  \epsilonck 
\bigr]^{-1} \, , 
\\
\label{eq:Glatt-Sigmac}
\Sigma_{c\bk}(\zomega) &= V^2\bigl[\omegaplus  -  \epsilonf
 -\Sigma_{f\bk}(\zomega)\bigr]^{-1} . 
\end{align}
\end{subequations}
For brevity, we will often omit momentum and/or frequency arguments.
The $f$-hybridization function $\Delta_f$ and 
the one-particle irreducible $f$ self-energy,  $\Sigma_f$, 
describe, respectively, the effects of hybridization and interactions on $f$ electrons. Their effects on $c$ electrons are described by $\Sigma_c$,
which is not one-particle irreducible and a function of $\Sigma_f$.
In particular, hybridization leads to so-called hybridization poles
in $\Sigma_{c}(\zomega)$, which in turn cause so-called hybridization 
gaps in the spectral functions $A_{c}(\omega) = - \frac{1}{\pi} \im 
G_{c}(\omegaR)$ (discussed in detail in later sections).%
\subsection{Two-site cellular DMFT}
\label{sec:two-site-cluster DMFT}

We study the PAM using a two-site CDMFT approximation,
considering a unit cell of two neighboring lattice sites as a cluster impurity and the rest of the lattice as a self-consistent bath. We choose to focus solely on solutions with SU(2) spin rotation symmetry, U(1) total charge symmetry and inversion symmetry, i.e.\ solutions which treat sites 1 and 2 as equivalent. Enforcing these symmetries may induce artificial frustration in some regions of the phase diagram;
in particular, they exclude the possibility of symmetry-breaking order such as antiferromagnetism. 
We have two reasons for nevertheless focusing only on non-symmetry-broken solutions. 
First, in some materials
the antiferromagnetic QCP (AF-QCP) and the KB--QCP do not coincide: in 
\YRS\ they can be shifted apart by applying chemical pressure~\cite{Friedemann2009}, and in \CCI\ they naturally lie apart \cite{Maksimovic2022}. 
This strongly suggests that the onset of antiferromagnetic order is not an
intrinsic property of the KB--QCP itself~\cite{Coleman2010}. (The question why the AF-QCP often coincides with the KB--QCP is interesting, but not addressed in this paper.) Second, in experimental studies, symmetry-breaking order is usually absent in the quantum critical region. It is therefore of interest to understand the properties of the KB--QCP and the NFL regime above it in
the absence of symmetry breaking. Having chosen to exclude
symmetry breaking, we refrain from studying the limit $V\to 0$, where
its occurrence is increasingly likely for energetic reasons. Studies of symmetry-broken phases are left for future work.

The CDMFT approximation for the PAM, excluding symmetry breaking,
leads to a self-consistent two-impurity Anderson model (2IAM)
defined by \cite{DeLeo2008a,DeLeo2008,Tanaskovic2011}
\begin{align}
\label{eq:2IAM}
 H_{\textrm{2IAM}} & = \sum_{i\sigma} \epsilonf 
f_{i\sigma}^{\dagger}f_{i\sigma}^\pdag 
+ \sum_{i} 
U f^\dagger_{i\uparrow} f^\pdag_{i\uparrow} 
f^\dagger_{i\downarrow} f^\pdag_{i\downarrow}
\\
\nonumber
 & \quad + \sum_{i  \sigma} V  \bigl(c_{i\sigma}^{\dagger} f_{i\sigma}^\pdag + \mathrm{h.c.}\bigr) 
- \! \sum_{ij\sigma} \, c_{i\sigma}^{\dagger} 
(t \tau^x + \mu \mathbbm{1})_{ij} 
c_{j\sigma}^\pdag
\\ 
 &  \quad
 - \sum_{ij \lambda \sigma} V_{ij\lambda}  \bigl(c_{i\sigma}^{\dagger} a_{\lambda j \sigma}^\pdag + \mathrm{h.c.}\bigr) +
 \sum_{\lambda i \sigma} E_{\lambda i } a^{\dagger}_{\lambda i \sigma}
a_{\lambda i \sigma}^\pdag  \, . \nonumber
\end{align}
Here $i \in\{1,2\}$ labels the two cluster sites
in the ``position basis'', 
and $\mathbbm{1} =  \binom{1\;0}{0\; 1}$, $\tau^x = \binom{0\;1}{1\;0}$.
There are two spinful baths, with 
annihilation operators $a_{\lambda i \sigma}$.  Both baths 
hybridize with both cluster sites, whose assumed equivalence  implies $V_{ij\lambda} = V_{ji\lambda}$.
These couplings, chosen to be real, and the bath energies, $E_{i\lambda}$,
together define the $c$-hybridization function 
\begin{align}
\label{eq:Deltac}
(\Delta_c)_{i j}(\zomega) & = \sum_{\lambda l} \frac{V_{il\lambda} V_{jl\lambda}}{\omegaplus - E_{\lambda l }} \, .
\end{align}

The cluster correlators of the 2IAM are $2\times 2$ matrix 
functions. In the cluster position basis they are given by 
\begin{subequations}
\label{eq:Gcluster}
\begin{flalign}
\label{eq:Gcluster-Gf}
G_f(\zomega) &= \left[ \omegaplus-\epsilonf
 -\Delta_f(\zomega)-
\Sigma_f(\zomega)\right]^{-1} \! ,  
\hspace{-1cm} &
\\
\label{eq:Gcluster-Gc}
\quad \;\; G_c(\zomega) &= \left[ \omegaplus+\mu 
+t\cdot \tau^x 
-\Delta_c(\zomega)-\Sigma_c(\zomega)\right]^{-1}  \! , 
\hspace{-1cm} &
\\[1mm]
\label{eq:Gcluster-fc}
G_\fc (\zomega) & = 
\Sigma_c (\zomega) G_c (\zomega) / V = \Delta_f(\zomega) G_f (\zomega)/V \, , 
\hspace{-1cm} &
\\
\label{eq:Gcluster-Deltaf}
\Delta_f(\zomega) &= V^2\left[ \omegaplus +\mu
+t\cdot \tau^x 
-\Delta_c(\zomega)\right]^{-1} \! , 
\hspace{-1cm} &
\\
\label{eq:Sigmac}
\Sigma_c(\zomega) &= V^2\left[ \omegaplus - \epsilonf
-\Sigma_f(\zomega)\right]^{-1} \! .
\hspace{-1cm} &
\end{flalign}
\end{subequations}
These have to be solved self-consistently, 
by iteratively computing $\Sigma_f$ via an impurity solver 
and re-adjusting the dynamical mean field $\Delta_c$
(see App.~\ref{app:CDMFT} and Refs.~\cite{Kotliar2001,Maier2005}). 

By $\mr{SU}(2)$ spin symmetry the 
 hybridization function $\Delta_{c}$ is spin-diagonal and spin-independent. 
The same is true for $\Delta_f$, which is fully determined by $\Delta_c$.
 Moreover $1 \! \leftrightarrow \! 2$ inversion symmetry ensures that they are linear combinations of $\mathbbm{1}$
and $\tau^x$. 
They can therefore be diagonalized independently of $\omega$ 
using the Hadamard transformation $U_H = \frac{1}{\sqrt 2}\binom{1 \; \pminus 1}{1 \; -1}$, which maps $\tau^x$ to $\tau^z = \binom{1 \; \pminus 0 }{0 \; -1}$. It is thus convenient to correspondingly
transform $H_{\textrm{2IAM}}$, expressing it through  
the bonding/antibonding operators
\begin{equation}
f_{\pm,\sigma}=
\tfrac{1}{\sqrt{2}}\left(f_{1 \sigma} \pm f_{2 \sigma} \right)  ,
\qquad 
c_{\pm,\sigma}=
\tfrac{1}{\sqrt{2}}\left(c_{1 \sigma} \pm c_{2 \sigma} \right)  .
\end{equation}
After reperiodization (discussed in App.~\ref{app:ssec:reperiodization}), these modes represent  Brillouin zone regions centered at $\Gamma=(0,0,0)$ and $\Pi=(\pi,\pi,\pi)$, respectively ~\cite{Stanescu2006a,Stanescu2006}. The labels $\alpha\in\{+,-\}$ on $f_{\alpha\sigma}$ and $c_{\alpha\sigma}$ will thus be called ``momentum'' labels.
The $\pm$ modes are only coupled via the Coulomb interaction term, 
which can only change the total charge in each channel by 0 or $\pm 2$. 
This implies two $\mathbb{Z}_2$ symmetries: 
the number parity operators $\hat P_\pm = \hat N_\pm \! \mod 2$
for the $+$ and $-$ channels are both conserved, with 
eigenvalues $P_{\pm} \in \{0,1\}$. 

\subsection{Numerical renormalization group}
\label{sec:NRG-method}

We solve the 2IAM using the full-density-matrix NRG~\cite{Weichselbaum2007,Weichselbaum2012,Lee2016}. Following Wilson \cite{Wilson1975,Bulla2008}, the bath’s continuous spectrum is discretized logarithmically and 
the model is mapped onto a semi-infinite Wilson chain. We represent the impurity $f$ and $c$ orbitals by sites  $n=-1$ and $0$, respectively, and the bath by sites 
$n \ge 1$. The hopping amplitudes for $n\ge 1$ decay exponentially, 
$\sim \Lambda^{-n/2}$, where $\Lambda >1$ is a discretization parameter.
This energy-scale separation is exploited 
to iteratively diagonalize the model, adding one site at a time while discarding
high-energy states. For a ``length-$N$'' chain (i.e.\ one with largest site index $N$),
the lowest-lying eigenergies have spacing $\sim \Lambda^{-N/2}$. By increasing $N$, one can thus zoom in on 
ever lower energy scales.

We set up the Wilson chain 
in the momentum basis, in which the $\pm$ modes are coupled only via the interaction term on site $-1$. To reduce
computational costs, we use an interleaved chain \cite{Mitchell2014,Stadler2016} of alternating $+$ and $-$ orbitals.
(Interleaving slightly lifts degeneracies, if present, 
 of the sites being interleaved---but this is not an issue here, since the $\pm$ modes are non-degenerate due to $\tau^z$ contributions to hopping terms. Indeed, we have double-checked, especially close to the QCP, 
that our interleaved results are reproduced when using a computationally more costly standard Wilson chain geometry.) We exploit the SU(2) spin, U(1) charge and both $\mathbbm{Z}_2$ parity symmetries in our NRG calculations using the QSpace tensor library~\cite{Weichselbaum2012a,Weichselbaum2020}, further reducing computational costs. Together with interleaving, this allows us to achieve converged data using a fairly small NRG discretization parameter of $\Lambda=3$ while keeping $N_{\textrm{keep}} \leq 20,000$ SU(2) multiplets. Because spectra close to the QCP can be quite sensitive to z-shifting, we refrained from z-averaging.

For further methodological details on achieving
DMFT self-consistency and reperiodization, see App.~\ref{app:CDMFT}.

%
\begin{figure}[t!]
\centerline{\includegraphics[width=\linewidth]{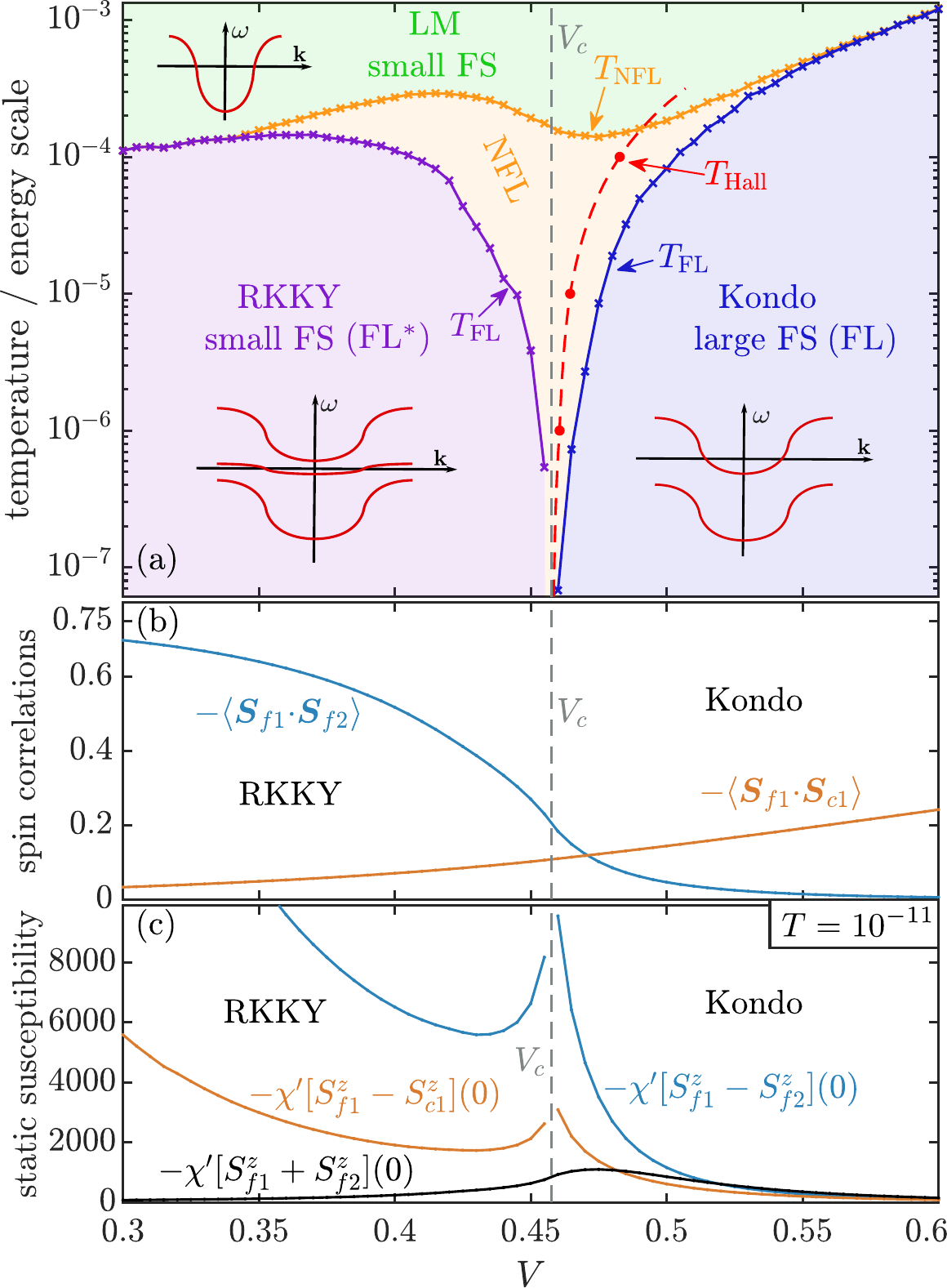}}
\caption{%
(a) Paramagnetic phase diagram of the PAM as a function of the $c$-$f$ hybridization $V$ and temperature $T$ (on a log scale).
At $T=0$, there are two distinct phases, 
the RKKY phase and the Kondo phase, separated by a QCP at $\Vc = 0.4575(25)$. At $T>0$, we find four different regimes, labeled LM (local moment), 
RKKY, Kondo and NFL, connected via smooth crossovers. For the first three, 
associated band structures are depicted schematically in insets.  
 (See Fig.~\ref{fig:Ak_lin_FS} and Figs.~\ref{fig:A_vs_V}--\ref{fig:A_vs_T_V046} for the CDMFT band structure results.) The crossovers are charaterized by the temperature scales $\TNF$ (orange solid line) and $\TFL$ (purple and blue solid lines) below which NFL and FL behaviors emerge, respectively. 
The scales $\TNF (V)$ and $\TFL (V)$ were determined by analyzing dynamical susceptibilities of the self-consistent 2IAM at $T=0$, as explained in Sec.~\ref{sec:dynamical-susceptibilities} and App.~\ref{app:Chi_App}. The red dots marked $T_{\mr{Hall}}$ (and the guide-to-the-eye red dashed line) indicate how the crossover from a large FS in the Kondo phase to the small FS in the RKKY phase evolves with temperature.
$T_{\mr{Hall}}$ was determined
by analyzing the Hall coefficient similar to Ref.~\onlinecite{Paschen2004} (see Fig.~\ref{fig:Hall_PAM} and App.~\ref{app:THall_PAM}).
(b) Equal-time intersite $f$-$f$ spin correlation $\braket{\vec{S}_{f1}\!\cdot\!\vec{S}_{f2}}$ and local $c$-$f$ spin correlation $\braket{\vec{S}_{f1}\!\cdot\!\vec{S}_{c1}}$, and (c) static susceptibilities for the total $f$ spin $S_{f1}^z + S_{f2}^z$, the intersite staggered $f$-$f$ spin $S_{f1}^z - S_{f2}^z$ and the local staggered $f$-$c$ spin $S_{f1}^z - S_{c1}^z$ of the effective 2IAM, all plotted at $T=10^{-11}$ as a function of $V$.
}
\label{fig:Escale}
\end{figure}

\section{Phase diagram}%
\label{sec:phasediagram}
Based on a detailed study of the dynamical properties of various local operators, described in Sec.~\ref{sec:dynamical-susceptibilities},
we have established a phase diagram for the PAM, shown in Fig.~\ref{fig:Escale}(a). While its generic structure has been known for a long time~\cite{Doniach1977,Senthil2003,Coleman2007,Coleman2010,Yamamoto2010},
 we reach orders of magnitude lower temperatures and better energy resolution than previously possible and characterize the various regimes through a detailed analysis of real-frequency correlators. We first focus on zero temperature, 
involving two distinct phases separated by a QCP,
 then discuss finite-temperature behavior, involving smooth crossovers between several different regimes. 
\textit{Zero temperature:}
At $T=0$, we find two phases when tuning $V$, separated by a QCP at $\Vc=0.4575(25)$.
For $V > \Vc$, we find a Kondo phase, where the $f$ and $c$ bands are hybridized to form a two-band structure and the correlation between $f$-orbital local moments in the effective 2IAM is weak, as shown by the blue line in Fig.~\ref{fig:Escale}(b) and discussed below.
This phase is a FL, with a Fermi surface (FS) whose volume satisfies the Luttinger sum rule~\cite{Oshikawa2000,Luttinger1960} when counting both the $f$ and $c$ electrons (see Fig.~\ref{fig:LT_PAM}, to be discussed later).
We henceforth call a FS \textit{large} or \textit{small}
if its volume accounts for both $c$ and $f$ electrons, or only
$c$ electrons, respectively. The Kondo phase is adiabatically connected to the case of $U = 0$ and $V > 0$ and is thus a normal FL.
For $V < \Vc$, we find a RKKY phase, where the local moments at nearest neighbours have strong antiferromagnetic correlations (see blue line in Fig.~\ref{fig:Escale}(b) and its discussion below), while $\mr{SU}(2)$ spin symmetry is conserved by construction.
This phase, too, is a FL, with a small FS 
accounting only for $c$ electrons. While this phase thus appears to violate Luttinger sum rule, it still obeys a more general version of that rule~\cite{Dzyaloshinskii2003,Senthil2003,Hazra2021}: we find a surface of poles of $\Sigma_{f\bk}(\zomega = 0)$ (Luttinger surface), which, together with the FS, accounts for the total particle number. We will discuss this in detail in Secs.~\ref{sec:Ak_vs_w} and~\ref{sec:Luttinger}. 
In the RKKY phase, the FS coincides with the FS of the free $c$ band but the effective band structure differs from that of a free $c$ band:
there are \textit{three} bands (see Fig.~\ref{fig:Ak_lin_FS} and \Fig{fig:A_vs_V} below, to be discussed later), including a narrow QP band crossing the Fermi level. This narrow QP band is responsible for the FL behavior we observe in the RKKY phase. Based on Ref.~\onlinecite{Fabrizio2023} revealing that Luttinger surfaces may define spinon Fermi surfaces, we conjecture that the RKKY phase is a fractionalized FL (\FLstar)~\cite{Senthil2003,Senthil2004}. We will explore this conjecture in more detail in future work.

Figure~\ref{fig:Escale}(b) shows the equal-time $f$-$f$ intersite spin correlators $\braket{\vec{S}_{f1}\!\cdot\!\vec{S}_{f2}}$ and the local $c$-$f$ $\braket{\vec{S}_{f1}\!\cdot\!\vec{S}_{c1}}$ of the effective 2IAM at $T=10^{-11}$, plotted as functions of $V$. $\braket{\vec{S}_{f1}\!\cdot\!\vec{S}_{f2}}$ smoothly evolves from $\simeq 0$ at $V=0.6$ deep in the Kondo phase to $\simeq -0.75$ deep in the RKKY phase. On the other hand, the absolute value of $\braket{\vec{S}_{f1}\!\cdot\!\vec{S}_{c1}}$ smoothly decreases when going from the Kondo phase to the RKKY phase. This shows that the Kondo phase  is dominated by local $c$-$f$ correlations, indicative of spin screening, and only has weak non-local $f$-$f$ spin correlations. By contrast, the RKKY phase is dominated by non-local antiferromagnetic $f$-$f$ spin correlations and has only weak local $c$-$f$ correlations. Note that equal-time spin correlations of the self-consistent 2IAM are continuous across the QCP and  do not show non-analytic behavior at $\Vc$. Rather, we will see below that the QCP is characterized by a zero crossing of the effective bonding $f$ level (see discussion of Fig.~\ref{fig:QP_para} below). This is accompanied by a sharp jump of the FS and the appearance of a dispersive pole in the $f$ self-energy (see the discussion of Figs.~\ref{fig:A_vs_V},~\ref{fig:SEk_vs_V} and~\ref{fig:LT_PAM} below).
Figure~\ref{fig:Escale}(c) shows static susceptibilities for 
the total $f$-spin $S^z_{f1}+S^z_{f2}$, the staggered intersite $f$-$f$-spin $S^z_{f1}-S^z_{f2}$, and the staggered local $f$-$c$-spin $S^z_{f1}-S^z_{c1}$, plotted versus $V$ at $T=10^{-11}$. (Susceptibilities
are defined in \Eq{eq:dynamicalchi} below.) While the total $f$-spin susceptibility evolves smoothly across the QCP, 
both staggered susceptibilities, which are related to intersite $f$-$f$ singlet formation and Kondo singlet formation, respectively, show
singular behavior near the QCP. This suggests that the latter arises from a competition between intersite $f$-$f$ singlet formation and Kondo singlet formation. Further, both staggered susceptibilities become very large deep in the RKKY phase, reflecting the tendency of this phase towards antiferromagnetic order.
\textit{Finite temperature:}
When the temperature is increased from zero, both FL phases cross over, at a $V$-dependent scale $\TFL(V)$, to an intermediate NFL critical regime, characterized by the absence of coherent QP (see Fig.~\ref{fig:A_vs_T_V046} and Sec.~\ref{sec:FiniteSizeSpectra}). Importantly, the scale $\TFL(V)$ vanishes when  $V$ approaches $\Vc$ from either side, thus the NFL regime extends all the way down to $T=0$ at the QCP. With increasing temperature, the NFL regime crosses over, at a scale $\TNF(V)$ (larger than $\TFL(V)$) , to a local moment (LM) regime, which is adiabatically connected to $V=0$. There, free $c$ electrons are decoupled from $f$ orbital local moments,  resulting in a one-band structure. The crossover scales $\TNF$ and $\TFL$ can be extracted from an analysis of dynamical susceptibilities at $T=0$ (see Sec.~\ref{sec:dynamical-susceptibilities} and App.~\ref{app:Chi_App}).
To make qualitative contact with experimental results on \YRS~\cite{Paschen2004}, we also show a scale $T_{\mr{Hall}}$, which marks the 
crossover between large and small FS based on analyzing the Hall coefficient in a way which closely resembles the analysis done in Ref.~\onlinecite{Paschen2004} (see Fig.~\ref{fig:Hall_PAM} and App.~\ref{app:THall_PAM}). In qualitative agreement with the experimental data of Ref.~\onlinecite{Paschen2004}, Fig.~3, $T_{\mr{Hall}}$ depends on the tuning parameter  ($V$ in our case, $B$-field in Ref.~\onlinecite{Paschen2004}) and bends towards the Kondo side of the phase diagram.

\section{Two-stage screening}
\label{sec:two-stage screening}

The presence of two crossover scales, $\TNF$ and $\TFL$, implies that the 
evolution, with decreasing energy, from unscreened local $f$ moments to a fully-screened FL regime evolves through two stages. In this section
we study this evolution from two perspectives, focusing first on NRG finite-size spectra (\Sec{sec:FiniteSizeSpectra}), then on the dynamical properties of various local susceptibilities at $T=0$ 
(\Sec{sec:dynamical-susceptibilities}).

\subsection{Finite size spectra \label{sec:FiniteSizeSpectra}}%
%

\begin{figure}[t!]
\centerline{\includegraphics[width=\linewidth]{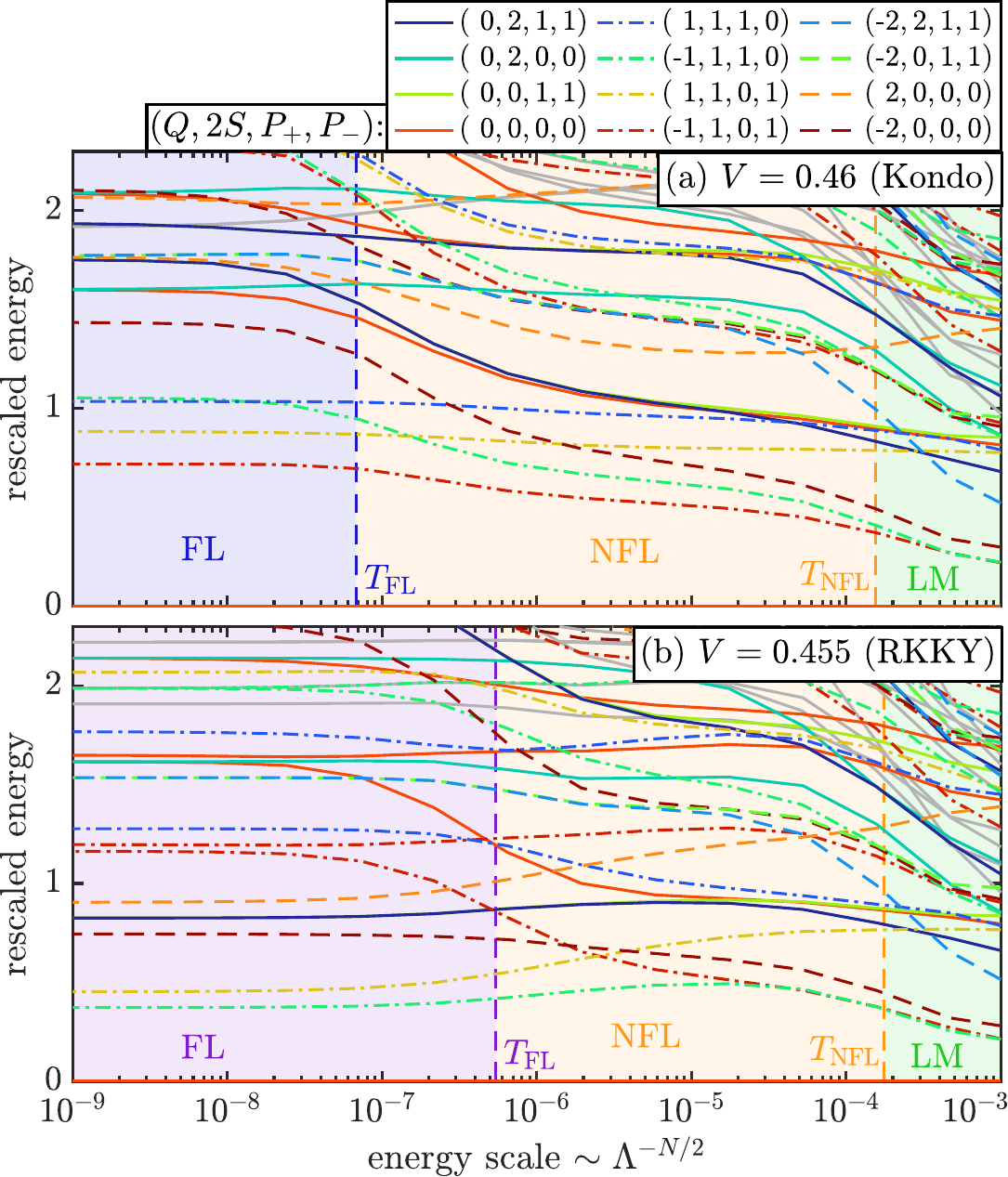}}
\caption{NRG flow diagrams for the self-consistent effective 2IAM at $T=0$, in (a) the Kondo phase at $V=0.46$  and (b) the RKKY phase at $V=0.455$, both close to the QCP. The rescaled eigenenergies are plotted as functions of the energy scale $\Lambda^{-N/2}$, for odd values of $N$. 
The vertical dashed lines indicate the scales $\TFL$ and $\TNF$. Quantum numbers for selected energy levels are indicated in the legend. Energy levels with total charge $|Q|=0$, $1$ or $2$ are shown using solid, dash-dotted or dashed lines, respectively.}
\label{fig:FiniteSize}
\end{figure}

%
We begin our discussion of the physical properties of the different regimes shown in Fig.~\ref{fig:Escale}(a) by studying NRG energy-level flow diagrams of the self-consistent effective 2IAM. Such diagrams show the (lowest) rescaled eigenenergies, $\Lambda^{N/2} E_i(N)$, of a length-$N$ Wilson chain 
as a function of the energy scale $\Lambda^{-N/2}$.  Conceptually, the $E_i(N)$
form the finite-size spectrum of the impurity plus bath
in a spherical box of radius $R_N \propto \Lambda^{N/2}$, centered on the
impurity \cite{Wilson1975,Delft1998b}: with increasing $N$, the finite-size level spacing, $\sim 1/R_N$, decreases exponentially. The resulting 
flow of the finite-size spectrum is stationary ($N$ independent)
while $\Lambda^{-N/2}$ lies within an energy regime governed by
one of the fixed points, but changes when $\Lambda^{-N/2}$ traverses a
crossover between two fixed points. 
We label eigenenergies by the conserved quantum numbers $(Q, 2S, P_+, P_-)$, where $Q$ is the total charge relative to the ground state, $S$
the total spin, and $P_\pm \in \{0,1\}$ the number parity eigenvalues
in the $\pm$ sectors.
\textit{Kondo phase:}
Fig.~\ref{fig:FiniteSize}(a) shows the NRG flow diagram for the self-consistent 2IAM at $T=0$ and $V=0.46$, which is in the Kondo phase
close to the QCP. The ground state has quantum numbers $(0,0,0,0)$. 
As already indicated in the discussion of Fig.~\ref{fig:Escale}(a), we find a FL at low energy scales and a NFL at intermediate energy scales. In the FL region below $\TFL>\Vc$, the low-energy many-body spectrum can be constructed from the lowest particle and hole excitations. These come with quantum numbers $(\pm1,1,1,0)$ and $(\pm1,1,0,1)$ for the bonding and anti-bonding channel, respectively, 
with the $P_{\pm}$ quantum numbers identifying the channel containing
the excitation. The low-energy many-body spectrum can then be generated by stacking these single-particle excitations, leading to towers of equally spaced energy levels, characteristic for FL fixed points~\cite{Bulla2008}. This directly shows that the Kondo phase is a FL featuring a QP spectrum at low energies.
At intermediate energy scales between $\TFL$ and $\TNF$, the effective 2IAM flows through the vicinity of a NFL fixed point. Our calculations strongly suggest that 
this NFL fixed point also governs the low-energy behavior of
the QCP at  $T=0$ and $V=\Vc$.  We thus identify this NFL fixed point with the critical fixed point of the QCP in the two-site CDMFT approximation. As will be pointed out in subsequent sections and summarized in Sec.~\ref{sec:other_models}, this fixed point shares several similarities with the NFL fixed points of the two-channel Kondo model (2CKM)~\cite{Andrei1984,Tsvelick1984,Tsvelick1985,Sacramento1989,Sacramento1991,Affleck1991,Affleck1991a},  the two impurity Kondo model (2IKM)~\cite{Jones1988,Jones1989,Affleck1992,Affleck1995,Jones2007,Mitchell2012a,Mitchell2012} and the 2IAM \textit{without} self-consistency~\cite{Sakai1990,Sakai1992,Fabrizio2003,DeLeo2004}, which is closely related to the 2IKM.  One may therefore argue that it is not surprising to find such a NFL fixed point also in a self-consistent solution of the 2IAM. On the other hand, the NFL fixed points of the 2IKM and the 2IAM are known to be \textit{unstable} to breaking $\pm$ mode degeneracy or particle-hole symmetry~\cite{Sakai1990,Sakai1992,Silva1996,Jones2007}. Further, 
for the 2IKM and the 2IAM, the RKKY interaction has to be inserted by hand as a direct interaction because
if $\pm$ mode symmetry and particle-hole symmetries are present (as needed to make the NFL fixed point accessible), then these symmetries
 prevent dynamical generation of an antiferromagnetic RKKY interaction~\cite{Fye1994}.  
It has therefore been argued that this NFL fixed point is artificial and not observable in real systems~\cite{Fye1994,Lechtenberg2017,Eickhoff2020}. From this perspective,
the behavior found here for our effective \textit{self-consistent} 2IAM 
is indeed unexpected and remarkable: although it lacks particle-hole symmetry or $\pm$ mode degeneracy and we do not insert the RKKY interaction by hand, it
evidently \textit{can} be tuned close to a QCP controlled by a 
NFL fixed point with 2IKM-like properties. 

We note in passing
that self-consistency is crucial to reach the NFL fixed point---we checked that naively tuning $V$ without self-consistency leads to a continuous crossover without a QCP
(see also App~\ref{app:Hybfun} for more details). It is not entirely clear to us why the self-consistency stabilizes the NFL fixed point, though we suspect the Luttinger sum rule~\cite{Luttinger1960,Oshikawa2000} to play a crucial role. We will discuss the Luttinger sum rule in more detail in Sec.~\ref{sec:Luttinger}.
We also remark that for the non-self-consistent 2IKM 
mentioned above, frequency-independent or only weakly frequency-dependent hybridization functions were used in the analyses concluding that its NFL fixed point requires some special symmetries. 
By contrast, for the self-consistent 2IAM studied here, the self-consistent hybridization functions acquire a rather strong frequency dependence in the vicinity of the KB--QCP and appear to become singular at the QCP itself (see App~\ref{app:Hybfun}).
Regarding the energy level structure of the self-consistent 2IAM, we did \emph{not} find obvious similarities to the NFL fixed point of the 2IKM, i.e.\ the level structure seems quite different. 
\textit{RKKY phase:}
Fig.~\ref{fig:FiniteSize}(b) shows the NRG flow diagram for the self-consistent 2IAM at $V=0.455<\Vc$, which is in the RKKY phase close to the QCP. Again, a NFL is found at intermediate energies and a FL fixed point at low energies. This directly establishes that the RKKY phase, too, is a FL described by a QP spectrum at low energies. However, note that close to the QCP, the level structure of the FL fixed point in the RKKY phase ($V=0.455$) is quite different from that in the Kondo phase ($V=0.46$). This suggests that these Fermi liquids are not smoothly connected. Indeed, 
we will see in Sec.~\ref{sec:Luttinger} that their FS volumes
differ. This  implies different scattering phase shifts and hence different NRG eigenlevel structures, consistent with the fact that the level structures of Figs.~\ref{fig:FiniteSize}(a) and \ref{fig:FiniteSize}(b) differ strikingly in the FL regimes on the left. 

\begin{figure*}[tb!]
\includegraphics[width=\textwidth]{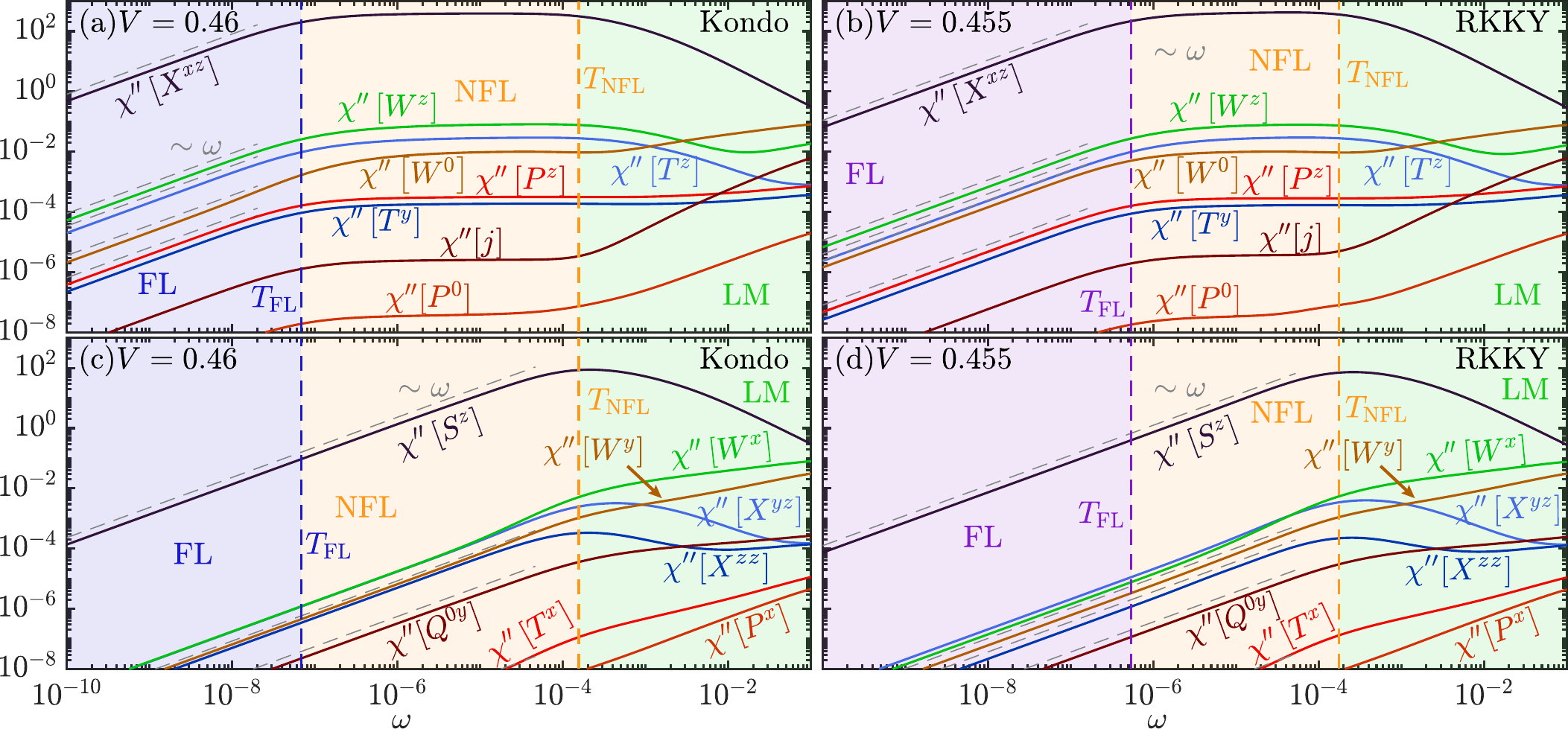}
\caption{%
Dynamical impurity susceptibilities $\chi''(\omega)$ for various operators defined in Eq.~\eqref{eq:operators}, computed in (a,c) the Kondo phase, at $V = 0.46$, and (b,d) the RKKY phase at $V=0.455$, both at $T = 0$, close to the QCP. For the susceptibilities collected in (c,d), the $\chi''$ curves exhibit a maximum around $\TNF$; for those in (a,b), they instead exhibit a plateau between $\TFL$ and $\TNF$. This plateau indicates that the FL is reached in a two-stage screening process, leading to the emergence of NFL behavior at intermediate energy scales. Grey dashed lines are guides-to-the-eye 
for $\sim \omega$ behavior.
\label{fig:Chi_QCP}
}
\end{figure*}

%

%
\subsection{Dynamical susceptibilities} \label{sec:dynamical-susceptibilities}
%
%

The characteristic level structure of RG fixed points 
governs the behavior of dynamical properties at $T=0$,
causing striking crossovers at the scales $\TFL$ and $\TNF$. In this section, we extract these from the dynamical susceptibilities of local operators.

Let $O$ be a local operator acting non-trivially only on the cluster impurity in the self-consistent 2IAM. 
We define its dynamical susceptibility as
\begin{align}
\label{eq:dynamicalchi}
\chi[O](\omegaR) = - \mi \int_0^\infty \mathrm{d} 
t \, e^{\mi \omegaRinExponent t}
\langle [O(t),O^\dagger(0)]\rangle \, , 
\end{align}
$\langle \cdot \rangle = \mr{Tr} (\rho \, \cdot)$ denotes a thermal expectation value.
When $\omega$ lies within an energy range governed by a specific
fixed point, the imaginary part of such a susceptibility
typically displays power-law behavior, $\chi'' [O] (\omega) = - \frac{1}{\pi} \im \chi[O](\omegaR) \sim \omega^{\alpha}$. 
When $\omega$ traverses the crossover region between fixed points, the exponent $\alpha$ changes, indicating a change in the degree of screening of the local fluctuations described by $O$. 
A log-log plot of $\chi''$ vs.\ $\omega$ thus 
consists of straight lines with slope $\alpha$ in regions
governed by fixed points, connected by peaks or kinks (see Fig.~\ref{fig:Chi_QCP}). 
We thus define the crossover scales $\TNF$ and $\TFL$ via the position of these kinks, as described below. (A systematic method for determining the kink positions
is described in App.~\ref{app:Chi_App}.) When discussing finite-temperature properties 
in later sections, we will see that the scales so obtained also serve as crossover scales separating low-, intermediate- and high-temperature regimes.

\begin{figure}[bt!]
\includegraphics[width=\linewidth]{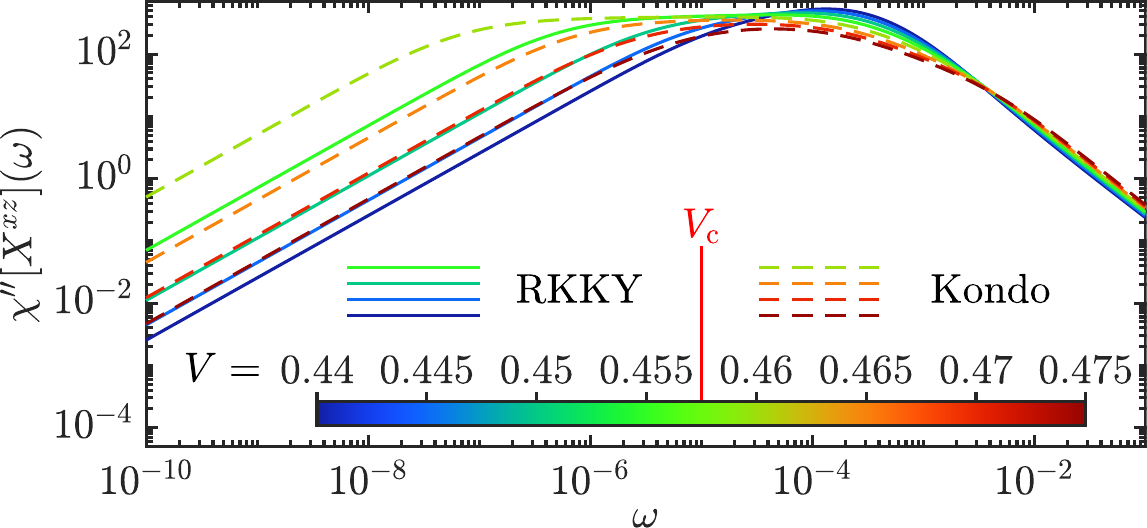}
\caption{%
Evolution of $\chi''[X^{xz}]$ with $V$ across the QCP at $T=0$. 
Solid lines: RKKY phase; dashed lines: Kondo phase.
As $V \to \Vc$ from either above or below
such that $\TFL \to 0$, the dashed and dotted plateaus both extend to ever lower scales (they would coincide at $\Vc$, where $ \TFL = 0$), demonstrating that the KB--QCP is continuous. 
\label{fig:Xxz_QCP}
}
\end{figure}

We have computed $\chi''[O](\omega)$ for the following local cluster operators, defined in the momentum-spin basis, with indices $\alpha \in \{+,-\}$ and $\sigma \in \{\uparrow, \downarrow\}$:
\begin{subequations}
\label{eq:operators} 
\begin{align}
\label{eq:operators_b}
T^{a} &= \tfrac{1}{2} f_{\alpha\sigma}^\dagger 
\tau_{\alpha\alpha'}^a  \delta_{\sigma\sigma'} 
f_{\alpha'\sigma'}  & &(\textrm{momentum})
\\ 
S^{b} &= \tfrac{1}{2} f_{\alpha\sigma}^\dagger 
\delta_{\alpha\alpha'}  \sigma_{\sigma\sigma'}^b 
f_{\alpha'\sigma'}  & &(\textrm{spin}) 
\\
X^{ab} &= \tfrac{1}{2} f_{\alpha\sigma}^\dagger 
\tau_{\alpha\alpha'}^a \sigma_{\sigma\sigma'}^b
f_{\alpha'\sigma'}  & &(\textrm{spin-momentum}) 
\\
W^{a} &= f_{\alpha\sigma}^\dagger 
\tau_{\alpha\alpha'}^a \delta_{\sigma\sigma'} c_{\alpha'\sigma'} + \textrm{h.c.}   & &(\textrm{hybridization})
\\ 
\label{eq:operators_P}
P^{a} &= \tfrac{1}{\sqrt{2}} f^\pdag_{\alpha\sigma} 
\tau^a_{\alpha\alpha'}  \mi \sigma^y_{\sigma\sigma'}  
f^\pdag_{\alpha'\sigma'}   & &(\mathrm{singlet\, pairing})
\\
Q^{ab} &=  f^\pdag_{\alpha\sigma} f^\pdag_{\alpha'\sigma'} \hat{\tau}^{a}_{\alpha\alpha'} \hat{J}^{b}_{\sigma\sigma'}  & &(\mathrm{triplet\, pairing}) 
\\
j &= -\mi t e (c^\dagger_{1\sigma} c^\pdag_{2\sigma} - \textrm{h.c}) & &(\textrm{current}) 
\end{align}
\end{subequations}
Here, sums over repeated indices are implied,  $\tau^a$ and $\sigma^b$ are Pauli matrices in the momentum and spin sectors, respectively, 
$\mi \sigma^y = \binom{\pminus 0 \; \; 1}{-1 \; \; 0}$, and $\hat{J}^{b}$ are SU(2) generators in the triplet representation. 
These operators can also be expressed in the position spin basis
via the Hadamard transformation $U_H$, which maps $\tau^x \to \tau^z$, $\tau^y \to -\tau^y$ and $\tau^z \to \tau^x$. For example, $T^{z}$ 
can be expressed as 
\begin{equation}
T^{z} = \sum_{\alpha,\alpha'=\pm} \tfrac{1}{2} 
f_{\alpha\sigma}^\dagger 
 \tau_{\alpha\alpha'}^z 
 f_{\alpha'\sigma}  = 
 \sum_{i,j=1}^{2} \tfrac{1}{2} 
 f_{i\sigma}^\dagger 
 \tau_{ij}^x 
 f_{j\sigma} \, ,
\end{equation}
describing $f$ hopping between sites 1 and 2.
Similarly, $S^z$ describes the total $f$-electron spin, 
$X^{xz} = S^z_{f1} - S^z_{f2}$ the staggered magnetization, 
{$W^{z}= \sum_{ij} f_{i\sigma}^\dagger \tau^{x}_{ij} c_{j\sigma} + \mr{h.c.}$} 
nearest-neighbor $f$-$c$ hybridization, and $P^{z}$ $f$-electron nearest-neighbor singlet pairing.
Fig.~\ref{fig:Chi_QCP} shows various $\chi''$ susceptibilities 
at $T=0$ for the choices $V=0.46$ [(a,c)] and $V=0.455$ [(b,d)], in the Kondo and RKKY phases close to the QCP, respectively.
For $\omega <\TFL$, all $\chi''$'s decrease linearly with decreasing $\omega$, indicative of FL behavior. Hence all local fluctuations are fully screened in that energy window, leading to well-defined Fermi-liquid QPs.
By contrast, for $\TFL < \omega < \TNF$ only the $\chi''$'s in panels (c) and (d) (e.g.\ $\chi''[S^{z}]$) decrease with decreasing $\omega$, 
while the ones in panels (a) and (b) (e.g.\ $\chi''[T^{z}]$, $\chi''[X^{xz}]$,  $\chi''[W^{z}]$ and  $\chi''[P^{z}]$) all traverse plateaus. These plateaus are reminiscent of 
those found for $\chi_{\mathrm{2CK}}''[S^{z}]$ of the overscreened, 
$S=\tfrac{1}{2}$ two-channel Kondo model (2CK),  and 
for $\chi_{\mathrm{2IKM}}''[S_1^{z}-S_2^{z}]$ of the two-impurity Kondo model (2IKM) in their respective NFL regimes (see Fig.~\ref{fig:2CK_vs_PAM} below).
This implies that the total spin in the $\pm$ basis is screened, whereas the momentum, spin-momentum, pairing and hybridization fluctuations are \textit{overscreened}, yielding the intermediate NFL. 
The $\chi''$ curves at $T = 0$ in Fig.~\ref{fig:Chi_QCP} clearly demonstrate that when $V$ is close to $\Vc$, the screening process evolves through two stages, characterized by $\TNF$ and $\TFL$. Precisely \textit{at} the critical point, where $\TFL\!=\!0$, the plateaus would extend all the way to zero. Conversely, when $V$ is tuned away from the QCP, $\TFL$ tends toward $\TNF$ (see Fig.~\ref{fig:Escale}(a)). The two scales merge for $|V - \Vc| \gtrsim 0.1$, where the $\chi''$ plateaus have shrunk to become mere peaks, shown for $\chi''[X^{xz}]$ in Fig.~\ref{fig:Xxz_QCP}.
The curves in Fig.~\ref{fig:Xxz_QCP} further illustrate that the KB--QCP is continuous because $\chi''[X^{xz}]$ evolves smoothly without any discontinuities across the KB--QCP. We find similar behavior for the other dynamical susceptibilities shown in Fig.~\ref{fig:Chi_QCP}.
In a companion paper~\cite{Gleis2023a}, we show that those $\chi''$ which exhibit a plateau [those shown in Fig.~\ref{fig:Chi_QCP}(a,b)]
exhibit logarithmic $\omega/T$ scaling in the NFL region while the corresponding static susceptibilities are singular at the KB--QCP, where the NFL region extends down to $T=0$. The fact that many static susceptibilities with different symmetries diverge at the QCP suggests that many different, possibly competing symmetry breaking orders may be
possible in the vicinity of the QCP. Which order prevails (if any) will be carefully studied in future work.
%

\section{Single-particle properties -- preliminaries}
\label{sec:single-particle-preliminaries}

The fact that $\TFL \to 0$ at the QCP (\Fig{fig:Escale}(a)) indicates a breakdown of the FL and QP concepts at the KB--QCP. Experimental evidence for such a breakdown is found in the sudden reconstruction of the FS~\cite{Paschen2004,Friedemann2010}, and the divergence of the effective mass~\cite{Custers2003} at the KB--QCP. It is to date not fully settled how this should be understood~\cite{Coleman2001,Kirchner2020}.
In the next two sections, \Sec{sec:SingleParticleCluster} and \ref{sec:Ak_vs_w},  we revisit such questions, exploiting the ability of CDMFT--NRG to explore very low temperatures and frequencies. This will allow us to clarify the behavior of spectral functions and self-energies in unprecedented detail. We discuss their cluster versions for the self-consistent 2IAM in \Sec{sec:SingleParticleCluster}, and the corresponding lattice functions for the PAM in \Sec{sec:Ak_vs_w}.

In the present section we set the stage for this analysis by summarizing, for future reference, some well-established considerations regarding single-particle properties. We first recall standard expressions for low-frequency  expansions of correlators and self-energies in the PAM, and the definition of its Fermi surface (\Sec{sec:self-energy-expansions}). Even though our focus here is on the PAM, we note that low-frequency expressions similar to those reviewed in \Sec{sec:self-energy-expansions} can be obtained for the Kondo lattice using slave particles~\cite{Senthil2003,Senthil2004,Si2014,Kirchner2020}. In \Sec{sec:Kondo-breakdown}, we discuss possible scenarios how the parameters appearing in the low-frequency expansions in \Sec{sec:self-energy-expansions} behave in the vicinity of a KB--QCP.

\subsection{Fermi-liquid expansions, Fermi surface, Luttinger surface}
\label{sec:self-energy-expansions}

In what follows, we will often refer to the low-energy expansions,
 applicable in 
 conventional FL phases, of the the functions $G_{x\bk} (\zomega)$ and $\Sigma_{x\bk}(\zomega)$ ($x=f,c$) defined in \Eqs{eq:Glatt}. We list them here for future reference. 

\textit{Fermi-liquid expansions:}
 For $\zomega$ below a characteristic FL scale, $|\zomega| \ll \TFL$, 
 and for $\bk$ close to the FS, the self-energies can be expanded as 
 \cite{Abrikosov1963}
\begin{align}
\label{eq:Sigmax_lin_omega}
\Sigma_{x\bk}(\zomega) = \re\Sigma_{x\bk}(0) + \zomega
\re \Sigma'_{x\bk}(0) + \delta \Sigma_{x\bk} (\zomega)
\, ,
\end{align}
where $\Sigma'(\zomega) = \partial_\zomega \Sigma(\zomega)$
and $\delta\Sigma_{x\bk}(\zomega)$ is of order of $\mc{O} (\zomega^2/\TFL^2)$.
Moreover, analyticity of $G_{x\bk}(z)$ in the upper half-plane requires $\im \, \Sigma_{x\bk}(\zomega) < 0$ for $\im\, z >0$.
For $z=\omegaR$ this implies $\im \, \Sigma_{x\bk}(\omegaR) < 0$ 
and $\re \Sigma'_{x\bk}(0) \le 0$ (the latter follows since
 $\delta \Sigma_{x\bk}(\zomega) \sim \zomega^2$ implies $\im \, \Sigma_{x\bk}(0) = 0$).

The expansion coefficients determine the so-called free QP energies,  weights and the effective hybridization, 
\begin{subequations}
\label{subeq:Zestar}
\begin{align}
\label{eq:epsilonxk}
\epsilonxk^\ast 
&= Z_{x\bk}\bigl[\epsilonxk 
+\re\Sigma_{x\bk}(0)\bigr] \, ,
\\ 
\label{eq:Zxk}
Z_{x\bk} & = \bigl[1-\re \Sigma'_{x\bk}(0) \bigr]^{-1} \, ,
\\ 
\label{eq:Vstark}
V^{\ast}_{\bk} & = \sqrt{Z_{f\bk}} V \, ,
\end{align}
\end{subequations}
with $\epsilonxk = \epsilonf$ or $\epsilonck$ for $x=f,c$. 
Since $\re \, \Sigma'_{x \bk}(0) \le 0$, 
the QP weights satisfy $Z_{x\bk} \le 1$.
The QP energies and weights,
in turn, appear in the low-frequency expansions of 
 $\Sigma_c$ and  $G_c$ (cf.~\Eqs{eq:Glatt}): 
\begin{subequations}
\label{subeq:SigmaGc_through_Sigmaf}
\begin{align}
\label{eq:Sigmac_through_Sigmaf}
\Sigma_{c\bk}(\zomega) & = 
\frac{(V^{\ast}_{\bk})^2}{\omegaplus-\epsilon_{f\bk}^\ast}  + 
\mathcal{O}(\zomega^2/\TFL^2)  \, , 
\\
\label{eq:Sigmac_through_Gigmaf}
G_{c\bk}(\zomega) & = \frac{Z_{c\bk}}{\omegaplus-\epsilon_{c\bk}^\ast}  + 
\mathcal{O} (\zomega^2/\TFL^2) \, .
\end{align}
\end{subequations}
Evidently, the low-energy expansion of $\Sigma_c$ is
fully determined by that of $\Sigma_f$, with 
\begin{align}
\label{eq:Sigma_c-coefficients-through-Sigma_f}
  \re \, \Sigma_{c \bk}(0) &= - \frac{V^{\ast 2}_{\bk}}{\epsilonfk^\ast}, \quad 
  \re \, \Sigma'_{c \bk}(0) = - \frac{V^{\ast 2}_{\bk}}{\epsilonfk^{\ast 2}} \, , \\
  \label{eq:Z_c-through-Sigma_f}
  Z_{c\bk} &= 
  \left[ 1+ \frac{V^{\ast 2}_{\bk}}{\epsilonfk^{\ast 2}}\right]^{-1} 
  = \frac{\epsilonfk^{\ast 2}}{\epsilonfk^{\ast 2} + V^{\ast 2}_{\bk}} \, .
\end{align}
These expressions make explicit how $c$-$f$ hybridization effects $c$ electrons: the sign of the energy shift $ \re \, \Sigma_{c\bk}(0)$ is determined by and opposite to that of $\epsilonfk^\ast$;  and $\re \, \Sigma'_{c \bk}(0)$ changes from zero to negative, thereby decreasing $Z_{c\bk}$ below $1$ and causing $c$ electron mass enhancement~\cite{Abrikosov1963}. Since $\epsilonfk^{\ast} \sim Z_{f\bk}$ and $V^{\ast}_{\bk} \sim \sqrt{Z_{f\bk}}$, Eq.~\eqref{eq:Z_c-through-Sigma_f} implies $Z_{c\bk} \sim Z_{f\bk}$ when $Z_{f\bk} \ll 1$.

\textit{Fermi surface:} 
\label{sec:definition-Fermi-surface}
Next, we recall the definition of the FS. For the PAM, this is not entirely trivial, since the correlators $G_c$ and $G_f$ are not independent, but coupled through \Eqs{eq:Glatt}. 

We focus on $T=0$ (else the FS is not sharply defined).
If there is no hybridization, 
$V=0$, the situation is simple: the partially-filled $c$ band is metallic and the half-filled $f$ band a Mott insulator. 
Then, the  
FS  comprises essentially only $c$ electrons and is defined by the conditions
\cite{Nishikawa2018,Abrikosov1963}
\begin{equation}
  \label{eq:define-Fermi-surface}
  \epsilon^\ast_{c\bk}  = 0 \, ,  \qquad 
  Z_{c \bk} > 0 \, , 
  \qquad \im \,\Sigma_{c\bk}(0) = 0  \, .
\end{equation}
The first condition identifies the FS as the locus of $\bk$ points in the Brillouin zone for which the free QP energy vanishes; the second states that the QP occupation should exhibit an abrupt jump when this surface is crossed ($Z_{c \bk}$ governs the size of this jump); and the third requires the QP scattering rate to vanish at the FS. Together, they imply that the FS is the locus of $\bk$ points at which $G^{-1}_{c\bk}(0) = 0$.

More generally, for $V \neq 0$, we start from the matrix form, $G_\bk(\zomega)$, of the combined $f$ and $c$ correlators [\Eq{eq:Gmatrix}]. Then, the FS   is defined as the locus of $\bk$ points for which some eigenvalue of $G^{-1}_\bk(0)$ vanishes. 
Thus, all points on the FS satisfy the condition $\det \! \left[ G_\bk^{-1}(0) \right] = 0$, or
\begin{align}
\epsilonck 
 \bigl(\epsilonf 
 +  \Sigma_{f\bk}(0) \bigr) -V^2 = 0 \, . 
 \label{eq:detG=0}
\end{align}
For $V \to  0$, 
either the first or second factor on the left must vanish,
implying $\epsilonck 
= 0$ or $\epsilonf + 
\Sigma_{f\bk}(0) = 0$, respectively. 
The first condition defines the bare FS for the $c$ band; the second
condition is never satisfied for the case of present interest,  where the bare $f$ electrons form a half-filled Mott insulator with $\mu$ lying within the gap. 

If $V$ is non-zero, \Eq{eq:detG=0} implies that \textit{both} of the following inequalities hold on the FS (assuming that $\Sigma_{f\bk}(0)$ does not diverge there):
\begin{align}
  \label{eq:finite-V-conditions}
  \epsilonck 
  \neq 0 \, ,  \quad  \epsilonf 
  + \Sigma_{f\bk}(0) \neq 0 \, . 
\end{align}
Thus, the bare and actual FS do not intersect.
Dividing \Eq{eq:detG=0} by the first or second factor,  we obtain
 ~\cite{DeLeo2008,DeLeo2008a}
\begin{align}
\label{eq:Gfk_Gck_0}
G_{f \bk}^{-1}(0) = 0 \, ,  
\qquad 
G_{c \bk}^{-1}(0) = 0 \, .
\end{align}
 These two conditions are equivalent, in that one implies the other, via
 \Eq{eq:detG=0}. Moreover, 
the second inequality in \eqref{eq:finite-V-conditions}  ensures that $\Sigma_{c\bk}(0)$ does not diverge, hence the  second condition  in \eqref{eq:Gfk_Gck_0} implies \Eqs{eq:define-Fermi-surface}.  We thus conclude that \Eqs{eq:define-Fermi-surface} define the FS also for nonzero $V$.

\textit{Luttinger surface:}
A second surface of importance is the \textit{Luttinger surface} (LS)
\cite{Dzyaloshinskii2003,Fabrizio2020,Skolimowski2022}. For the PAM it is defined \cite{DeLeo2008a} as the locus of $\bk$ points for which $\Sigma_{f\bk}(z)$ has a pole at $z=0$. This definition, together with Eqs.~(\ref{eq:Glatt-Gf},\ref{eq:Glatt-Sigmac}), implies that the following relations hold on the  LS: 
\begin{align}
\label{eq:DefinitionLS}
|\Sigma_{f\bk}(0)| = \infty \, ,  \quad 
G_{f\bk}(0) = 0\, , \quad \Sigma_{c\bk}(0) = 0 \, . 
\end{align}
The first relation just restates the definition of the LS; the second should be contrasted with the relation $G_{f\bk}(0) = \infty$ holding on the FS;
and the third implies, via  \eqref{eq:epsilonxk}, that 
$\epsilonck^\ast = Z_{c\bk} \epsilonck$, i.e.\ on the LS, the renormalized
dispersion is obtained from the bare one purely by rescaling, without any shift. If the LS coincides with the bare FS, then the FS, too, coincides with the bare $c$-electron FS ($\epsilon_{c\bk} = 0$).

Note that $|\Sigma_{f\bk}(\omega)|$ can only diverge 
at isolated frequency values, not on extended frequency intervals, since the frequency integral over its spectral function must be finite. Terefore, when $|\Sigma_{f\bk}(0)|$ diverges, $|\Sigma'_{f\bk}(0)|$ diverges too. By Eq.~\eqref{eq:Zxk},
it follows that $Z_{f\bk} = 0$ on the LS.

The behavior of $Z_{c\bk}$ depends on how strongly $\Sigma_{f\bk}(\zomega)$
diverges for $\zomega \to 0$.  For example, suppose $\Sigma_{f\bk}(\zomega)
\sim \zomega^{-\alpha}$ for some $\alpha >0$. Then, 
\Eq{eq:Glatt-Sigmac} implies $\Sigma_{c\bk} \sim \zomega^{\alpha}$, 
$\Sigma'_{c\bk} \sim \zomega^{\alpha-1}$. Thus, we obtain 
$Z_{c\bk} \sim \zomega^{1-\alpha} \to 0$ or  $\in (0,1)$ or $=1$ for $\alpha <1$
 or $=1$ or $>1$, respectively, i.e.\ the $c$-QP weight may or may not be renormalized.
We show results for the FS and LS in \Sec{sec:Ak_vs_w}, and discuss their volumes together with Luttinger's sum rule in \Sec{sec:Luttinger}.

\subsection{Kondo breakdown}
\label{sec:Kondo-breakdown}

\begin{figure}[t!]
\centerline{\includegraphics[width=\linewidth]{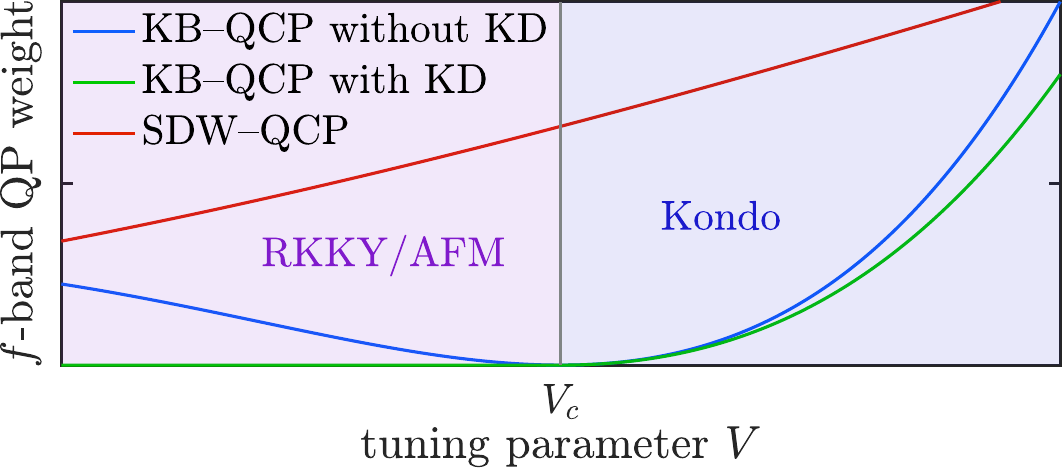}}
\caption{%
Sketch of the generic $f$-band QP weight in different scenarios for quantum criticality.
In the conventional SDW--QCP (red), the QP weight is finite at the QCP. At a KB--QCP, it is zero at the QCP. In a type (1) scenario with KD (green), it remains zero in the RKKY phase; in a type (2) scenario without KD (blue), it is finite there.
\label{fig:QPweight_sketch} }
\end{figure}

In the introduction, we have qualitatively described the Kondo breakdown scenarios that have been proposed to characterize the KB--QCP.
For future reference, we here distinguish KB scenarios of two types: (1) a KB--QCP \textit{with} Kondo destruction (KD), defined below, and (2) a KB-QCP \textit{without} KD.
In both scenarios, the $f$ electron quasiparticle weight $Z_f$ decreases to zero when approaching the KB--QCP from the Kondo side. However, 
in (1) $Z_f$ remains zero on the RKKY side (i.e.\ all Kondo correlations have been destroyed, hence the monicker ``KD''), whereas in (2), $Z_f$ is non-zero on the RKKY side, too (i.e.\ some Kondo correlations survive there). (Thus, our nomenclature distinguishes between KB, which happens only \textit{at} the critical point, and KD, which, in a type (1) scenario, happens throughout the RKKY phase.) We emphasize that (2) is different from the Hertz--Millis type SDW--QCP where the QP weight is also non-zero at the QCP. Fig.~\ref{fig:QPweight_sketch} sketches the different scenarios. Most of the scenarios for the KB--QCP proposed in the literature are of type (1), with KD. By contrast, we find a KB--QCP of type (2), without KD. We summarize below the concepts used to describe the HF problem and in particular how type (1) and (2) scenarios differ.
%

%

%
(i) \textit{Hybridization gap:} The hybridization of $c$ and $f$ electrons leads to a well-developed pole in $\Sigma_{c\bk}(\zomega)$, 
called hybridization pole ({\HP}), lying at energies well above the FL scale $\TFL$. It manifests itself as a strong peak 
in $- \im \Sigma_{c\bk}(\zomega)$, causing the corresponding $c$-electron spectral functions to exhibit distinctive gaps or pseudogaps, called hybridization gaps ({\HG}s). It occurs irrespective of whether the $T=0$ phase is Kondo or RKKY correlated and is present for temperatures both above and below $\TFL$.  The formation of a \HG has been observed in many experiments, as reviewed in the introduction. Note that while in the non-interacting PAM ($U=0$), the \HG is positioned at $\epsilon_{f\bk}^{\ast}$, this is not the case in the interacting case: the \HG forms at scales which can much larger than the FL scale (in our case, it forms at $\TNF$, see next section), i.e. $\epsilon_{f\bk}^{\ast}$ and the position of the \HG are renormalized differently by interactions. 

(ii) \textit{Kondo phase:}
In the Kondo phase, $Z_{f\bk}$ is non-zero for all $\bk$. 
The presence of Kondo correlations is 
phenomenologically described \cite{Si2014,Kirchner2020}
by Eq.~\eqref{eq:Sigmac_through_Sigmaf}. This situation can phenomenologically be interpreted as arising through an effective hybridization with strength $V^{\ast}_{\bk}$ (sometimes referred to as ``amplitude of static Kondo correlations''~\cite{Si2014,Kirchner2020}) of $c$ electrons with an effective $f$ band with dispersion $\epsilon^\ast_{f\bk}$.
This effective hybridization shifts the $c$-electron Fermi surface from its $V=0$ form, defined by
$\epsilonck 
= 0$,
to a form defined by $\epsilon^{\ast}_{c\bk} = 0$, i.e.
\begin{equation}
\label{eq:kF}
\epsilonck 
- \frac{V^{\ast 2}_{\bk}}{\epsilon^{\ast}_{f\bk}} = 0 \, .
\end{equation}
Therefore the FS volume changes, reflecting the influence of $f$ orbitals.
Within the Kondo phase, $V^\ast_{\bk}$ remains non-zero but continuously approaches zero as the KB--QCP is approached. 
Moreover, in the Kondo phase  the ratio $\left(V^\ast_{\bk}\right)^2\! /\epsilon^{\ast}_{f\bk}$ remains essentially constant as long as $V^\ast_{\bk}$ is finite because both $(V_\bk^\ast)^2 \sim Z_{f\bk}$ and $\epsilon^\ast_{f\bk} \sim Z_{f\bk}$ (c.f. Eq.~\eqref{eq:epsilonxk}). Therefore, the FS will remain basically unchanged in the Kondo phase, even very close to the KB--QCP.

Two comments are in order. First, in general, the first term in \Eq{eq:Sigmac_through_Sigmaf} usually does not represent an actual pole of $\Sigma_{c\bk}$: it was derived assuming $|\zomega| \ll \TFL$
and $\bk$ close to the FS, whereas $\epsilon^\ast_{f\bk}$ typically lies outside that window (i.e.\ for $\zomega \to \epsilon^\ast_{f\bk}$, \Eqs{eq:Sigmax_lin_omega} and \eqref{eq:Sigmac_through_Sigmaf} no longer apply). Second, $\epsilon^\ast_{f\bk}$ is 
not directly related to the hybridization gaps, as already mentioned in point (i) above: the latter are determined by pseudogaps of $G_{c\bk}$ of \eqref{eq:Glatt-Gc}, and since these lie at high energies of order $\pm \TNF$, their 
positions are not governed by
\Eq{eq:Sigmac_through_Sigmaf}, but rather by the general form \eqref{eq:Glatt-Sigmac} of $\Sigma_{c\bk}$ (see Sec.~\ref{sec:Ak_vs_w} below). 
(iii) \textit{Kondo breakdown:} As summarized in Refs.~\cite{Si2014,Kirchner2020}, the following behavior is expected when approaching the KB--QCP from the Kondo phase: $V^\ast_\bk$, or equivalently $Z_{f\bk}$, decreases continuously to zero, hence the low-energy hybridization becomes weaker, while the ratio $\left(V^\ast_{\bk}\right)^2\!/\epsilon^{\ast}_{f\bk}$ and thus the FS remain constant, and different from the bare $c$-electron FS. Since $Z_{c\bk} \sim Z_{f\bk}$ if $Z_{f\bk} \ll 1$ [c.f. Eq.~\eqref{eq:Z_c-through-Sigma_f} and its discussion], e.g. close to the KB--QCP, both QP weights are expected to continuously decrease to zero when approaching the KB--QCP.
At the KB--QCP, $V^\ast$ vanishes, i.e.\ low-energy hybridization and thus Kondo correlations break down. 

(iv) \textit{RKKY with Kondo destruction:}
In the type (1) KD scenario~\cite{Si2014,Kirchner2020},
$V^\ast_\bk$,  or equivalently $Z_{f\bk}$, \textit{remains} zero in the RKKY phase, i.e.\ Kondo correlations remain absent, i.e.\ they have been destroyed. By \Eq{eq:kF}, that implies the FS reduces to the bare $c$-electron one, accounting for $c$ electrons only. All in all, the FS jumps across the KB--QCP due to Kondo destruction. 
Since $Z_{f\bk} = 0$ in this scenario, Eq.~\eqref{eq:Sigmax_lin_omega} for the $f$-electrons and therefore also Eq.~\eqref{eq:Sigmac_through_Sigmaf} does not apply anymore. 
Eq.~\eqref{eq:Sigmax_lin_omega} may however still apply for the $c$-electrons [see also the discussion below Eq.~\eqref{eq:DefinitionLS}], leading to QP mass enhancement due to the existence of $c$-$f$ hybridization at finite frequencies. This is sometimes referred to as ``dynamical Kondo correlations''~\cite{Gegenwart2008,Cai2019}.
The type (1) KD scenario emerges from the Kondo lattice model both in a slave-particle~\cite{Senthil2003,Senthil2004} 
or an EDMFT treatment~\cite{Si2001,Si2003,Zhu2003,Glossop2007}.

(v) \textit{RKKY without Kondo destruction:}
Here, we describe the type (2) scenario of a Kondo breakdown without Kondo destruction in the RKKY phase. We emphasize that also in this scenario, the QP weights become zero at the KB--QCP and the FS is small in the RKKY phase. Nevertheless, in the RKKY phase the $f$-electron QP weight is non-zero at the FS. In the type (2) scenario, $Z_{f\bk}$ becomes zero only at a LS [see Eq.~\eqref{eq:DefinitionLS}], where the $f$-electron self-energy diverges.
If the LS does not coincide with the FS (the converse would require significant fine-tuning), this implies non-zero $Z_{f\bk}$ at the FS and $c$-$f$ hybridized QPs also in the RKKY phase.

The type (2) scenario described above is not so unusual:
there is growing evidence that Mott insulators described beyond the single-site DMFT approximation 
generically feature momentum-dependent Mott poles~\cite{Pairault1998,Pudleiner2016,Wagner2023},
with a singular part of the $f$ self-energy of the form~\cite{Wagner2023},
\begin{equation}
\label{eq:SE_Mott}
\Sigma_{f,\mr{singular}} \sim \frac{1}{\omegaplus - \epsilon^\ast_{\Sigma\bk}} \, .
\end{equation}
Here, $\epsilon^\ast_{\Sigma\bk}$ is related to the free dispersion with renormalized parameters 
and opposite sign of the hopping amplitudes~\cite{Wagner2023}. Such a self-energy therefore
features a LS, defined by $\epsilon^\ast_{\Sigma\bk} = 0$.
It has been suggested that the LS is the defining feature of Mott phases and that this feature is stable to perturbations~\cite{Huang2022,Zhao2023,Horava2005}.
The LS of a Mott phase should therefore in principle be 
stable to small hybridization with a metal, provided the hybridization strength is not too large, resulting in a OSMP.
$Z_{f\bk}$ is only zero at the LS where $\Sigma_f$ diverges. If the LS and the FS do not coincide, it follows that $Z_{f\bk}$ is non-zero at the FS.

A comment is in order regarding single-site DMFT or EDMFT, where the Mott pole is not momentum dependent and the QP weight is 
zero throughout the whole BZ. In single-site DMFT, this phase is not stable to inter-orbital hybridization~\cite{Medici2005,Kugler2022}. As a result,
single-site DMFT will always describe a Kondo phase at $T=0$~\cite{Medici2005}. By contrast, OSMPs described by EDMFT seem to be 
stable to inter-orbital hybridization~\cite{Yu2017}, leading to the type (1) scenario described above.
In our own CDMFT-NRG studies of the PAM,
 we find a KB scenario of type (2). In \Sec{sec:SingleParticleCluster}, we will establish this by a detailed study the low-energy behavior of the self-consistent effective 2IAM. 
 There, the role of $\bk$ is taken by $\alpha =\pm$, i.e.\ we will show that
\begin{equation}
\label{eq:static_Kondo_cluster}
\Sigma_{c\alpha}(\zomega) = \frac{V^{\ast2}_{\alpha}}{\omegaplus -\epsilon_{f\alpha}^\ast} +  \mathcal{O}\bigl(z^2/\TFL^2\bigr)
\end{equation}
is valid, with $V_\alpha^\ast \neq 0$ on both sides of the QCP.
When $V$ approaches $\Vc$ from either side, $V^\ast_\alpha$ approaches 0, 
leading to a breakdown of Kondo correlations at the KB--QCP. 
We will find that the FS reconstruction at $\Vc$ is caused by a sign change of $\epsilon^{\ast}_{f+}$, as explained in the \Sec{sec:SingleParticleCluster}. The consequences for the lattice model are established in Sec.~\ref{sec:Ak_vs_w}, and for the Luttinger sum rule in  Sec.~\ref{sec:Luttinger}.
%

%
\begin{figure*}[t!]
\centerline{\includegraphics[width=\textwidth]{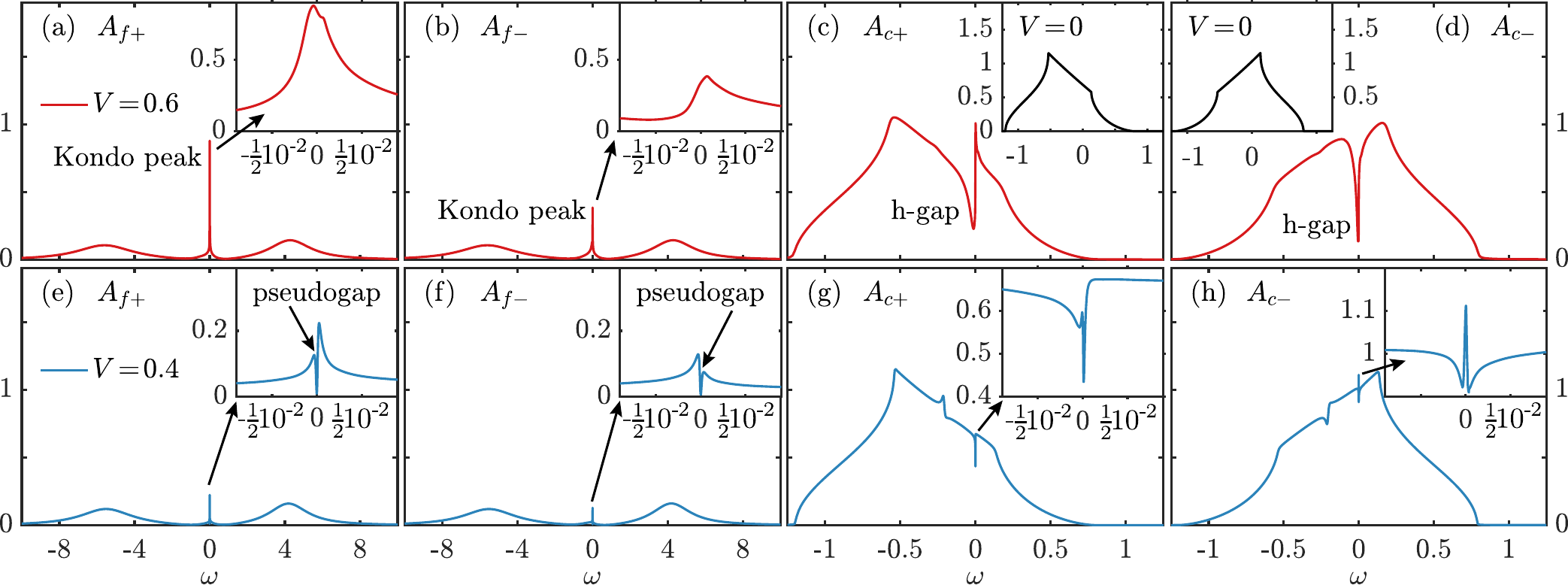}}
\caption{%
Cluster spectral functions at $T = 0$, 
for $V=0.6$ (top row, deep in the Kondo phase)
and for $V=0.4$ (bottom row, deep in the RKKY phase).
Panels (a,b,e,f) and (c,d,g,h) show the $f$ and $c$-electron spectral functions, respectively.
The insets in
 (c,d) show the $c$-electron spectral functions at $V=0$; the insets in all other panels show a zoom into the low-frequency region.
\rule[-5mm]{0mm}{0mm} 
}
\label{fig:A_vs_V_T1e-11_Cluster_lin} 
\end{figure*}

\section{Single-particle properties -- cluster \label{sec:SingleParticleCluster}}%

In this section we discuss the single-particle properties of the self-consistent 2IAM, focusing on  the spectral functions and  retarded self-energies of both $c$ and $f$ orbitals.
A discussion of the corresponding momentum-dependent lattice properties follows in Sec.~\ref{sec:Ak_vs_w}.
We will argue that the KB quantum phase transition is a continuous orbital selective Mott transition~(OSMT) at $T=0$. The Kondo phase is a normal metallic phase
while in the RKKY phase, the $f$ electrons are in a Mott phase, i.e.\ this phase is an orbital selective Mott phase~(OSMP).
The defining feature of the OSMP is
a momentum-dependent pole in the $f$-electron self-energy, \textit{not} a single-particle gap. 
The RKKY phase is thus \textit{not} an orbital selective Mott \textit{insulator}.
Indeed, we find that even in the OSMP, the $f$ electrons exhibit a finite QP weight due to finite hybridization with the $c$ electrons.

In sections~\ref{sec:SpectralFunctionsCluster} and~\ref{sec:SpectralFunctionsCluster_OSMT}, we discuss spectral functions and self-energies at $T=0$ on the real-frequency axis, exploiting the capabilities of 
NRG to resolve exponentially small energy scales. 
We then investigate QP properties in more detail in Sec.~\ref{sec:QP_properties}: we clearly show that both the $c$ and $f$-electron QP weights are finite in both
the Kondo and RKKY phases, but vanish at the KB--QCP.
Finally, in Sec.~\ref{sec:SpectralFunctionsCluster_Tdep}, we discuss finite temperature properties, showing that $c$-$f$ hybridization is already fully developed around $\TNF$, whereas QP coherence
and self-energy poles are only fully formed at $\TFL$. 

\subsection{Cluster spectral functions at $T=0$: overview\label{sec:SpectralFunctionsCluster}}%
%

In this subsection, we
provide a phenomenological overview over
the cluster spectral properties 
of the self-consistent 2AIM at $T= 0$ as functions of $V$;
details and physical insights follow in Sec.~\ref{sec:SpectralFunctionsCluster_OSMT}.
We adopt the $\pm$ basis, where $G_f$ and $G_c$ from Eq.~\eqref{eq:Gcluster} are both diagonal, and study  $A(\omega) = -\tfrac{1}{\pi} \im G(\omegaR)$ for both $f$ and $c$ orbitals.  (When referring to $A_f$ below, we mean both components, $A_{f\pm}$, and likewise for $A_c$.)
Our results for $A_{f}$ and $A_c$ are shown  in Fig.~\ref{fig:A_vs_V_T1e-11_Cluster_lin}
on a linear frequency scale to provide a coarse overview.
We enumerate some of 
their characteristic features,
proceeding from high- to low-frequency features.

(i) \textit{Hubbard peaks, band structure:}
Figures~\ref{fig:A_vs_V_T1e-11_Cluster_lin}(a,b,e,f) and (c,d,g,h) show the spectral functions $A_f(\omega)$ and $A_c(\omega)$
for different $V$ on a linear frequency scale, for the ranges $\omega \in[-10,10]$
and $\omega \in[-1.25,1.25]$, respectively,
containing all significant spectral weight. 
$A_f$ has two Hubbard bands around $\omega \simeq \pm 5 = \pm U/2$. They are almost independent of $V$. Moreover, 
they show the same structure for $A_{f+}$ and $A_{f-}$, implying
that these high-energy features are momentum independent.
By contrast, $A_c$ has no Hubbard bands since the $c$ electrons 
do not interact,
with spectral weight only in the range of the non-interacting bandwidth, $\omega \in[-1.2,0.8]$. Its shape mimics that obtained for
$V=0$ (insets of Figs.~\ref{fig:A_vs_V_T1e-11_Cluster_lin}(c,d)), reflecting the bare $c$-electron band structure, except for some sharp structures at intermediate and low frequencies, discussed below.

(ii) \textit{Center of $c$ band:}
For $A_c$, the highest-frequency sharp feature furthest from $\omega =0$ lies
at $\omega \simeq -0.2$, the middle of the bare 
$c$-electron band. 
This feature 
is prominently developed deep in the RKKY phase at $V=0.4$ (Fig.~\ref{fig:A_vs_V_T1e-11_Cluster_lin}(g,h))
but almost invisible deep in the Kondo phase at $V=0.6$ (Fig.~\ref{fig:A_vs_V_T1e-11_Cluster_lin}(c,d)). It
is due to scattering of $c$ electrons by antiferromagnetic fluctuations and
reflects a tendency towards antiferromagnetic order in the RKKY phase.
Though our CDMFT setup excludes such order, it
does find strong antiferromagnetic 
correlations in the RKKY phase (see $\braket{\vec{S}_{f1}\!\cdot\!\vec{S}_{f2}}$ in Fig.~\ref{fig:Escale}(b)), causing enhanced scattering
of $c$ electrons at the band center.

(iii)  \textit{Kondo peaks, hybridization gaps:}
Deep in the Kondo phase at $V=0.6$ (red lines), 
$A_f$ shows a sharp Kondo peak near $\omega \simeq 0$, while $A_c$ has a distinct dip, known as hybridization  gap~(\HG, see also Sec.\ref{sec:single-particle-preliminaries}), 
at a small, negative value of $\omega$. Both these features are indicative of strong $c$-$f$ hybridization and coherent QPs. 
By contrast,  deep in the RKKY phase at $V=0.4$ (blue lines),
the Kondo peak in $A_f$ has disappeared,
giving rise to a pseudogap (see the insets of Fig.~\ref{fig:A_vs_V_T1e-11_Cluster_lin}(e,f)),
and the \HG in $A_c$ has become very weak.
Nevertheless, we will see in Sec.~\ref{sec:SpectralFunctionsCluster_OSMT} that even deep in the RKKY phase, $c$-$f$ hybridization is present even at low energies
and the $f$-electron QP weight is finite in this phase.
This leads to a sharp peak inside the h-gaps of $A_c$ (see the insets of Fig.~\ref{fig:A_vs_V_T1e-11_Cluster_lin}(g,h)).
As will be discussed in more detail in Sec.~\ref{sec:Ak_vs_w}, this sharp feature reflects a narrow QP band with a FS close to $\Pi = (\pi,\pi,\pi)$.

(iv) \textit{Momentum dependence:}
The spectral functions and self-energies show several qualitative and/or quantitative differences between the $+$ and $-$ channels
(different Kondo peak heights, different \HG shapes, etc.)
 Such channel asymmetries reflect the fact that our system is electron-doped---the Fermi surface lies closer to the $\Gamma$ point than the $\Pi$ point, causing a stronger  $c$-$f$ hybridization (encoded in 
$\Sigma_{c \pm}$) for bonding than anti-bonding orbitals.
This asymmetry leads to different  behavior for momenta near $\Gamma = (0,0,0)$ and $\Pi = (\pi,\pi,\pi)$.
However, these asymmetries in the \textit{spectral functions} are not necessarily indicative of non-local correlations.
Especially deep in the Kondo phase, the channel asymmetries are mostly due to the single-particle dispersion and are not non-local \textit{self-energy} effects.
We discuss this in more detail in Sec.~\ref{sec:SpectralFunctionsCluster_OSMT}.

%

\begin{figure*}[t!]
\centerline{\includegraphics[width=\textwidth]{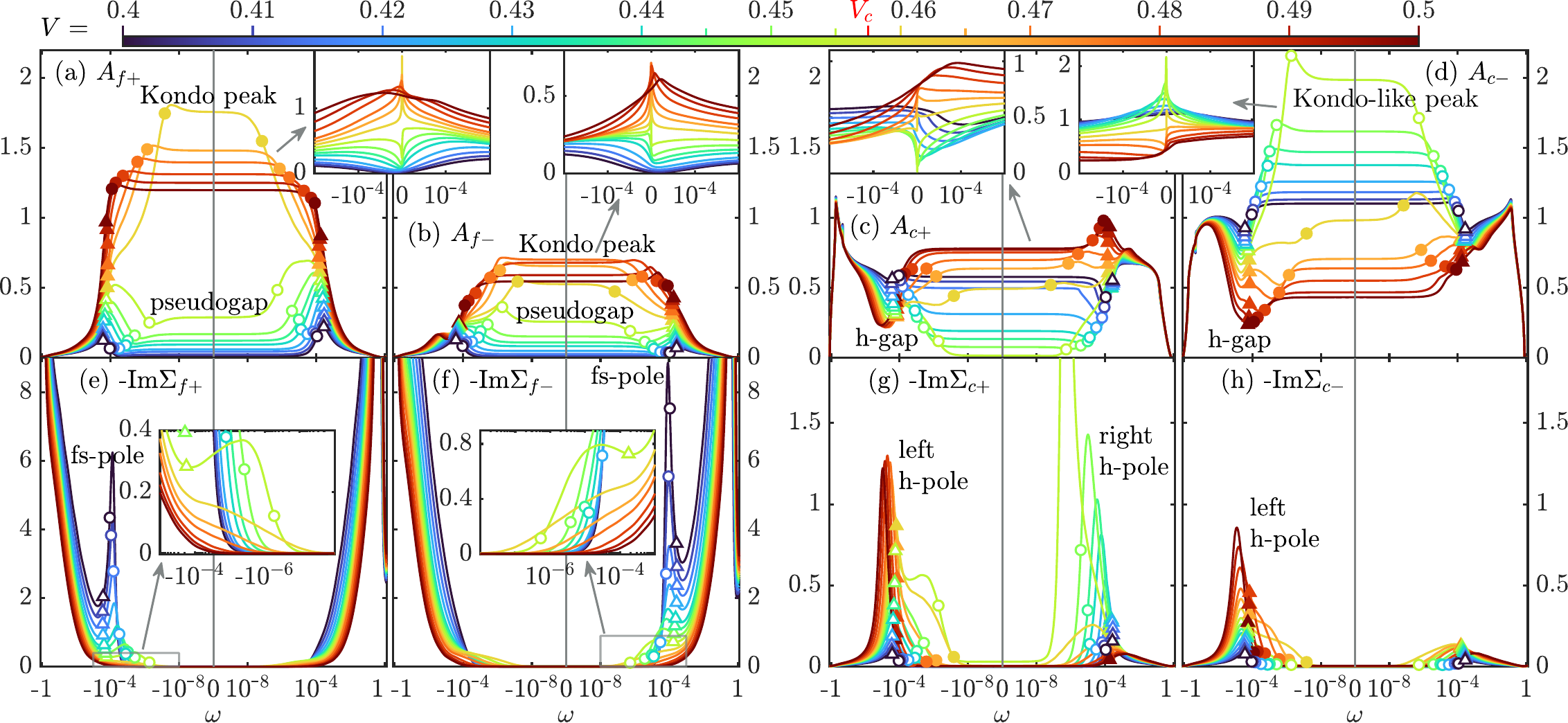}}
\caption{%
Evolution of cluster spectral functions $A_{x\alpha}(\omega)$ and retarded self-energies $\Sigma_{x\alpha}(\omegaR)$ at $T=0$ as $V$ is tuned across the QCP. Colored curves correspond to $V$ values marked by ticks on the color bar. 
A symmetric log scale with $10^{-10}< |\omega| < 1.25 $ is used.
On such a scale, the plateaus seen for all curves for very low frequencies, 
$|\omega| < 10^{-8}$, demonstrate that no new features arise 
in that range. Triangles and circles mark the crossover scales $\pm\TNF$ and $\pm\TFL$, respectively, 
with filled (open) symbols identifying curves in the Kondo (RKKY) phase.
The insets in (a-d) show the spectral functions on a linear frequency scale for $|\omega|<2\!\cdot\! 10^{-4}$.}
\label{fig:A_vs_V_T1e-11_Cluster_log} 
\end{figure*}

\subsection{Cluster spectral functions and self-energies at $T=0$: details\label{sec:SpectralFunctionsCluster_OSMT}}%

Next, we discuss the main spectral features relevant to KB physics, 
referring to Fig.~\ref{fig:A_vs_V_T1e-11_Cluster_log}. It shows both the spectral functions $A_{x\alpha} (\omega)$ and retarded self-energies $\Sigma_{x\alpha} (\omegaR)$ using a symmetric logarithmic frequency scale with $|\omega| > 10^{-10}$. Figures~\FigAlog(a--d) show this evolution 
for $A_f$ and $A_c$ while Figs.~\FigAlog(e--h) show the corresponding self-energies.

(i) \textit{Self-energy poles for $\Sigma_{f}$:}
The most important feature is the pole in $\Sigma_{f}$~(denoted as \SP), which is present in the RKKY phase but not in the Kondo phase, see Fig.~\ref{fig:A_vs_V_T1e-11_Cluster_log}(e,f).
This \SP in the RKKY phase is indicative of Mott physics~\cite{DeLeo2008,DeLeo2008a} present in the $f$ band but not in the $c$ band
(see also the discussion in Sec.~\ref{sec:Kondo-breakdown}). 
This brings us to one of our main conclusions: \textit{the RKKY phase is an OSMP and the KB quantum phase transition is an OSMT}.
Moreover, the \SP continuously disappears when approaching the KB--QCP from the RKKY phase. This further shows that the KB quantum phase transition is a \textit{continuous} OSMT (see also Fig.~\ref{fig:Xxz_QCP} and its discussion). We also note that we found no coexistence region, which further underpins our conclusion of a continuous QPT.

Our conclusion that the KB is an OSMT matches the conclusion of previous CDMFT plus ED studies of the PAM using the same parameters~\cite{DeLeo2008a,DeLeo2008}. Nevertheless, the considerably improved accuracy of our NRG impurity solver compared to the ED impurity solver used there yields new conceptional insights and reveals new emergent physics.
 
The \SP in the RKKY phase is positioned at 
a negative frequency for $\Sigma_{f+}$ (at $\omega \simeq -\TFL$) and at a positive frequency for $\Sigma_{f-}$ (at $\omega \simeq \TFL$). 
Therefore, its position depends on momentum, i.e.\ it is \textit{dispersive}. A dispersive \SP is a generic feature of Mott phases in finite dimensions $d<\infty$~\cite{Pairault1998,Wagner2023} (see also Sec.~\ref{sec:Kondo-breakdown}).
By contrast
in the $d\to\infty$ limit (or equivalently in the single-site DMFT approximation)
where the self-energy and thus also the \SP is momentum independent, the OSMP is \textit{not} stable against interorbital hopping~\cite{Kugler2022} (i.e.\ against finite $c$-$f$ hybridization $V$ in the present context).
The  momentum resolution provided by 2-site CDMFT, though coarse, is therefore a crucial ingredient to stabilize the OSMP. 
We will show in Sec.~\ref{sec:Ak_vs_w} that after reperiodization of the CDMFT self-energy, the dispersive nature of the \SP leads to a 
reperiodized self-energy with a continuous $\bk$-dependence of the form of Eq.~\eqref{eq:SE_Mott}. Based on the results of Ref.~\onlinecite{Fabrizio2023}, the \SP\ can be associated with emergent spinon excitations; its emergence  therefore suggests a fractionalization of the $f$-electron. Since the position of the sf-pole is momentum dependent, the emergent spinon is dispersive.
Even though the dispersive \SP\ is the most pronounced momentum-dependent feature of $\Sigma_f$, 
more subtle momentum-dependent features are responsible for the NFL physics close to the QCP:
In the Kondo phase, shoulder-like structures show up in $\Sigma_f$ below $\TFL<|\omega| < 10^{-4}$ [see insets of Figures~\ref{fig:A_vs_V_T1e-11_Cluster_log}(e,f)].
These are more pronounced for $\Sigma_{f-}$ than for $\Sigma_{f+}$, leading to momentum-dependent scattering rates 
at the corresponding energy scales.

The features of $A_f$~[Figs.~\FigAlog(a,b)], $A_c$~[Figs.~\FigAlog(c,d)] and $\Sigma_c$~[Figs.~\FigAlog(g,h)] can largely be understood in terms of a continuous OSMT.
In the following, we enumerate and describe the main spectral and self-energy features and discuss their connection to the presence or absence of {\SP}s in $\Sigma_f$.
We follow the evolution, with decreasing $V$, from the Kondo phase through the  QCP into the RKKY phase,
noting the following salient features:

(ii) \textit{From Kondo peak to pseudogap for $A_{f\pm}$:} 
In the Kondo phase, the Kondo peak of $A_{f+}$ lies slightly below $- \TFL$, that of $A_{f-}$ slightly below $\TFL$.
As $V$ decreases towards $\Vc$,  the Kondo peaks of both $A_{f+}$ and $A_{f-}$ shift towards zero and become higher and narrower [Figs.~\FigAlog(a,b)], leaving behind shoulder-like structures at $\pm \TNF$ (marked
by triangles). 
The Kondo peak of $A_{f+}$ is higher than that of $A_{f-}$, reflecting the fact that in the Kondo phase, the FS is positioned closer to the $\Gamma$ point than to the $\Pi$ point.
When $V$ crosses $\Vc$,  the \KP abruptly changes into a pseudogap, flanked by the two shoulders. 
The emergence of the pseudogap in $A_f$ is caused by the appearance of {\SP}s in $\Sigma_f$ in the RKKY phase.
A further decrease of $V$ deepens the pseudogap
because the poles in $\Sigma_f$ become stronger.
The pseudogap never becomes a true gap (except for $V=0$) because the poles of $\Sigma_f$ are positioned away from $\omega =0$.
Thus, the QP weight of the $f$ electrons is finite even in the RKKY phase (see also Sec.~\ref{sec:QP_properties}).
This is one of the crucial differences to the findings of Refs.~\onlinecite{DeLeo2008,DeLeo2008a}; there,
a charge gap in $A_f$ was found due to the poor energy resolution of the ED impurity solver used in these studies.

(iii) \textit{Pseudogap for $A_{c+}$,
Kondo-like peak for $A_{c-}$:} 
Once $V$ drops below $\Vc$, $A_{c+}$ rapidly develops a
pronounced pseudogap around $\omega =0$, which weakens (becomes less pronounced) when $V$ is decreased further.
This pseudogap emerges because $\epsilon_{f+}^\ast$~(see Eq.~\eqref{eq:static_Kondo_cluster}) continuously changes sign at the KB--QCP, from $\epsilon_{f+}^\ast<0$ in the Kondo phase to $\epsilon_{f+}^\ast>0$ in the RKKY phase (see Fig.~\ref{fig:QP_para}(b) below). Due to the low energy form of $\Sigma_{c+}$ shown in Eq.~\eqref{eq:static_Kondo_cluster}, this leads to a \HP in $\Sigma_{c+}$ 
 which is close to $\omega=0$ in the vicinity of the KB--QCP and whose position changes sign across the QCP, in the same way as $\epsilon_{f+}^\ast$. This \HP is clearly visible in Fig.~\FigAlog(g), where we show $-\im\Sigma_{c+}$. We discuss the sign change of $\epsilon_{f+}^\ast$ in more detail in Sec.~\ref{sec:QP_properties},
where we also show that this sign change is intricately connected to the emergence of the \SP in $\Sigma_{f+}$. The sign change of $\epsilon_{f+}^\ast$ and therefore also the pseudogap in $A_{c+}$ is therefore an 
integral part of the OSMT. We further show in Sec.~\ref{sec:Luttinger} that the sign change of $\epsilon_{f+}^\ast$ is ultimately tied to a reconstruction of the FS.

 In striking contrast, $A_{c-}$ develops a Kondo-like peak around $\omega = 0$, whose peak height increases rapidly as $V$ drops below $\Vc$, and then decreases when $V$ is decreased further. The emergence of such a peak  for \textit{delocalized} electrons is rather unexpected, which is why we call it ``Kondo-like'' (in contrast
to ``Kondo peak'' for localized electrons).
This sharp peak suggests that close to the KB--QCP, the $c$-electrons become more localized, i.e.\ 
that their Fermi velocity is strongly renormalized downward due to the momentum dependence of $\Sigma_c$.

 (iv) \textit{Hybridization poles for $\im \Sigma_{c}$:}
The \HG at negative frequencies for $A_{c}$ is
caused by a corresponding peak in $\im \Sigma_{c}$ [Figs.~\FigAlog(g,h)]. It reflects a $c$ self-energy pole, to be called left hybridization pole (left \HP).
The frequency location of the \HG and left \HP
is comparable in magnitude to the NFL scale. 
When $V$ is reduced towards and past $\Vc$
into the RKKY phase, the left \HP in $\im \Sigma_{c}$ weakens and almost disappears, causing the same for the 
\HG in $A_{c}$. 
At the same time, for the bonding channel a peak in $-\im \Sigma_{c+}$ 
 at positive frequencies emerges in the RKKY phase. It corresponds to
 an additional pole of $\Sigma_{c+}$, to be
called right \HP, located at $\simeq \TFL$.
It causes the \HG in $A_{c+}$ around $\omega=0$ close to the KB--QCP. By contrast, for the anti-bonding channel 
$\im \Sigma_{c-}$ does not have a second pole---its very weak
peak on the right in fact is the tail of its left \HP
(This becomes more clear from the temperature dependence of $\Sigma_c$ and will be explained in more detail in Sec.~\ref{sec:SpectralFunctionsCluster_Tdep}.)
(v) \textit{Low-energy Fermi liquid:}
For all considered values of $V$, the 
$\omega=0$ quantities $\im\Sigma_{f}(\izeroplus)$ and $\im\Sigma_{c}(\izeroplus)$ all vanish. This implies FL behavior at the lowest energy scales for all $V \neq \Vc$ (consistent with
the results of Sec.~\ref{sec:FiniteSizeSpectra}). 
Moreover,  $A_{f}(0)$
and $A_{c}(0)$ never vanish, even in the RKKY phase. Thus  the pseudogap in $A_{f}$ never becomes a true gap, implying that $f$ electrons keep contributing to the QPs constituting the low-energy FL.

 \subsection{Quasiparticle properties \label{sec:QP_properties}}
 In this subsection, we substantiate our claims made in the previous subsection regarding QP weights and $\epsilon_f^\ast$. 
 In particular, we seek to show that 
the low-frequency form Eq.~\eqref{eq:static_Kondo_cluster} applies to $\Sigma_{c}$ on both sides of the KB--QCP. 
Note that Eq.~\eqref{eq:static_Kondo_cluster} is meaningful only
 if the low-energy physics shows FL behavior; our study of
 finite-size spectra in Sec.~\ref{sec:FiniteSizeSpectra} confirmed that this is the case.
We define cluster QP weights and effective level positions for both $f$ and $c$ electrons 
by replacing $\bk\to\alpha$ in Eqs.~\eqref{eq:epsilonxk}: 
 \begin{subequations}
 \begin{align}
  \label{eq:efast}
 \epsilon^\ast_{x\alpha} &= Z_{x\alpha}
 \bigl[\epsilon_{x\alpha}
 +\re \Sigma_{x\alpha}(0)\bigr] , \\
  Z_{x\alpha} &= \bigl[1-  \re \Sigma'_{x\alpha}(0) \bigr]^{-1} , 
 \hspace{1cm} (x=f,c) \hspace{-0.5cm}
 \end{align}
 \end{subequations}
with $\epsilon_{f\alpha} = \epsilonf -\mu$ and $\epsilon_{c \alpha} = - \alpha t - \mu$.
The $f$-electron QP weight is non-zero as long as $\Sigma'_{f\alpha}(0)$ is not infinite, i.e.\ as long as $\Sigma_{f\alpha}$ has no pole at 
$\zomega=0$. In the Kondo phase,
there are no low-frequency poles in $\Sigma_{f\alpha}$ at all, while the poles appearing in the RKKY phase are shifted away from 
$\zomega=0$.
%

 \begin{figure}[tb!]
 \centerline{\includegraphics[width=\linewidth]{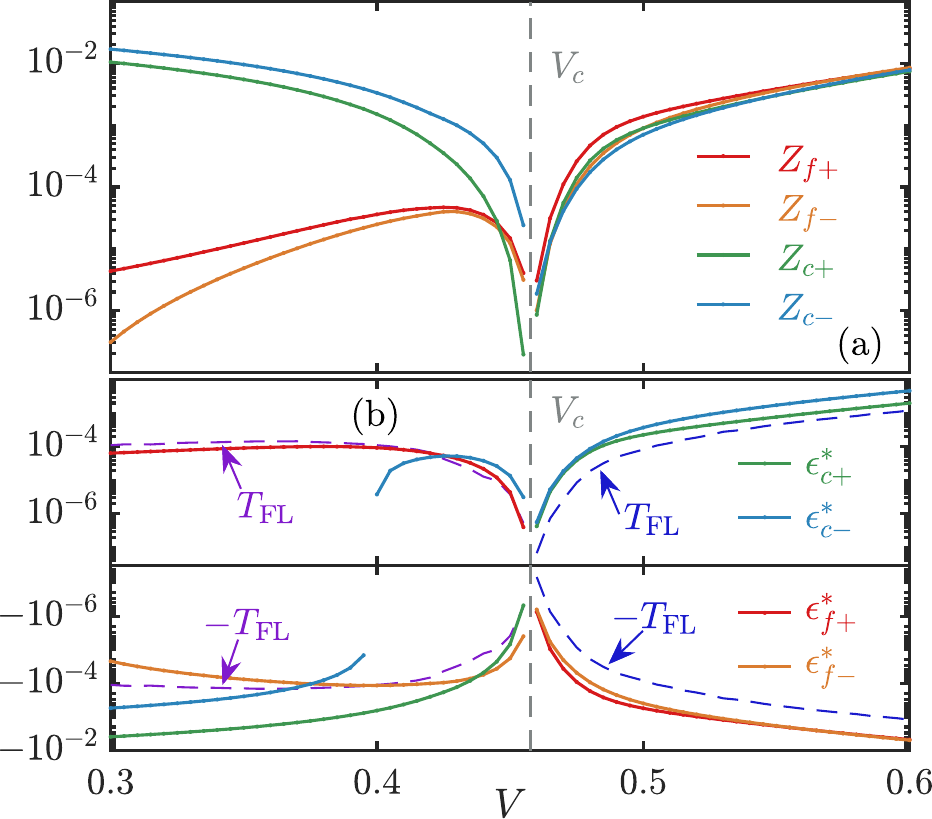}}
 \caption{%
 (a) QP weights and (b) effective level positions of the self-consistent effective 2IAM, computed 
directly from the NRG finite size spectra at $T=0$. Blue and purple dashed lines in panel (b) mark $\pm\TFL$ and $\pm\TFLstar$ for reference, respectively. 
}
 \label{fig:QP_para} 
 \end{figure}
 
\begin{figure}[b!]
\centerline{\includegraphics[width=\linewidth]{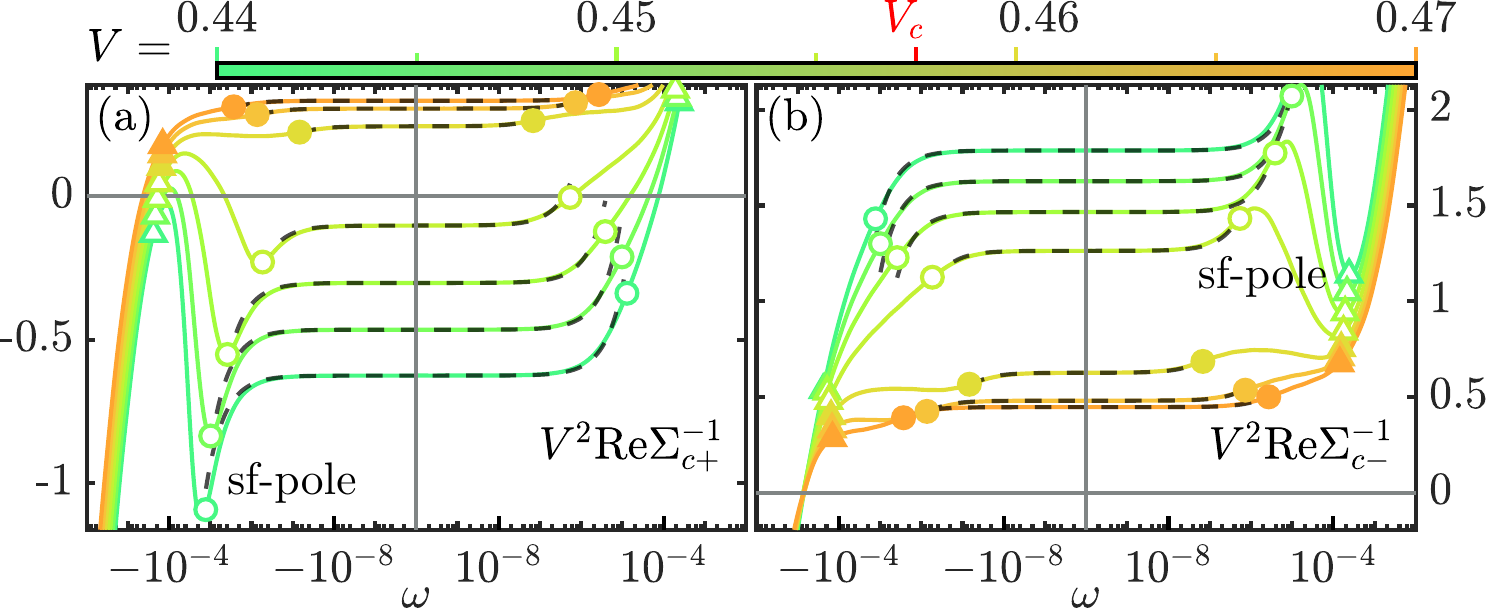}}
\caption{%
Evolution of $V^2\re \Sigma^{-1}_{c\pm}(\omegaR)$ at $T=0$
as $V$ is tuned in a narrow range across the QCP. As $V$ is lowered, non-monotonic behavior emerges, whose double-wiggle structure (local maximum followed by local minimum) reflects the \SP{s} of $-\re \Sigma_{f \pm}$ (c.f. \Eq{eq:tildeE}). Circles, triangles and tick marks on the color bar have the same meaning as in Fig.~\ref{fig:A_vs_V_T1e-11_Cluster_log}. Black dashed lines indicate linear behavior around $\omega=0$.
}
\label{fig:SECinv_vs_V_T1e-11_Cluster_log} 
\end{figure}

 %
\begin{figure*}[bt!]
\centerline{\includegraphics[width=\textwidth]{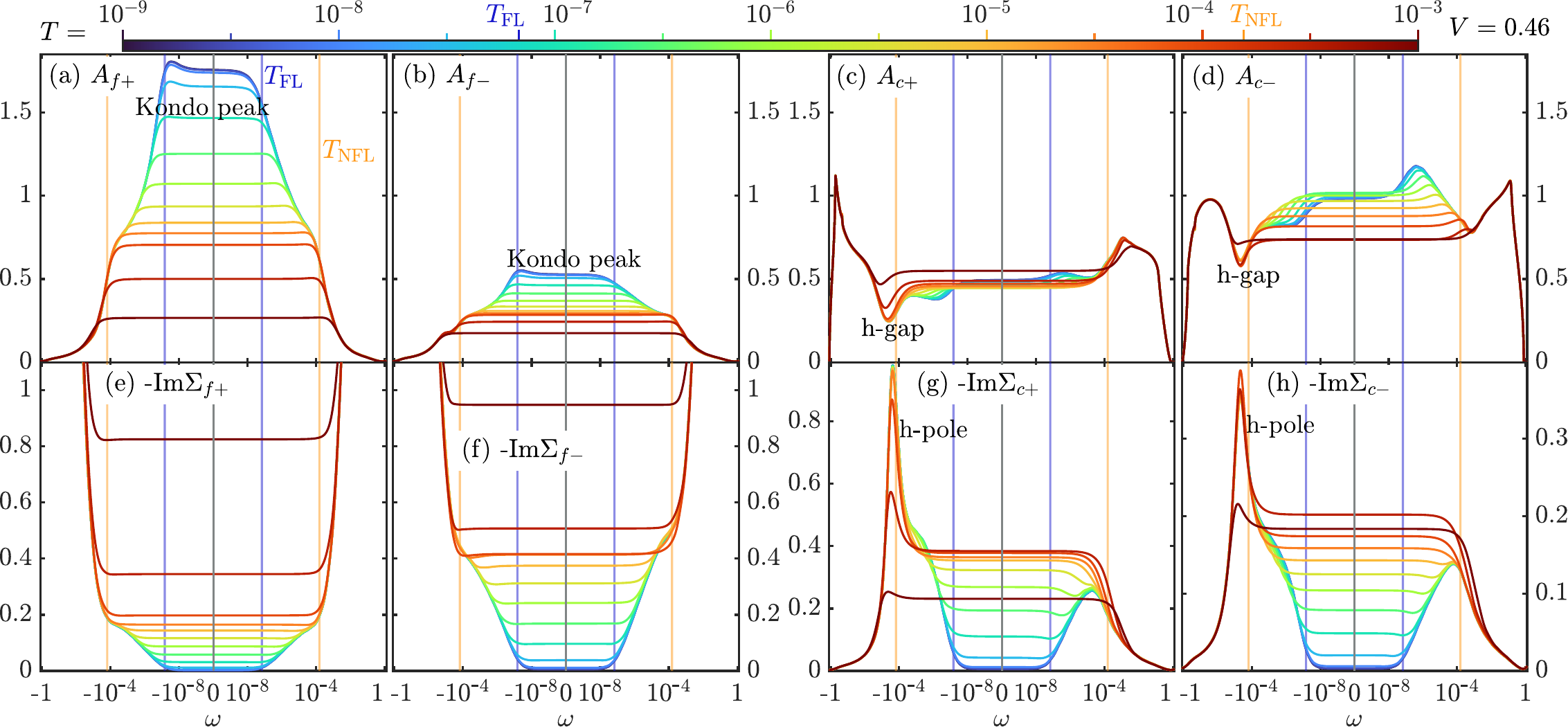}}
\caption{%
Evolution of cluster spectral functions and retarded cluster self-energies 
$\Sigma_{x\alpha}(\omegaR)$ at $V=0.46>\Vc$
(Kondo phase at $T=0$) 
as temperature is increased. Colored curves correspond to $T$ values
marked by ticks on the color bar. A symmetric log scale with $10^{-10} < |\omega| < 1.25$ is used.
Vertical blue and orange lines mark $\pm\TFL$ and $\pm\TNF$, respectively. \vspace{3mm}}
\label{fig:A_vs_T_V046_Cluster_log} 
\end{figure*}

 %
 The QP weights $Z_{x\alpha}$ and effective level positions $\epsilon_{x\alpha}$ can be extracted directly from NRG finite size spectra~\cite{Hewson1993,Hewson2004}, 
 which avoids fitting frequency dependent data. 
Figure~\ref{fig:QP_para} shows them all.
We see that both $Z_{f}$ and $Z_{c}$ vanish
at the KB--QCP, as expected.
However, $Z_{f}$ is finite not only in the Kondo phase \textit{but also in the RKKY phase}. This further substantiates our claim that the $f$ electrons contribute to the low energy QP even for $V<\Vc$.
The key difference between the RKKY and Kondo phases therefore is \textit{not} zero versus non-zero $Z_{f}$
--- instead, 
the key difference turns out to be the sign of $\epsilon^\ast_{f+}$.
Note that our discussion below is applicable for the \textit{electron-doped} case considered in this work;
a corresponding discussion of the hole-doped case would follow along the same lines, but with the signs of $\epsilon^\ast_{f\alpha}$
flipped and the role of bonding and anti-bonding cluster orbitals interchanged.
Both $\epsilon^\ast_{f+}$ and $\epsilon^\ast_{f-}$ are negative in the Kondo phase, 
see Fig.~\ref{fig:QP_para}(b). 
The same is true for $U=0$, 
and in this sense the non-interacting limit is adiabatically connected to the Kondo phase.
 However, while $\epsilon^\ast_{f-}$ remains negative in the RKKY phase, $\epsilon^\ast_{f+}$ changes sign at the KB--QCP and becomes positive.
Now, \Eqs{eq:Sigmac} and \eqref{eq:efast} imply
 \begin{subequations}
 \begin{flalign}
 \label{eq:tildeE}
V^2 \Sigma^{-1}_{c\alpha}(\zomega)
 &= \omegaplus - \epsilonf 
 - \Sigma_{f\alpha}(\zomega) \, ,
\\ 
\label{eq:signchange}
\mathrm{sgn} \, \epsilon_{f\alpha}^\ast &= - \mathrm{sgn} \, \mathrm{Re}\Sigma^{-1}_{c\alpha}(0) \, . 
\end{flalign}%
\end{subequations}  
Thus, the sign change in $\epsilon^\ast_{f+}$ 
is also visible in Fig.~\ref{fig:SECinv_vs_V_T1e-11_Cluster_log}(a) as a sign change of 
$V^2 \re \Sigma^{-1}_{c+}(0)$. 
In Sec.~\ref{sec:Luttinger}, where we perform a careful analysis of the Luttinger sum rule for the PAM, we will show explicitly that this
sign change of $\epsilon^\ast_{f+}$ leads to a 
jump of the FS volume corresponding to exactly one electron per site.
%

%
\begin{figure*}[t!]
\centerline{\includegraphics[width=\textwidth]{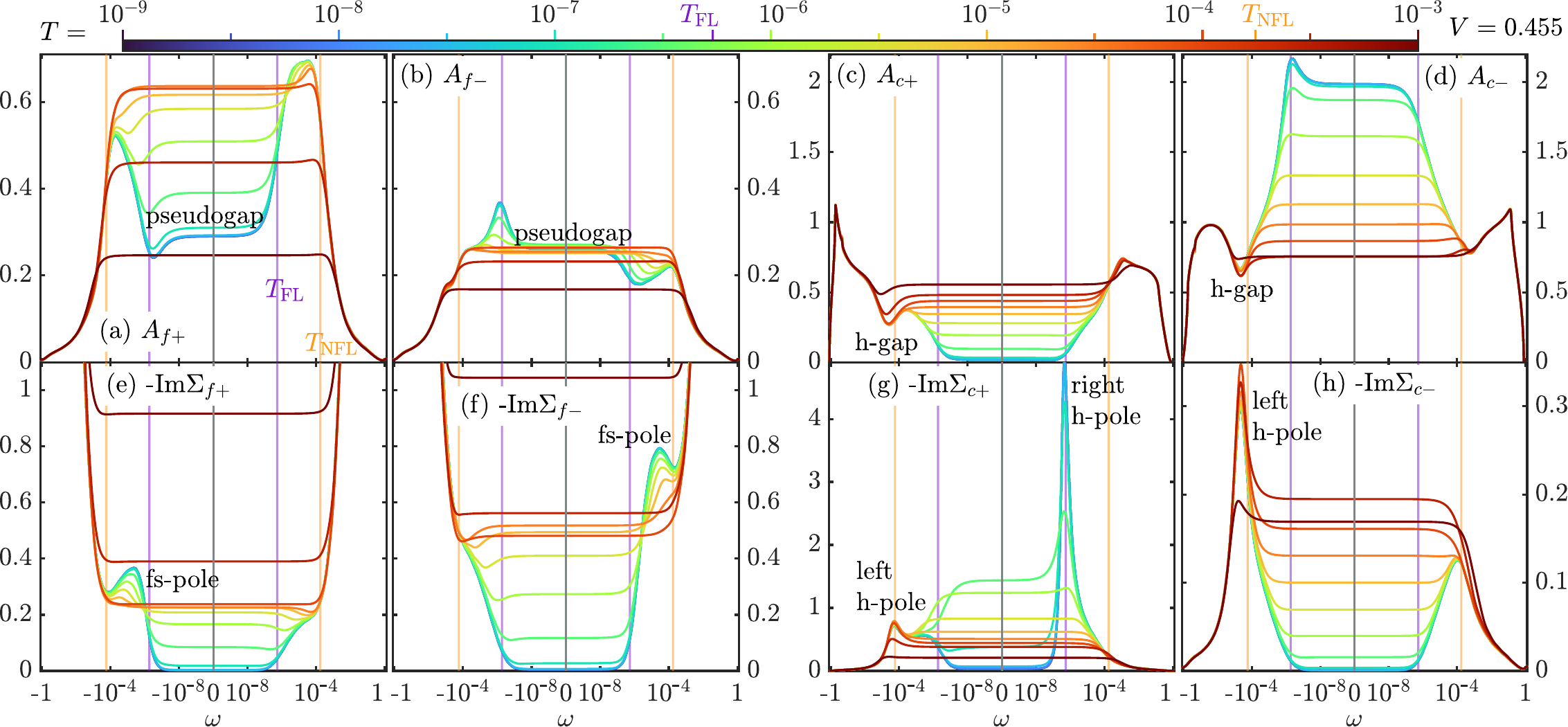}}
\caption{%
Same as Fig.~\ref{fig:A_vs_T_V046_Cluster_log}, but for 
$V=0.455 < \Vc$ (RKKY phase at $T=0$).} 
\label{fig:A_vs_T_V0455_Cluster_log} 
\end{figure*}

%
Figure~\ref{fig:SECinv_vs_V_T1e-11_Cluster_log}(a) also reveals how the sign change of $\epsilon^\ast_{f+}$ is connected to the \SP
in $\Sigma_{f+} (\omegaR)$. 
Close to the KB--QCP, in both the Kondo and RKKY phases, when $\omega$ is increased (starting from large negative),
$V^2 \re \Sigma^{-1}_{c+}(\omegaR)$ increases through zero around $\omega\simeq -\TNF$. This sign change results in the left \HP in 
$\Sigma_{c+}$, visible in Fig.~\ref{fig:A_vs_V_T1e-11_Cluster_log}(g).
(Deeper in the RKKY phase, $V^2 \re \Sigma^{-1}_{c+}(\omegaR)$ does not actually change sign at $\omega\simeq -\TNF$ but nevertheless becomes almost zero, again causing the left \HP of $\Sigma_{c+}$.)
This feature is adiabatically connected to the $U\to0$ case, where it is also present. 
It is an intermediate energy feature which characterizes the onset of NFL behavior. 
(We will show in Sec.~\ref{sec:SpectralFunctionsCluster_Tdep} that the left \HP in $\Sigma_c$ forms around $T\simeq \TNF$, irrespective of $V>\Vc$ or $V<\Vc$.)
In the Kondo phase, the sign of $V^2 \re \Sigma^{-1}_{c+}(\omegaR)$ changes only once with increasing $\omega$, remaining positive for $\omega > -\TNF$ and in
particular at $\omega=0$, so that $\epsilon_{f+}^\ast$ is negative.
By contrast, in the RKKY phase the initial increase 
with $\omega$ in $V^2\re\Sigma_{c+}^{-1}(\omegaR)$  for $\omega$ large negative \textit{is counteracted by the \SP\ in $\Sigma_{f+}(\omegaR)$,}
which induces a double-wiggle structure in 
$V^2\re\Sigma_{c+}^{-1} (\omegaR)$ near $\omega \simeq -\TFL$.
As a result, $V^2\re\Sigma_{c+}^{-1}(\izeroplus)$ is negative and $\epsilon_{f+}^\ast$ positive in the RKKY phase. The KB-QCP lies in between, 
at $\epsilon_{f+}^\ast=0$.
To summarize: In the Kondo phase, the sign change of $V^2 \re \Sigma^{-1}_{c}(\omegaR)$ around $\omega \simeq -\TNF$ leads to the left \HP\ and to $V^2 \re \Sigma^{-1}_{c}(0)>0$, implying $\epsilon_{f+}^\ast<0$. 
In the RKKY phase,
the formation of the \SP in $\Sigma_{f+}(\omegaR)$ at energy scales between $-\TNF$ and $-\TFL$ results in $V^2 \re \Sigma^{-1}_{c+}(0)<0$ and
therefore $\epsilon^\ast_{f+} > 0$.
%

\subsection{Temperature dependence close to $\Vc$}%
\label{sec:SpectralFunctionsCluster_Tdep}
%
We next discuss the temperature dependence of cluster spectral functions and self-energies close to the QCP at $\Vc=0.4575$. We first discuss the case 
$V =0.46 >\Vc$ (Fig.~\ref{fig:A_vs_T_V046_Cluster_log}), then $V = 0.455 < \Vc$
(Fig.~\ref{fig:A_vs_T_V0455_Cluster_log}), which at $T=0$ yield the Kondo and RKKY phases, respectively. For each we proceed from high to low temperatures.

$V = 0.46$, $T \gtrsim \TNF$: When the temperature is lowered in the LM regime from $T=10^{-3}$  towards $\TNF$, the onset of $c$-$f$ hybridization leads
to the emergence of 
a hybridization gap in $A_{c} (\omega)$ (Fig.~\ref{fig:A_vs_T_V046_Cluster_log}(c,d)) and a left \HP in $-\im \Sigma_{c}(\omegaR)$ (Fig.~\ref{fig:A_vs_T_V046_Cluster_log}(g,h)),
all at $\omega\simeq -10^{-4} \simeq - \TNF$.
This triggers the onset
of screening, signified by increased spectral weight in $A_{f}$ around $\omega=0$ (Fig.~\ref{fig:A_vs_T_V046_Cluster_log}(a,b))
and a decrease of $-\im\Sigma_{f}$ at $\omega=0$ (Fig.~\ref{fig:A_vs_T_V046_Cluster_log}(e,f)). However, at $\TNF$, no coherent QP have formed yet: both $-\im\Sigma_{f}$ and $-\im\Sigma_{c}$ have significant spectral weight around $\omega \simeq 0$, implying strong scattering for electrons near the chemical potential. 

$V = 0.46$, $\TFL \lesssim T \lesssim \TNF$:
When the temperature is lowered further towards $\TFL$, screening becomes
stronger: both $-\im\Sigma_{f}$ and $-\im\Sigma_{c}$ at $\omega \simeq 0$ decrease, albeit slowly, as $\sim\ln(T)$. (In Figs.~\ref{fig:A_vs_T_V046_Cluster_log}(e--h), $T$ values equally spaced on logarithmic scale yield 
$-\im\Sigma(\izeroplus)$ values equally spaced on a linear scale.) At the same time, coherent QPs begin to form: a sharp Kondo peak gradually forms in $A_{f}$, and narrow structures $A_{c}$ emerge.

$V = 0.46$, $T \lesssim \TFL$:
Below $\TFL$, coherent QP have formed ($\Sigma_{f}(\izeroplus)$
and $\Sigma_{c}(\izeroplus)$ approach zero), 
and the $T$-dependence of the spectral functions becomes weak. 

We next consider the case $V<\Vc$.

$V = 0.455$, $T \gtrsim \TNF$: 
In the LM regime, the behavior for $V<\Vc$ is similar to that for $V>\Vc$: 
as $T$ is lowered from $10^{-3}$ towards $\TNF$, a hybridization gap emerges in $A_{c}(\omegaR)$ (Fig.~\ref{fig:A_vs_T_V0455_Cluster_log}(c,d))
, and a left \HP in $-\im\Sigma_{c}$
(Fig.~\ref{fig:A_vs_T_V0455_Cluster_log}(g,h)), all at $\omega\simeq -10^{-4}$. Correspondingly, $A_{f}$ increases for frequencies near $\omega=0$, the temperature dependence at $T>\TNF$ is thus similar to $V>\Vc$, as expected. 

$V = 0.455$, $ \TFL \lesssim T \lesssim  \TNF$:
By contrast, below $\TNF$ the temperature dependence is quite different
from $V>\Vc$. When $T$ is lowered within 
the NFL regime from $\TNF$ towards $\TFL$, $\im \Sigma_{f}$
develops \SP{s} (Fig.~\ref{fig:A_vs_T_V0455_Cluster_log}(e,f)).  At the same time, $-\im\Sigma_{c+}$ develops a right \HP at a small positive frequency, $\omega \gtrsim \TFL$
(Fig.~\ref{fig:A_vs_T_V0455_Cluster_log}(g)), causing a strong increase of $-\im\Sigma_{c+}(0)$. 
The formation of the right \HP is associated with the formation of the \SP{s} in $\Sigma_f$, as discussed in Sec.~\ref{sec:QP_properties}.
On the other hand, $-\im\Sigma_{c-}$ behaves quite similar to its counterpart at $V=0.46$ in the NFL region, decreasing logarithmically around $\omega=0$. The appearance of \SP{s} in the $f$-electron self-energies is accompanied by the formation of 
(somewhat asymmetric) pseudogaps in $A_{f}$ around 
$\omega=0$ (Fig.~\ref{fig:A_vs_T_V0455_Cluster_log}(a,b)).  Moreover, the right \HP in $-\im\Sigma_{c+}$ in the NFL region causes a pseudogap in $A_{c+}$ in the same temperature window
(Fig.~\ref{fig:A_vs_T_V0455_Cluster_log}(c)). $A_{c-}$, on the other hand, develops a sharp peak around $\omega=0$ in the NFL region, similarly to the behavior of $A_{f+}$ for $V>\Vc$
(Fig.~\ref{fig:A_vs_T_V0455_Cluster_log}(d)). 

$V = 0.455$, $T \lesssim  \TFL$:
When $T$ is lowered below $\TFL$, the imaginary parts of all self-energies quickly tend to zero at $\omega=0$, as expected in a FL. (Thus,
the overall behavior of $-\im\Sigma_{c+}(\mi \zeroplus)$ with decreasing temperature is non-monotonic, first increasing in the NFL regime, then decreasing down to zero in the FL regime.)

\textit{No marginal FL phenomenology:}
In the NFL regime neither the $f$ nor the $c$ electron self-energies show marginal FL phenomenology. The latter would require $-\im\Sigma(\omegaR,T) \sim \max(|\omega|,T)$ \cite{Varma1989}. Instead, the imaginary parts of the self-energies have a much weaker, namely logarithmic frequency and temperature dependence. Nonetheless, the spectral parts of various susceptibilities all show the same phenomenological frequency dependence~\cite{Varma1989} in the NFL region, namely a plateau for $T \lesssim |\omega|
\lesssim \TNF$ and a $\sim\omega$ dependence for $\omega<T$~\cite{Varma1989}. This frequency dependence has already been shown above in Fig.~\ref{fig:Chi_QCP}; 
we discuss the temperature dependence in a companion paper~\cite{Gleis2023a}. There, we also emphasize that the susceptibilities are 
not governed by the self-energy alone, as would be the case, in diagrammatic parlance, when evaluating only the bubble contribution---instead, vertex corrections play a crucial role.
This especially also concerns the conductivity, which shows a $\sim 1/T$ dependence in the NFL region despite the $\sim\ln(T)$ dependence of $\im \Sigma(\izeroplus)$~\cite{Gleis2023a}. 
%

%
\begin{figure*}[hbt!]
\includegraphics[width=0.98\textwidth]{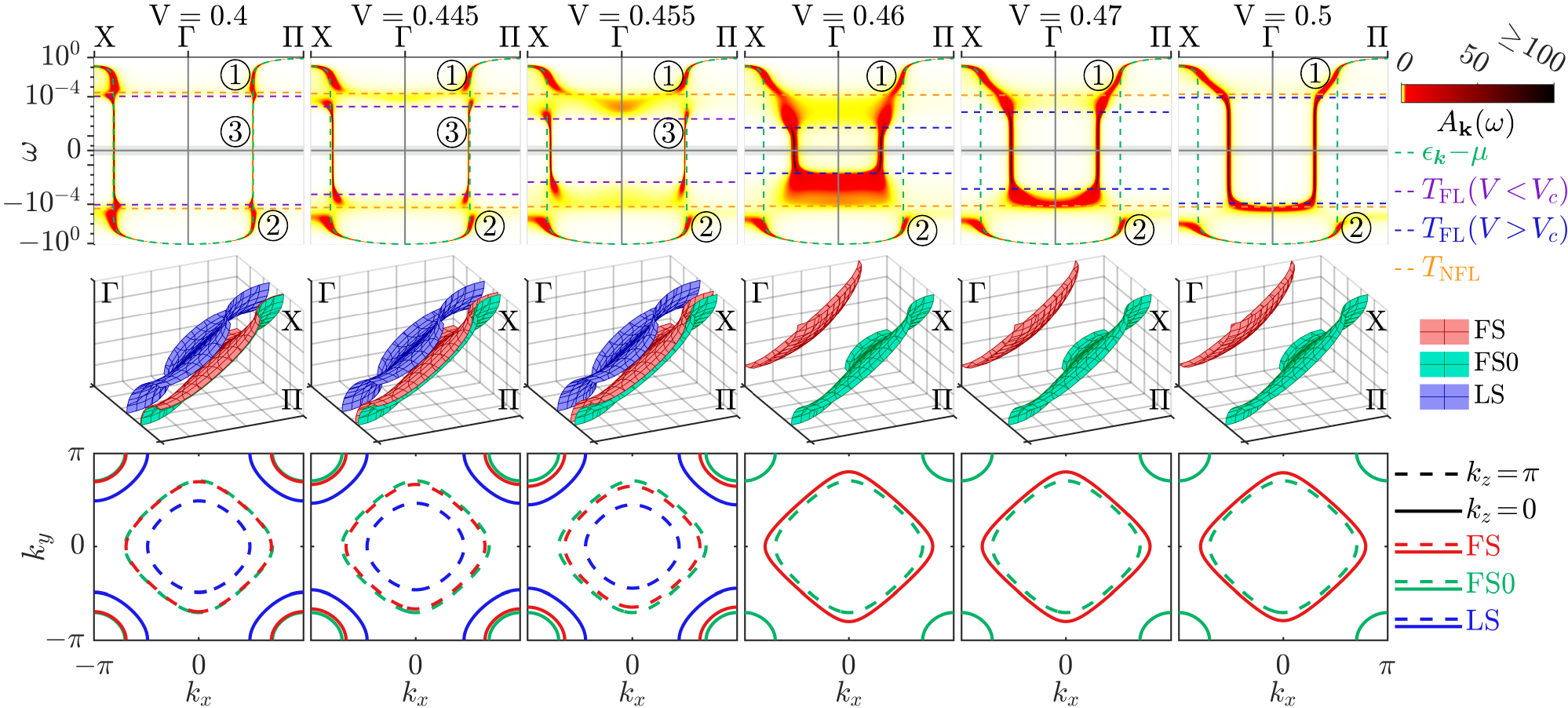}
\caption{%
Top row: Momentum resolved total spectral function $A_\bk(\omega) = A_{c\bk}(\omega) + A_{f\bk}(\omega)$ at $T=0$, for various $V$ crossing the QCP at $\Vc=0.4575(25)$, plotted from $X=(\pi,\pi,0)$ over $\Gamma=(0,0,0)$ to $\Pi=(\pi,\pi,\pi)$. $A_\bk(\omega)$ is shown on a logarithmic frequency scale for $|\omega|>10^{-9}$ and on a linear scale for $|\omega|<10^{-9}$ (grey shaded region). The green dashed line marks the $c$-electron dispersion at $V=0$. On the Kondo side ($V>\Vc$), $A_\bk(\omega)$ shows a two-band structure, marked \circled{1} and \circled{2}, as expected from adiabatic continuation from the $U=0$, with the upper band intersecting $\omega=0$ close to $\Gamma$. On the RKKY side ($V<\Vc$), $A_\bk(\omega)$ shows a three-band structure, marked \circled{1}, \circled{2} and \circled{3}, with the narrow middle band intersecting $\omega=0$ close to $\Pi$. Middle row: Corresponding FS (red)
where $\epsilonck^\ast = 0$, FS at $V=0$ (green, FS0) where $\epsilonck=0$, and LS where $\Sigma_{f\bk}(0) = \infty$ (blue), in an octant of the first Brillouin zone. For $V > \Vc$, the FS is centered around $\Gamma$. Crossing the QCP towards $V < \Vc$, the FS center point jumps to $\Pi$ and a LS emerges. Bottom row: Brillouin zone cuts showing the FS (red lines), FS0 (green lines) and LS (blue lines) at constant $k_z = 0$ (solid) or $k_z = \pi$ (dashed) versus $k_x$ and $k_y$.
}
\label{fig:A_vs_V}
\end{figure*}
%

\section{Single particle properties -- lattice \label{sec:Ak_vs_w}}%

Having focused on the single-particle properties of the effective 2IAM in the previous section, we now discuss how these translate to the lattice model. Our analysis builds on  that of De Leo, Civelli and Kotliar \cite{DeLeo2008a,DeLeo2008}, but with our better energy resolution, we uncover much additional detail and new emergent physics at low energies. In particular, we obtain a detailed understanding of the Fermi surface reconstruction occurring when traversing the KB--QCP.

\subsection{Reperiodization}
\label{sec:reperiodization}

Since the CDMFT artificially breaks translation invariance, the $c$ and $f$ electron self-energies have to be reperiodized 
before computing lattice spectral functions . To this end, we reperiodize the cluster cumulant 
\cite{Stanescu2006,Stanescu2006a}
\begin{equation}
\label{eq:Mc}
M(\zomega) = \bigl[ \omegaplus+\mu-\Sigma_c(\zomega)\bigr]^{-1}
\, .
\end{equation}
It may be viewed as a $c$-electron propagator excluding bare nearest-neighbor hopping, and in an expansion around the $t=0$ limit~\cite{Stanescu2004} takes the role of the cluster self-energy. Its reperiodized version $M_\bk(\zomega)$ is defined as
\begin{equation}
M_\bk(\zomega) = M_{11}(\zomega) + M_{12}(\zomega) \sum_{a=x,y,z}  \tfrac{1}{3} \cos(k_a) \, .
\end{equation}
Here, $M_{11}$ and $M_{12}$ are the local and nearest-neighbor cluster cumulants, respectively, and the latter is accompanied by the 
cosine factors arising when diagonalizing a non-interacting hopping 
Hamiltonian. At the points $\Gamma=(0,0,0)$ and $\Pi=(\pi,\pi,\pi)$, 
the lattice cumulants reduce to the cluster ones, 
$M_+ = M_{11} + M_{12} = M_\Gamma $ and 
$M_- = M_{11} - M_{12} = M_\Pi $. 
This reflects the correspondence, mentioned in Sec.~\ref{sec:ModelMethods}, of the 
BZ points $\Gamma$, $\Pi$ and the  bonding, anti-bonding cluster orbitals. The $\bk$-dependent self-energies are then defined via the relations
\begin{subequations}
\label{subeq:Mck}
\begin{align}
\label{eq:Mck}
M_\bk(\zomega) &= \bigl[\omegaplus + \mu-\Sigma_{c\bk}(\zomega)\bigr]^{-1} \\
\Sigma_{c\bk}(\zomega) &= V^2\left[\omegaplus -\epsilonf^0+\mu-\Sigma_{f\bk}(\zomega)\right]^{-1} \, . 
\end{align}
\end{subequations}
We will see in Sec.~\ref{sec:Luttinger} that the CDMFT solution to the PAM must fulfill a generalized version of the Luttinger sum rule
--- this sum rule is verifiable via cluster quantities only, i.e.\ without reperiodization. The reperiodization scheme above is however not guaranteed to preserve this property. We therefore slightly modify the above scheme by adjusting $\mu$ in Eq.~\eqref{eq:Mc} and Eq.~\eqref{eq:Mck} 
during the reperiodization only, such that the reperiodized self-energies fulfill this generalized Luttinger sum rule in the same way as the corresponding cluster quantities. We describe our reperiodization procedure and evaluate its validity in detail in App.~\ref{app:ssec:reperiodization}.

\subsection{Lattice spectral functions}
\label{sec:lattice-spectral-functions}

Figure~\ref{fig:A_vs_V} (top row) shows the evolution with $V$ of the ($T=0$) momentum-dependent spectral function $A_\bk(\omega)=A_{f\bk}(\omega)+A_{c\bk}(\omega)$ and of the FS.
(Corresponding retarded self-energies $\im \Sigma_{f\bk} (\omegaR)$ and $\im\Sigma_{c\bk}(\omegaR)$ are shown in Fig.~\ref{fig:SEk_vs_V} below.) To highlight all relevant energy scales and 
the changes of $A_\bk(\omega)$ at low frequencies close to the QCP, we use a logarithmic frequency grid for $|\omega|>10^{-9}$
and a linear grid for $|\omega|<10^{-9}$ (grey shaded region) to cross $\omega=0$. 
Figure~\ref{fig:A_vs_V} (bottom row) shows three different
surfaces, defined in \Sec{sec:self-energy-expansions}: the free Fermi surface $V=0$ (FS0, green), where 
$\epsilonck = 0  $; the actual Fermi surface (FS, red), 
where $\re \, G_{x\bk}^{-1}(\izeroplus)  = 0$ 
for both $x=f,c$ (numerically,
we use $\re \, G_{c\bk}^{-1}(\izeroplus)  = 0$); 
and the Luttinger surface (LS, blue), 
where $|\Sigma_{f\bk}(0)| = \infty$, $G_{f\bk}(0) = 0$ and $\Sigma_{c\bk}(0) = 0$ (numerically, we use $\re\Sigma_{c\bk}(\izeroplus) = 0$).  
To summarize: 
\begin{subequations}
\label{eq:FS_LS_FS0_def}
\begin{flalign}
\label{eq:FS0_def}
&\text{FS0:} \hspace{-4mm} 
& 
\epsilon_{c\bk} & = 0 \, , \hspace{-3mm}   & 
\\
\label{eq:FS_def}
&\text{FS:} \hspace{-4mm}
& \re \, G_{x\bk}^{-1}(\izeroplus) &=\! 0 \hspace{-2mm}
& \Leftrightarrow &  & \hspace{-3mm} 
\left\{\!\begin{array}{lr}
\det G^{-1}_{\bk}(\izeroplus) \\
\epsilon_{c\bk}^\ast \\
 \epsilon_{f\bk}^\ast - \frac{(V_{\bk}^{\ast})^2}{\epsilon_{c\bk}}
\end{array}\!\!\right\}
 & =  0
\, , & 
\\
\label{eq:LS_def}
&\text{LS:} \hspace{-4mm} 
& | \Sigma_{f\bk}(0) | & = \infty  \hspace{-3mm}
&\Leftrightarrow & & 
\Sigma_{c\bk}(\izeroplus) &  = 0 \, . & 
\end{flalign}
\end{subequations}

Next, we discuss some salient features of Fig.~\ref{fig:A_vs_V}.

\textit{Band structures:}
In the Kondo phase, $A_\bk(\omega)$ has a two-band structure~\cite{Martin1982} (marked \circled{1} and \circled{2} in Fig.~\ref{fig:A_vs_V}) with a well-defined QP peak in band \circled{1}, as expected in this phase. 
The dispersion around $\omega=0$ is clearly shifted away from the $V=0$ dispersion (indicated by a dashed green line in Fig.~\ref{fig:A_vs_V}).
As $V$ is lowered towards the QCP at $\Vc$, the upper band develops a broad region of incoherent spectral weight in the NFL energy window $\TFL<|\omega|<\TNF$.
Interestingly, in the RKKY phase, the spectral function shows a three-band structure
(marked \circled{1}, \circled{2} and \circled{3} in Fig.~\ref{fig:A_vs_V}):
The top and bottom bands (\circled{1} and \circled{2}) do not cross $\omega=0$. In addition, there is a third, narrow middle band (\circled{3})
of width $\sim \TFL$ crossing $\omega=0$ near $\Pi$. 
The dispersion around $\omega=0$ of band \circled{3} almost matches the $V=0$ dispersion. Band \circled{3} is separated from the other two bands by a region of incoherent spectral weight in the NFL window $\TFL<|\omega|<\TNF$.
The additional bandgap in the RKKY phase comes from a dispersive \SP\
in $\Sigma_{f}$ (see Figs.~\ref{fig:A_vs_V_T1e-11_Cluster_log} and~\ref{fig:SEk_vs_V} and their discussions), associated with orbital selective Mott physics. Ref.~\onlinecite{Fabrizio2023} suggests that the \SP\ may be interpreted as an emergent coherent spinon excitation~\cite{Fabrizio2023}. This suggests that the $f$-electron is fractionalized and the RKKY phase is a fractionalized FL~\cite{Senthil2003,Senthil2004}. We  will clarify this view in future work.
As discussed in Sec.~\ref{sec:Kondo-breakdown}, 
for non-single-site DMFT (as here, where we use 2-site CDMFT),
it is natural for a Mott insulator (the $f$ band in the present case)
to hybridize with a metal (the $c$ band) and contribute non-zero weight to the low-energy QP, resulting in an OSMP.
Because the $f$ band, which is in a (strongly correlated) Mott phase, contributes non-zero QP weight, the QP dispersion is strongly renormalized, 
leading to the narrow middle band.
Its narrow width ($\sim \TFL$)  for
$V < \Vc$ indicates a
large effective mass $m^\ast$, as observed experimentally e.g. in Refs.~\onlinecite{Gegenwart2002,Gegenwart2003}
and discussed in detail in Sec.~\ref{sec:Sommerfeld} below. Note that this strong-correlation effect 
occurs even though the actual FS lies very close to the free $c$-electron Fermi surface FS0, for reasons discussed below.
%

\begin{figure*}[hbt!]
\includegraphics[width=0.98\textwidth]{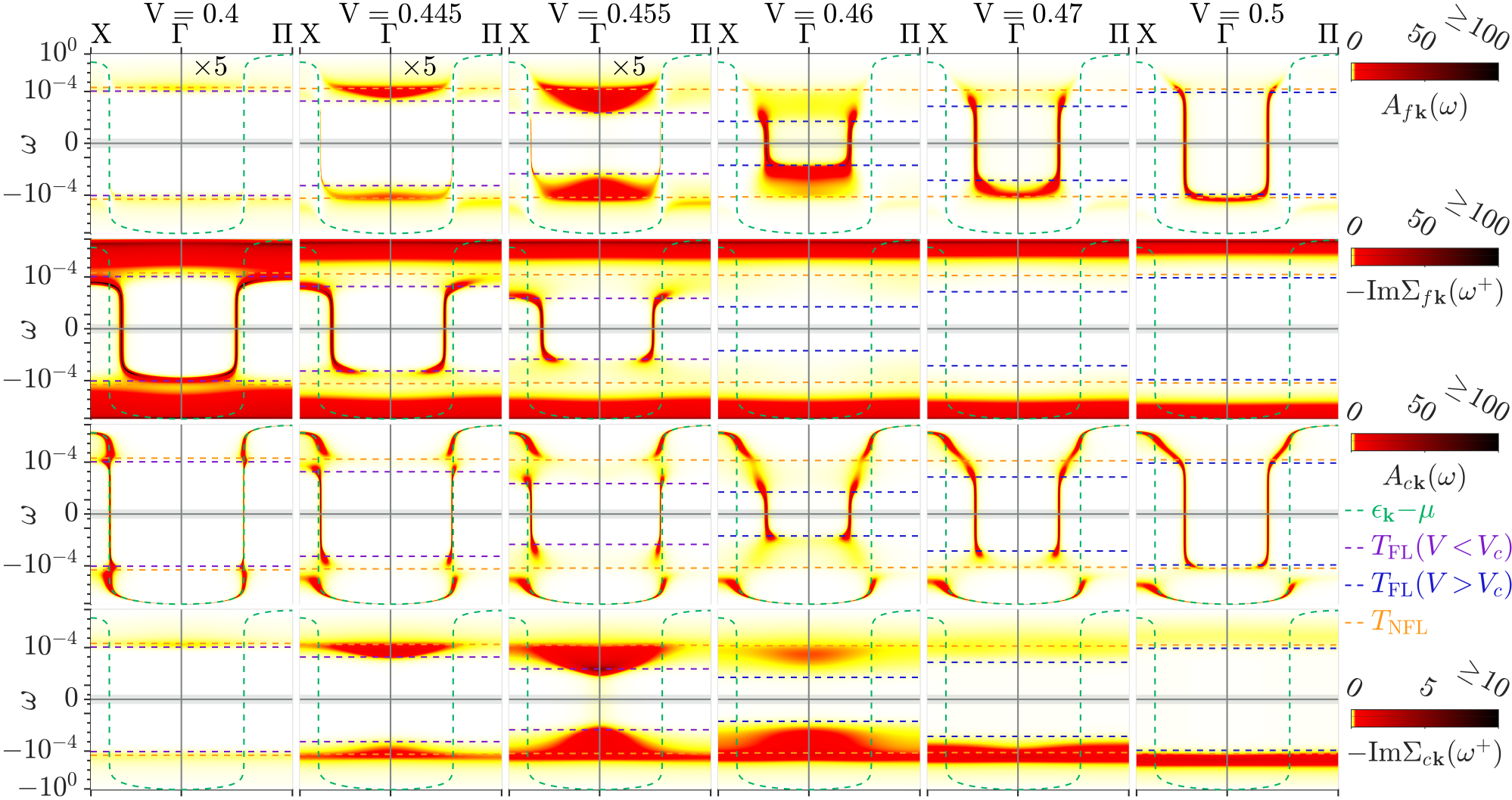}
\caption{%
Momentum resolved $f$ and $c$ spectral functions and
imaginary parts of retarded self-energies, $A_{f\bk}(\omega)$ (top row),
 $-\im \Sigma_{f \bk}(\omegaR) $ (second row), 
$A_{c\bk}(\omega)$ (third row), and   
$-\im \Sigma_{c\bk}(\omegaR)$ (bottom row) at $T=0$, corresponding to the spectral functions shown in Fig.~\ref{fig:A_vs_V}. 
The three left-most panels ($V<V_c$) of the top row
show $5\times A_{f\bk}(\omega)$ to improve the visibility of $f$-electron spectral features in the OSMP. Note that in the RKKY phase, the spectral weight of $A_{f\bk}(\omega)$ close to $\omega=0$ is underestimated by our reperiodization scheme, as discussed in App.~\ref{app:ssec:reperiodization}. 
We have scaled it by a factor of $5$ to improve visibility. As discussed in Sec.~\ref{sec:SingleParticleCluster}, there is 
a non-zero $f$-electron contribution to the FS throughout the RKKY phase.
}
\label{fig:SEk_vs_V}
\end{figure*}

\textit{Fermi surface reconstruction:}
In the Kondo phase, the FS (red) is electron-like and centered around the $\Gamma=(0,0,0)$-point, in contrast to the hole-like free FS0 (green) of the $c$ electrons, centered at $\Pi = (\pi,\pi,\pi)$. 
The FS depends only very weakly on $V$ because it is constrained by the Luttinger sum rule: 
the latter relates the density $n_f+n_c$ to the FS volume (see Sec.~\ref{sec:Luttinger} for details), which depends only very 
weakly on $V$ (see Fig.~\ref{fig:LT_PAM}).
As the QCP is crossed, the FS undergoes a sudden reconstruction and becomes centered around $\Pi$, positioned close to FS0. This leads to a jump of the Hall coefficient, as discussed with Fig.~\ref{fig:Hall_PAM}. The FS reconstruction
is accompanied by the emergence of a LS (blue), which accounts for the change in the FS volume (see Sec.~\ref{sec:Luttinger}).
The emergence of a LS is the hallmark property of a Mott phase and it has been shown that the LS is stable to small perturbations~\cite{Huang2022}. 
This emphasizes again our claim that the RKKY phase is an OSMP.
In the RKKY phase, the FS again depends only very weakly on $V$
due to the Luttinger sum rule constraint (the weak $V$-dependence is again due to a weak $V$-dependence of the filling).
%

\begin{figure*}[tb!]
\includegraphics[width=\textwidth]{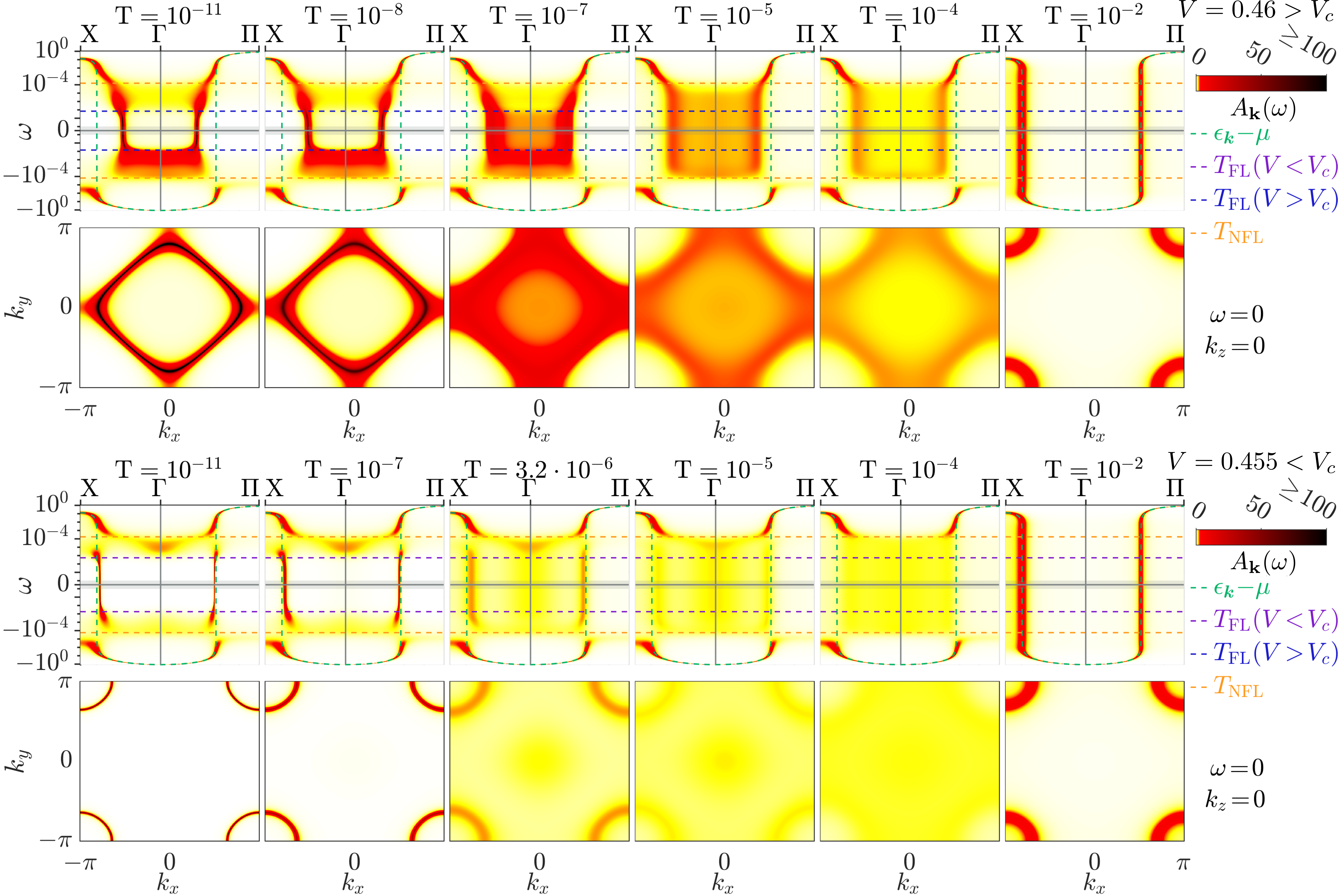}
\caption{%
Temperature dependence of $A_\bk(\omega)$ at $V=0.46>\Vc=0.4575$ (rows 1 and 2) and $V=0.455<\Vc$ (rows 3 and 4). The layout of rows 1 and 3 mirrors that of the top row of Fig.~\ref{fig:A_vs_V}. Rows 2 and 4 show cuts at $\omega=0$ and $k_z=0$. These illustrate how  the sharp FS at $T=0$, large for $V>\Vc$ and small for $V < \Vc$, dissolves as $T$ increases into the NFL regime and finally evolves to a (temperature-broadened) small FS in the LM regime at high $T$.}
\label{fig:A_vs_T_V046}
\end{figure*} 
\subsection{Lattice self-energies}
\label{sec:lattice-self-energies}

To get a better understanding how the features of the spectral functions in Fig.~\ref{fig:A_vs_V} emerge, Fig.~\ref{fig:SEk_vs_V} shows the imaginary parts of the momentum-dependent $c$- and $f$-electron self-energies. In the Kondo phase at $V>\Vc$, $-\im \Sigma_{f\bk}(\omegaR)$ shows only weak momentum dependence. It is large at high frequencies but vanishes towards $\omega=0$, consistent with the presence of coherent QP. Interestingly, its structure is almost independent of $V$ in the Kondo regime. However, as we have seen in Sec.~\ref{sec:SingleParticleCluster}, the 
$f$-electron self-energy becomes slightly non-local in the Kondo regime close to the QCP (though this is not easily visible in Fig.~\ref{fig:SEk_vs_V}). 

Crossing the QCP to the RKKY phase at $V<\Vc$, 
the non-dispersive high frequency structure remains essentially the same as in the Kondo phase. However, additionally
a sharp dispersing pole emerges in $\Sigma_{f\bk}$ at low frequencies for $|\omega|<\TFL$.
As discussed in Sec.~\ref{sec:Kondo-breakdown} and~\ref{sec:SingleParticleCluster}, this dispersive pole in $\Sigma_{f\bk}$ in the RKKY phase is a clear sign of Mott physics present in the $f$ band.
Importantly, the fact that the pole is dispersive (c.f. Eq.~\eqref{eq:SE_Mott}) results from employing cluster DMFT instead of single-site DMFT. We are therefore not in the $d\to\infty$ limit where the OSMP (i.e.\ the RKKY phase) would be
unstable to finite $c$-$f$ hybridization $V$~\cite{Kugler2022}. 

The $c$ self-energy, shown in the bottom row of Fig.~\ref{fig:SEk_vs_V}, has most of its spectral weight at $\omega \simeq -\TNF$ deep in the Kondo phase. It is mostly momentum independent and signals the position of the hybridization gap. As $V$ is decreased towards $\Vc$, $-\im\Sigma_{c\bk}(\omegaR)$ becomes increasingly momentum dependent, with more spectral weight at $\Gamma$ than at $\Pi$. Further, at $V=0.46$ close to the QCP on the Kondo side, spectral weight starts to smear out significantly over the NFL region and significant weight appears at positive frequencies. This suggests that $f$ and $c$ begin to hybridize more strongly at positive frequencies. When the QCP is crossed to $V<\Vc$, the spectral weight of 
$\Sigma_{c\bk}$ is now stronger at $\omega>0$ 
than at $\omega < 0$, showing that $c$ and $f$ now hybridize more strongly at positive frequencies than in the Kondo regime. However, 
significant spectral weight does also remain at $\omega<0$, reflecting the presence of a left and right \HP, as discussed already in the previous section. This suggests that in the RKKY regime, the $f$ band is split apart by the pole in $\Sigma_{f\bk}$ and the $c$ electrons then hybridize with both of the resulting two $f$ bands, leading to a three-band structure. Thus, the $f$ electrons still hybridize significantly with the $c$ electrons in the RKKY phase close to the QCP, explaining intuitively the strong renormalization of $m^\ast$ in the RKKY regime mentioned earlier and observed in experiments~\cite{Krellner2009}. Finally, as $V$ is lowered further towards $V=0$, the overall magnitude of $-\im\Sigma_{c\bk}(\omegaR)$ decreases rapidly, suggesting that $f$ and $c$ bands continuously decouple when $V\to 0$.

\subsection{Finite temperature}

In Figure~\ref{fig:A_vs_T_V046} we show the temperature dependence of $A_\bk(\omega)$ both at $V=0.46>\Vc$ and $V=0.455<\Vc$. For $T<\TFL$, the spectral functions are mostly independent of temperature as expected. As $T$ crosses $\TFL$ into the NFL region, the incoherent features at $\TFL<|\omega|<\TNF$ are thermally broadened and the sharp QP features at $\omega=0$ are smeared out, indicating a thermal destruction of the QP. Deep in the NFL region at $T\simeq10^{-5}$, spectral weight around $\omega=0$ is completely incoherent and no FS with sharp QP excitations can be made out. At $T=10^{-2}>\TNF$ in the LM region, the features of $A_\bk(\omega)$ become sharp again around $\omega=0$ and a single band forms which coincides with the $V=0$ $c$-electron dispersion. In this temperature regime, the $f$ electrons can be viewed as free local moments decoupled from the free $c$ electrons. The interaction between $c$ electrons and $f$ moments then leads to scattering, slightly smearing out the features in the spectral function while leaving the qualitative picture of the LM region unaltered.
\section{Generalized Luttinger sum rule \label{sec:Luttinger}}
In this section, we discuss the FS reconstruction of the previous section from the perspective of the generalized Luttinger sum rule \cite{Luttinger1960,Oshikawa2000,Dzyaloshinskii2003,Coleman2005,Powell2005,Curtin2018,Nishikawa2018,Heath2020,Skolimowski2022}. It states that if a Fermi surface exists,
the density, $n$, of electrons in partially-filled bands can be
expressed as
$n = 2v_\fs + 2I_\Luttinger$,
where $v_\fs$ is the FS volume and  $I_\Luttinger$ is an integral
known as Luttinger integral. 
In case of a FL, the generalized Luttinger sum rule as stated above can be derived from 
a simple, exact decomposition of the Green's function based on Dyson's equation (we review this decomposition below).
The sum rule becomes useful when it is possible to formulate constraints on $I_\Luttinger$.

For instance, if the interacting and non-interacting ground states are adiabatically connected \cite{Luttinger1960,Luttinger1961}, perturbative arguments
 can be used to show that $I_\Luttinger = 0$, leading to the celebrated
 Luttinger sum rule,  $n = 2 v_\fs$ (also called Luttinger's theorem).
Subsequent work~\cite{Coleman2005,Powell2005} has shown that  $I_\Luttinger = 0$ is a consequence of U(1) charge conservation
in case $\Sigma$ is $\Phi$-derivable, i.e.\ if $\Sigma = \delta \Phi / \delta G$, where $\Phi$ is the Luttinger--Ward~(LW) functional.

Explicitly, consider a multi-band model with a U(1) total charge symmetry (we assume every band has the same gauge charge).
The Green's function $G_{\bk}(\zomega)$ is matrix valued, with entries $G_{\alpha\beta\bk}(\zomega)$ where $\alpha$ and $\beta$
label the bands. 
If the matrix-valued self-energy is $\Phi$-derivable, i.e.\ $\Sigma_{\bk}(\zomega) =  \delta \Phi / \delta G_{\bk}(\zomega)$,
then the Luttinger integral,
\begin{align}
\label{eq:I_Luttinger_generic}
I_\Luttinger = 
\frac{-1}{\pi} \im \!  \int_\bz \frac{\mr{d}\bk}{\mc{V}_\bz} \int_{-\infty}^0 \!\!\! \mr{d}\omega \, 
\mr{Tr} \left[G_{\bk} (\omegaR) \, \partial_\omega \Sigma_{\bk} (\omegaR) \right] 
\end{align}
equals zero, $I_\Luttinger = 0$.
Note that $I_\Luttinger = 0$
\textit{only} holds in the $T\to 0$ limit.

 This extends the applicability of Luttinger's theorem beyond the perturbative regime (see also Ref.~\cite{Oshikawa2000} for a different approach),
 but requires the existence of the LW functional $\Phi[G]$, at least in the vicinity of the physical Green's function $G$
 (note that $\Phi$ can be constructed non-perturbatively~\cite{Potthoff2006}).
However, it has been established that
in general,
the LW functional is multivalued~\cite{Kozik2015,Tarantino2017,Gunnarsson2017} and does not exist for certain physically relevant Green's functions~\cite{Dave2013,Heath2020},
which can lead to $I_\Luttinger \neq 0$~\cite{Dave2013,Heath2020,Skolimowski2022}.
A very instructive analysis of a situation where $I_\Luttinger = 0$  breaks down is provided in Refs.~\onlinecite{Curtin2018,Nishikawa2018} in terms
 of a fermionic two-impurity model, where the role of the Luttinger sum rule is taken by the Friedel sum rule~\cite{Martin1982}. This model exhibits a QPT from a Kondo-type to an RKKY-type phase, with the local density remaining constant while the free QP density changes abruptly. This violates the Friedel sum rule, and the violation was traced to a Luttinger integral abruptly becoming nonzero. 
In this section, we perform a similar analysis for the KB--QPT of the PAM.
We find that the Luttinger integral is numerically zero in both the RKKY and Kondo phases;
the FS reconstruction is due to the appearance of a LS in the RKKY phase and 
not due to a failure of Luttinger's theorem as formulated in Ref.~\onlinecite{Coleman2005}.
As a warm-up,
we first consider a single partially-filled band of conduction electrons 
and briefly recall the origin of the generalized Luttinger sum rule. 
Then we focus on the PAM and derive a generalized Luttinger sum rule for $n_c + n_f$
(there are no useful separate sum rules for only $n_c$ or only $n_f$,
as these are not conserved quantities).
It involves not only the FS, comprising
all $\bk$ points in the Brillouin zone at which $G_{f\bk}(\izeroplus)$ 
and $G_{c\bk}(\izeroplus)$ have poles, but also the Luttinger surface (LS), at which 
$\Sigma_{f\bk}(\izeroplus)$ diverges \cite{Dzyaloshinskii2003}.
We then express our results purely through cluster quantities that are directly available from cellular DMFT calculations without
requiring reperiodization or interpolation. Thereafter, we show that the discontinuous jump of the FS volume when crossing the QCP 
into the RKKY phase is accompanied by the emergence of a LS while $I_\Luttinger$ remains zero throughout.
Finally, we discuss this finding together with the Hall coefficient calculated from our data and relate it to experimental findings on \YRS\
and \CCI.

\subsection{Luttinger's theorem for a single band}
\label{sec:Luttinger_lattice}
We begin by recalling standard arguments leading
to Luttinger's theorem, following Refs.~\cite{Luttinger1960,Abrikosov1963,Coleman2005,Nishikawa2018}. 
We consider a one-band model at $T=0$, 
with propa\-gator $G_{\bk}(\zomega) = [\omegaplus - \epsilon_{\bk} 
- \Sigma_{\bk}(\zomega)]^{-1}$.
The average electron density can be expressed as 
\begin{align}
n &= \frac{-2}{\pi} \im \! \int_\bz \frac{\mr{d}\bk}{\mc{V}_\bz} \int \mr{d}\omega \, G_{\bk}(\omegaR) \, \theta (-\omega) \, , 
\label{eq:n_c}
\end{align} 
where the prefactor 2 accounts for spin, $\mc{V}_\bz$ is the volume of the Brillouin zone, and the step function $\theta(-\omega)$ is the zero-temperature limit of the Fermi function.

For the next step, we need the identity 
\begin{equation}
G_\bk  = \partial_{\zomega} \ln G^{-1}_\bk + G_\bk \, \partial_{\zomega} \Sigma_\bk \, ,
\label{eq:ln_G_c_omega}
\end{equation}
expressing the correlator through derivates. 
Using the latter 
in \Eq{eq:n_c}, together with $\im\ln G^{-1}_\bk = \arg G^{-1}_\bk$, we obtain
\begin{align}
\label{eq:standard-Luttinger}
n & =  2  v_\fs + 2 I_\Luttinger 
 \, , 
\\ 
\label{eq:define-vFS-standard}
v_\fs & = \int_\bz \frac{\mr{d}\bk}{\mc{V}_\bz} 
\frac{\delta_{\bk}}{\pi} \, , 
\\ \nonumber
\delta_{\bk}  & = 
- \im \!  \int_{-\infty}^{0} \!\!\! {\rm{d}}\omega \,  
\partial_\omega \ln G_{\bk}^{-1} (\omegaR) 
= - \Bigl[\arg G_{\bk}^{-1}(\omegaR) \Bigr]_{-\infty}^{0} \, , 
\\ 
\label{eq:LuttingerIntegral-Ic} 
I_\Luttinger & =  
\frac{-1}{\pi} \im \!  \int_\bz \frac{\mr{d}\bk}{\mc{V}_\bz} \int_{-\infty}^0 \!\!\! \mr{d}\omega \, 
G_{\bk} (\omegaR) \, \partial_\omega \Sigma_{\bk} (\omegaR) \, .
\end{align}
Here, $\delta_{\bk}$ is the phase shift of $G_{\bk}^{-1}(\omegaR)$ 
$\omega = -\infty$ and $0$; $v_\fs$ is a shorthand for its integral over the 
BZ;  and $I_\Luttinger$ is the Luttinger integral
(cf. Eq.~\eqref{eq:I_Luttinger_generic}).
Equation~\eqref{eq:standard-Luttinger} is the generalized Luttinger sum rule. 
It is an exact expression of the electron density $n$; no assumptions on FL behavior have been made yet. 
We will show below that in a FL, the average phase shift \Eq{eq:define-vFS-standard} is given by the FS volume.
As mentioned above,
it can be argued on rather general grounds that $I_\Luttinger = 0$ in many cases~\cite{Coleman2005}. 
Conditions are that (i) the interaction preserves the U(1) charge symmetry 
and (ii) $\Sigma$ is $\Phi$-derivable, i.e.\ $\Sigma = \delta\Phi/\delta G$, where $\Phi$ is the LW functional.
Condition (ii) breaks down if $\Sigma$ is sufficiently singular, i.e.\ 
if no
functional $\Phi$ exists whose variation
w.r.t.\ $G$ produces $\Sigma$~\cite{Dave2013,Heath2020};
in that case, $I_\Luttinger$ can become non-zero.
Now, \textit{if} a sharp FS exists, i.e.\ if 
the imaginary part of the retarded self-energy
vanishes at $\omega = 0$ (cf.~\Eq{eq:define-Fermi-surface}),
\begin{equation}
\im \, \Sigma_{\bk}(\izeroplus) = - \zeroplus ,
\label{eq:Im_Sigma_c=0}
\end{equation}
then the integral $v_\fs$ defined in \eqref{eq:define-vFS-standard} gives the FS volume. Let us 
recapitulate why this is the case. Condition \eqref{eq:Im_Sigma_c=0}
holds for regular Fermi liquids, and more generally in the perturbative regime considered by Luttinger and Ward~\cite{Luttinger1960,Luttinger1961}.
Now, \Eq{eq:Im_Sigma_c=0} implies  
$ \arg G_{\bk}^{-1} (\izeroplus) = 
\pi \theta \bigl( -\re \, G_{\bk}^{-1}(\izeroplus )\bigr)$, while  
$ \im \, \Sigma_{\bk} (\omegaR) < 0$ implies 
$ \arg G_{\bk}^{-1}(-\infty +  \izeroplus ) 
= \pi$. Therefore, the phase shift is 
\begin{align}
\nonumber
\delta_{\bk} & =  
\pi-  \pi \theta \bigl(-\re \, G_{\bk}^{-1}(\izeroplus) \bigr)
\\ 
&  
= \pi \theta \bigl(
- \epsilon_{\bk} \!-\!  \re \Sigma_{\bk}(\izeroplus) \bigr) 
= \pi \theta(-\epsilon^\ast_{\bk}) \, .
\label{eq:delta_Gc} 
\end{align}
By definition, the Fermi surface encloses all $\bk$ points in the
Brillouin zone having 
$\epsilon^\ast_{\bk} < 0$.
For these, $\delta_{\bk}/\pi$ equals $1$, for all others it vanishes. Hence,  \Eq{eq:define-vFS-standard} 
reduces to 
\begin{align}
  v_\fs &    
=  \int_\bz \frac{\mr{d}\bk}{\mc{V}_\bz} 
\theta (- \epsilon^\ast_{\bk})  \, ,
\label{eq:define-v_FS}
\end{align}
which is the FS volume measured  in units of $\mc{V}_{\mr{BZ}}$ (hence dimensionless). 
Moreover, in a normal FL,
$\Sigma$ is not singular, hence $I_\Luttinger = 0$ holds,
hence the generalized Luttinger sum rule \eqref{eq:standard-Luttinger}
reduces to the Luttinger sum rule for a FL, $n  = 2 v_\fs$,
relating the density to the FS volume.
\subsection{Generalized Luttinger theorem for the PAM}
\label{sec:generalized-Luttinger-theorem}

\begin{figure}[bt!]
\includegraphics[width=\linewidth]{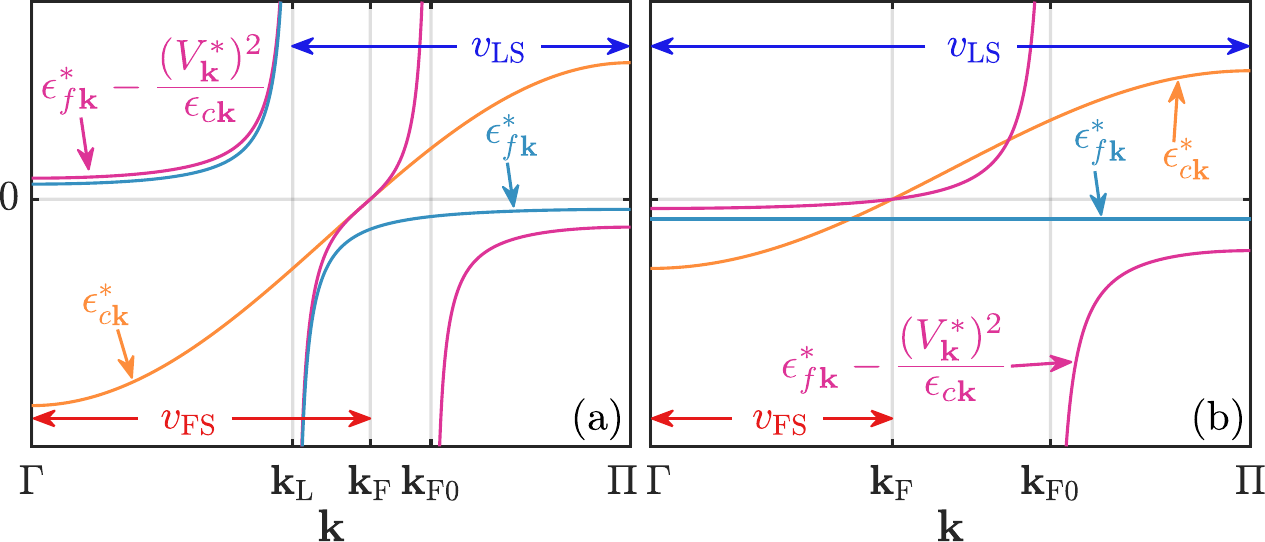}
\caption{%
Sketch of the qualitative behavior of 
$\epsilon_{f\bk}^{\ast}$, $\epsilon_{c\bk}^{\ast}$ and $\epsilon^{\ast}_{f\mathbf{k}} - 
(V^{\ast}_{\mathbf{k}})^2 / \epsilon_{c\mathbf{k}}$
in (a) the RKKY phase and (b) the Kondo phase. 
They are related to the Green's functions and the $c$-electron self-energy via
$\re\, G_{c\bk}^{-1}(0) = 
- \epsilon_{c\bk}^{\ast}/ Z_{c\bk}$, 
$\re\, G_{f\bk}^{-1}(0) = 
- \bigl[\epsilon^{\ast}_{f\mathbf{k}} - (V^{\ast}_{\mathbf{k}})^2/ \epsilon_{c\mathbf{k}}\bigr]/ Z_{f\bk}$ and $\re\, \Sigma_{c\bk}(0) = 
-(V^{\ast}_{\mathbf{k}})^2/\epsilon^{\ast}_{f\mathbf{k}}$. The FS (marked by $\bk_{\mr{F}}$) is defined by $\epsilon_{c\bk}^{\ast} = \epsilon^{\ast}_{f\mathbf{k}} - 
(V^{\ast}_{\mathbf{k}})^2/ \epsilon_{c\mathbf{k}} = 0$; its inside is defined by $\epsilon_{c\bk}^{\ast} < 0$. $|\epsilon_{f\bk}^{\ast}| = \infty$ ($\re\, \Sigma_{c\bk}(0) = 0$) defines the LS (marked by $\bk_{\mr{L}}$), with its inside defined by $\epsilon_{f\bk}^{\ast} < 0$. The function $\epsilon^{\ast}_{f\mathbf{k}} - 
(V^{\ast}_{\mathbf{k}})^2/\epsilon_{c\mathbf{k}}$
changes sign via a pole both at the LS (due to $\epsilon^{\ast}_{f\mathbf{k}}$) and at the free FS (marked by $\bk_{\mr{F0}}$) due to $\epsilon_{c\bk} = 0$.
In the Kondo phase, $\epsilon^{\ast}_{f\bk}$ remains negative everywhere in the BZ while in the RKKY phase,  $\epsilon^{\ast}_{f\bk}$ changes sign between $\Gamma$ and $\Pi$ [c.f. also Fig.~\ref{fig:QP_para}] via a pole at the LS.
As a result, $\epsilon^{\ast}_{f\mathbf{k}} - 
(V^{\ast}_{\mathbf{k}})^2/\epsilon_{c\mathbf{k}}$ changes sign twice in the Kondo phase (via a zero at $\bk_{\mr{F}}$ and a pole at $\bk_{\mr{F0}}$) and three times in the RKKY phase (via a pole at $\bk_{\mr{L}}$, a zero at $\bk_{\mr{F}}$ and a pole at $\bk_{\mr{F0}}$).
The pole at $\bk_{\mr{L}}$ shifts the position of $\bk_{\mr{F}}$ in the RKKY phase compared to its position in the Kondo phase.
Note that the distance between $\bk_{\mr{F}}$ and $\bk_{\mr{F0}}$ in (a) is exaggerated in the sketch above. 
}
\label{fig:Luttinger_PAM_sketch}
\end{figure}

The PAM describes hybridized $f$ and $c$ electrons, the former with local interactions, the latter without. Their correlators, $G_{f}$ and 
$G_{c}$, are coupled through \Eqs{eq:Glatt}. Importantly,
because $\Phi$ does not depend on propagators involving non-interacting orbitals~\cite{Abrikosov1963},
the LW functional
for the PAM
depends only on $G_f$, not on $G_c$ or $G_{fc}$.
Therefore, $\Sigma_f = \delta \Phi/\delta G_f$ is the only $\Phi$-derivable, proper (i.e.\ 1-particle irreducible) self-energy 
in the PAM.
$\Sigma_{c}$ knows about interactions only via 
its dependence on $\Sigma_{f}$ and hence is not a proper self-energy;
in particular $\delta \Phi/\delta G_c = \delta \Phi/\delta G_{fc} = 0$.
Analogous to the previous subsection, we first derive general formulas for the phase shifts. 
We then make assumptions on the self-energies compatible with our observations in Sec.~\ref{sec:SingleParticleCluster} and~\ref{sec:Ak_vs_w}.
These allow us to write down expressions in terms of the FS and LS volumes.
Our starting point again is an identity 
expressing correlators through derivatives. Equation~\eqref{eq:Gmatrix} implies
\begin{align}
\tr \, G_{\bk} &=
G_{f\bk} + G_{c\bk} = 
\tr \, G_{\bk} \partial_\zomega G_{\bk}^{-1} + G_{f\bk} \, \partial_\zomega \Sigma_{f\bk} \nonumber \\
&= 
\partial_\zomega \tr \, \ln G_{\bk}^{-1} + G_{f\bk} \, \partial_\zomega \Sigma_{f\bk} \nonumber \\
&=
\partial_\zomega \ln G_{c\bk}^{-1} +  \partial_\zomega \ln \Sigma_{c\bk}^{-1} + 
G_{f\bk} \, \partial_\zomega \Sigma_{f\bk} .
\label{eq:Gf_pSigmaf_pomega}
\end{align}
To derive the last equality,
note that the definitions \eqref{eq:Gmatrix} and \eqref{eq:Glatt} for the lattice correlators imply the relation
\begin{align}
\det G_{\bk}^{-1} &= (z-\epsilon_{c\bk})(z-\epsilon_f - \Sigma_{f\bk}) - V^2 \nonumber \\
&=  \Bigl(z-\epsilon_{c\bk} - \frac{V^2}{z-\epsilon_f - \Sigma_{f\bk}} \Bigr)(z-\epsilon_f - \Sigma_{f\bk}) \nonumber \\
& = G_{c\bk}^{-1} \Sigma^{-1}_{c\bk} / V^2 \, 
\end{align}
which, together with $\tr \, \ln G_{\bk}^{-1}= \ln \, \det G_{\bk}^{-1}$, yields \Eq{eq:Gf_pSigmaf_pomega}.
Integrating 
$G_f + G_c$ as in \Eq{eq:n_c},
\begin{align}
n_f \!+\! n_c &= \frac{-2}{\pi} \im \! \int_\bz \frac{\mr{d}\bk}{\mc{V}_\bz} \! \int_{-\infty}^0  \! \mr{d}\omega\, 
\bigl[G_{f\bk}(\omegaR) + G_{c\bk}(\omegaR) \bigr]  , 
\nonumber 
\end{align}
and using \eqref{eq:Gf_pSigmaf_pomega},
we obtain 
\begin{align}
n_f \!+\! n_c & = 2 v_\fs + 2 v_\ls + 2 I_\Luttinger \, , 
\label{eq:Luttinger_PAM} 
\end{align}
with ingredients defined in analogy to \Eqs{eq:define-vFS-standard}
to \eqref{eq:LuttingerIntegral-Ic}:
 \begin{flalign}
v_\fs &  = \int_\bz \frac{\mr{d}\bk}{\mc{V}_\bz} 
\frac{\delta_{c\bk} }{\pi} 
  \, , \quad 
  v_\ls   = \int_\bz \frac{\mr{d}\bk}{\mc{V}_\bz} 
\frac{\delta_{\smallsigma\bk} }{\pi} \, , 
\label{eq:define-v_FS-v_SS}
\\ 
\nonumber
\delta_{c\bk}  & = 
- \im \!  \int_{-\infty}^{0} \!\!\! {\rm{d}}\omega \,  
\partial_\omega \ln G^{-1}_{c\bk} (\omegaR) 
= - \Bigl[\arg G_{c\bk}^{-1}(\omegaR) \Bigr]_{-\infty}^{0} \, , 
\\ \nonumber
\delta_{\smallsigma \bk} & = 
- \im \!  \int_{-\infty}^{0} \!\!\! {\rm{d}\omega} \,  
\partial_\omega \ln \Sigma^{-1}_{c  \bk} (\omegaR) 
= + \Bigl[\arg \Sigma_{c\bk}^{+1}(\omegaR) \Bigr]_{-\infty}^{0} \, ,  
 \\ \label{eq:LuttingerIntegralc}
I_\Luttinger & =  
\frac{-1}{\pi} \im \!  \int_\bz \frac{\mr{d}\bk}{\mc{V}_\bz} \int_{-\infty}^0 \!\!\! \mr{d}\omega \, 
G_{f\bk} (\omegaR) \, \partial_\omega \Sigma_{f\bk} (\omegaR) \, . 
\end{flalign}
Here, $\delta_{c\bk}$ and $\delta_{\smallsigma \bk}$ describe 
the phase shifts of $G_{c\bk}^{-1}$ and $\Sigma_{c\bk}^{+1}$ 
between $\omega = -\infty$ and $0$;
$v_\fs$ and $v_\ls$ are shorthands for their integrals over the 
BZ; and 
$I_\Luttinger$ is the Luttinger integral, now involving only $f$ (no $c$) functions.
\Eqs{eq:Luttinger_PAM} to \eqref{eq:LuttingerIntegralc} 
are exact and, in the spirit of
Refs.~\onlinecite{Curtin2018,Nishikawa2018}, we will refer to 
\eqref{eq:Luttinger_PAM} as generalized Luttinger sum rule
for the PAM, or Luttinger-PAM sum rule, for short.
To the best of our knowledge, the existence of such a relation, 
involving the phase shifts not only of 
$G_{c\bk}^{-1}$ but also of $\Sigma_{c\bk}^{+1}$, 
has so far not been appreciated in the literature.
The form \eqref{eq:Luttinger_PAM} is specific to the PAM. For other 
multiband models, the arguments presented here will have to be suitably adapted.
Now, \textit{if} a sharp FS exists, i.e.\ if
\begin{align}
\label{eq:Im_Sigma_cf=0}
\im \, \Sigma_{c \bk}(\izeroplus) &= - \zeroplus \, ,  
\end{align}
then the integrals $v_\fs$ and $v_\ls$ defined in 
\Eqs{eq:define-v_FS-v_SS} give the FS and LS volumes, 
respectively. 
The argument proceeds as in the previous subsection.
Equation \eqref{eq:Im_Sigma_cf=0} implies  
$ \arg G_{c\bk}^{-1} (\izeroplus) = 
\pi \theta \bigl( -\re \, G_{c\bk}^{-1}(\izeroplus)\bigr)$
and
$ \arg \Sigma_{c\bk} (\izeroplus) \!=\! 
-\pi \theta \bigl( -\re \, \Sigma_{c\bk}(\izeroplus)\bigr)$, while  
$ \im \, \Sigma_{c\bk} (\omegaR) \!<\! 0$ implies 
$ \arg G_{c\bk}^{-1}(-\infty   + \izeroplus ) 
= \pi$ and $ \arg \Sigma_{c\bk}(-\infty  +\izeroplus ) = -\pi$. 
Therefore, the phase shifts in Eq.~\eqref{eq:LuttingerIntegralc} yield:
\begin{subequations}
\begin{flalign}
\nonumber
\delta_{c\bk} & =  
\pi-  \pi \theta \bigl(-\re \, G_{c\bk}^{-1}(\izeroplus) \bigr) &
\\ 
\label{eq:delta_Gc} 
&= \pi \theta \bigl(
- \epsilon_{c\bk} \!-\!  \re \Sigma_{c\bk}(\izeroplus) \bigr) 
= \pi \theta(-\epsilon^\ast_{c\bk}) \, , &
\\
\delta_{\Sigma\bk} & =  
\pi -  \pi \theta \bigl(- \re \, \Sigma_{c\bk}(\izeroplus) \bigr)
= \pi \theta \bigl( \re \Sigma_{c\bk}(\izeroplus) \bigr) . 
\hspace{-1cm} & 
\label{eq:delta_Sigmac} 
\end{flalign}
\end{subequations}
The phase shifts $\delta_{c\bk}$ and $\delta_{\Sigma\bk}$ are either $0$ or $\pi$. The jump between these values
occurs at the FS, defined by $\re \, G_{c\bk}^{-1}(\izeroplus) = 0$ 
and the LS, defined by $\re \Sigma_{c\bk}(\izeroplus) = 0$, respectively (see Eqs.~(\ref{eq:FS_def}) and Fig.~\ref{fig:Luttinger_PAM_sketch}).
Thus, \Eqs{eq:define-v_FS-v_SS} reduce to
\begin{align}
  v_\fs &  
=  \int_\bz \frac{\mr{d}\bk}{\mc{V}_\bz} 
\theta (- \epsilon^\ast_{c\bk})  \, ,
\quad v_\ls 
=  \int_\bz \frac{\mr{d}\bk}{\mc{V}_\bz} 
\theta (\re \Sigma_{c\bk}(\izeroplus))  \, ,
\label{eq:define-v_LS_PAM}
\end{align}
giving the FS and LS volumes in units of $\mc{V}_{\mr{BZ}}$.
Thus, if a sharp FS exists, the Luttinger-PAM sum rule \eqref{eq:Luttinger_PAM}
relates the density of $c$ and $f$ electrons to the FS and  
LS volumes $v_\fs$ and $v_\ls$ and the Luttinger integral $I_\Luttinger$.

%

The above arguments are directly applicable 
to our CDMFT+NRG results for the PAM, since 
the condition \eqref{eq:Im_Sigma_cf=0} is consistent with our 
 results for $0<V \neq \Vc$: indeed,
we find that 
the imaginary part of $\Sigma_c$ vanishes at $\omega=0$, both in the 
effective impurity model (see Sec.~\ref{sec:SingleParticleCluster}) and after reperiodization (see Sec.~\ref{sec:Ak_vs_w}).
Since $\Phi$ depends only on $G_f$, so that 
$\delta\Phi/\delta G_c = 0$ and $\delta\Phi/\delta G_{fc} = 0$,
$I_\Luttinger$ as defined in Eq.~\eqref{eq:LuttingerIntegralc}
has the same form as Eq.~\eqref{eq:I_Luttinger_generic}. We therefore expect
$I_\Luttinger = 0$ if the functional $\Phi$ exists in the vicinity of $G_f$,
but we will not concern ourselves here with general considerations 
about the existence of the LW functional.
 We have, however,  checked numerically that $I_\Luttinger = 0$ holds in our study of the PAM, for all considered values of $V$ at $T=0$, i.e.\ both in the Kondo and RKKY regimes. Thus we henceforth assume that $I_\Luttinger = 0$ throughout.
While $n_c$ and $n_f$ evolve smoothly with $V$, sudden jumps can occur for 
$v_\fs$ and $v_\ls$, which must compensate each other appropriately
if $I_\Luttinger$ remains zero.
In particular, by \Eq{eq:Luttinger_PAM} a 
jump in $v_\ls$ induces a 
jump in $v_\fs$, 
implying a FS reconstruction, even though $I_\Luttinger = 0$ remains unchanged. 
%

\subsection{Luttinger's theorem in CDMFT: computing $v_\fs$, $v_\ls$ and $I_\Luttinger$ without reperiodization} 
\label{sec:Luttinger_cluster}

%
The formulas derived in \Sec{sec:Luttinger_lattice} require explicit knowledge of the $\bk$-dependence of $\Sigma_f$ to compute $v_{\fs}$, $v_{\ls}$ and $I_\Luttinger$. As CDMFT artificially breaks translation invariance, we  have to reperiodize $\Sigma_f$ if we want to acquire knowledge on its $\bk$-dependence. Reperiodization is however a post-processing step which is to some extent ad-hoc. The specific choice of reperiodization will affect the values of the aforementioned quantities. 
In particular, the question whether $I_{\mr{L}} = 0$ holds can therefore not be answered conclusively when relying on some reperiodization scheme. 
In the following, we therefore provide and motivate formulas for $v_\fs$, $v_\ls$ and $I_\Luttinger$ which do not require reperiodization.
First, we note that the momentum integrals in the preceding part of this section represent a trace over all quantum numbers.
In case of translational invariance, it is convenient to use the momentum basis to perform this trace, as the Green's functions and self-energies are
diagonal in this basis.
In the 2-site CDMFT approach, it is however more convenient to represent them as $2\times2$ matrices depending on momenta $\bK$ in the
cluster BZ~(cBZ). This leads to the replacement
\begin{equation}
\int_{\mr{BZ}} \frac{\mr{d}\bk}{\mc{V}_{\mr{BZ}}} \to 
\frac{1}{2}\tr  \int_{\mr{\text{cBZ}}} \! \frac{\mr{d}\bK}{\mc{V}_{\mr{\text{cBZ}}}} 
\label{eq:k_latt_to_K_cl}
\end{equation}
in the formulas presented in \Sec{sec:generalized-Luttinger-theorem}. Here, 
 $\mc{V}_{\mr{\text{cBZ}}} = \mc{V}_{\mr{\text{BZ}}}/2$ the volume of the cBZ and $\tr$ is the trace of the $\bK$-dependent $2\times2$ matrices.
Now, within the CDMFT approximation, $\Sigma_f$ and therefore also $\Sigma_c$ are independent of $\bK$,
\begin{align}
\Sigma_{x\bK}(\zomega) = \Sigma_{x}(\zomega) 
\quad
(x = f,c) \, .
\end{align}
(This $\bK$-independence breaks the translation invariance.)
Moreover, the cluster propagators of Eq.~\eqref{eq:Gcluster}, defined as $\bK$-integrated objects,
are likewise $\bK$-independent:
\begin{equation}
G_{x}(\zomega) \equiv \int_{\mr{\text{cBZ}}}\frac{\mr{d}\bK}{\mc{V}_{\mr{\text{cBZ}}}} \, G_{x\bK}(\zomega)
\quad
(x = f,c) \, .
\label{eq:G-K-independent}
\end{equation}
Using \Eqs{eq:k_latt_to_K_cl} to \eqref{eq:G-K-independent},
the ingredients \eqref{eq:define-v_FS-v_SS} to \eqref{eq:LuttingerIntegralc} of the Luttinger-PAM sum rule \eqref{eq:Luttinger_PAM}  can now readily
be transcribed to obtain the following expressions:
 \begin{flalign}
 \label{eq:generalizedLuttinger_CDMFT}
v_\fs & = \tr \int_{\mr{cBZ}} \frac{\mr{d}\bK}{\mc{V}_{\mr{cBZ}}} 
\frac{\delta_{c\bK}}{2\pi} \, , \quad 
v_\ls = \tr \int_{\mr{cBZ}} \frac{\mr{d}\bK}{\mc{V}_{\mr{cBZ}}} 
\frac{\delta_{\smallsigma}}{2\pi} \, , 
\\
\nonumber
\delta_{c\bK}  & = 
- \im \!  \int_{-\infty}^{0} \!\!\! {\rm{d}}\omega \,  
\partial_\omega \ln G^{-1}_{c\bK} (\omegaR) 
\\ \nonumber
\delta_{\smallsigma} & = 
- \im \!  \int_{-\infty}^{0} \!\!\! {\rm{d}\omega} \,  
\partial_\omega \ln \Sigma^{-1}_{c} (\omegaR) 
= + \Bigl[\arg \Sigma_{c}^{+1}(\omegaR) \Bigr]_{-\infty}^{0} \, ,  
 \\
 \label{eq:LuttingerIntegral_CDMFT}
I_\Luttinger & =  
\frac{-1}{2\pi} \im \! \int_{-\infty}^0 \!\!\! \mr{d}\omega \, 
\tr\left[G_{f} (\omegaR) \, \partial_\omega \Sigma_{f} (\omegaR)\right] \, . 
\end{flalign}
Here, $v_\fs$, $v_\ls$
are expressed as traces of the cBZ integrals 
of the matrix-valued phase shifts $\delta_{c\bK}$, $\delta_{\Sigma}$. 
For the latter, which is  $\bK$-independent $\delta_{\smallsigma}$,
the integral is trivial.
For the $\bK$-dependent $\delta_{c\bK}$, we use
\begin{align}
\partial_\zomega \ln G^{-1}_{c\bK} (\zomega) 
=  
 G_{c\bK} (\zomega) (1-\partial_\zomega \Sigma_c(\zomega)) \, ,
 \end{align}
such that the $\bK$ integral yields a local cluster quantity,
\begin{align}
\label{eq:deltac_CDMFT}
\delta_c 
=\!
\int_{\mr{\text{cBZ}}}\!\frac{\mr{d}\bK}{\mc{V}_{\mr{\text{cBZ}}}} \delta_{c\bK}
\!=\! 
- \im \!  \int_{-\infty}^{0} \!\!\! {\rm{d}}\omega \,  
 G_{c} (\omegaR) (1\!-\!\partial_\omega \Sigma_c(\omegaR)) \, .
\end{align}
Note that in \Eq{eq:deltac_CDMFT}, $1-\partial_\zomega \Sigma_c(\zomega) \neq \partial_\zomega G_c^{-1}(\zomega) $ because 
$G_c(\zomega)$ is a $\bK$-integrated quantity, so that $G_c^{-1}(\zomega)$ contains
an additional hybridization term $\Delta_c(\zomega)$, c.f.\ \Eq{eq:Gcluster-Gc}.
Thus, Eqs.~\eqref{eq:generalizedLuttinger_CDMFT} reduce to
\begin{align}
v_\fs = \tfrac{1}{2\pi} \tr \,  \delta_c \, , 
\qquad
v_\ls = \tfrac{1}{2\pi} \tr \, \delta_\smallsigma \, .
\label{eq:FS_LS_CDMFT}
\end{align}
Equations \eqref{eq:FS_LS_CDMFT}, 
\eqref{eq:deltac_CDMFT} and \eqref{eq:LuttingerIntegral_CDMFT} 
achieve our stated goal of expressing $v_\fs$, $v_\ls$ and $I_\Luttinger$
purely through the local Green's functions and self-energies of the effective 2IAM.
They can hence can be computed without using reperiodization.
\subsection{Results}
%

%
\begin{figure}[bt!]
\includegraphics[width=\linewidth]{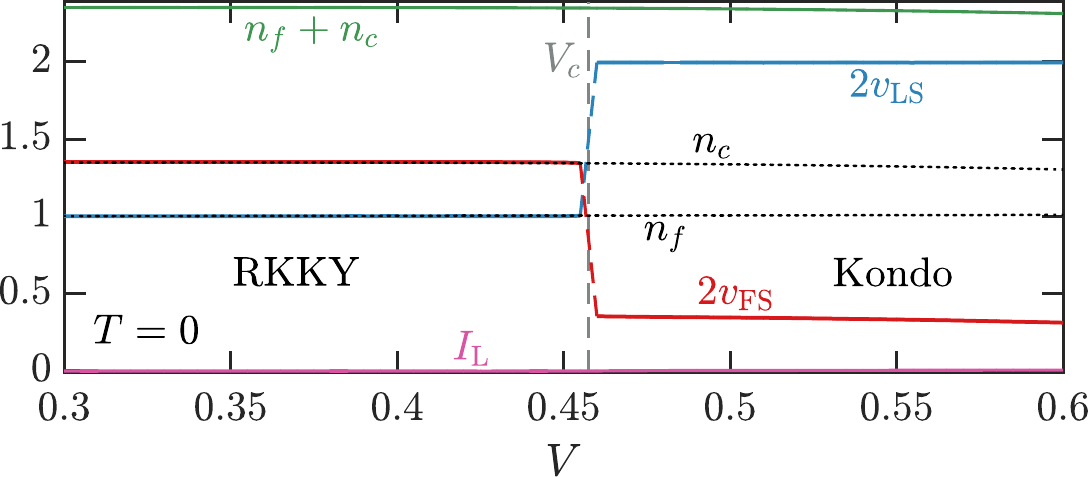}
\caption{%
FS and LS volumes, $v_\fs$ (red) and $v_\ls$ (blue), and the Luttinger integral $I_\Luttinger$ (pink), together with the particle numbers $n_f$ and $n_c$ (black dotted) and their sum (green), plotted as functions of $V$ at $T=0$.}
\label{fig:LT_PAM}
\end{figure}

%
Our CDMFT results for $v_\fs$, $v_\ls$ and $I_\Luttinger$ (obtained with the formulas from Sec.~\ref{sec:Luttinger_cluster}) are shown in Fig.~\ref{fig:LT_PAM},
together with the particle numbers $n_c$ and $n_f$. 
As mentioned before, we find $I_\Luttinger = 0$ for all considered values of $V$. 
We note that $I_\Luttinger = 0$ is an empirical numerical finding and may not hold if e.g. different fillings are considered~\cite{Altshuler1998,Rosch2007}; this will be explored in more detail in future work. Further, we find that the particle numbers $n_c$ and $n_f$ evolve  smoothly across the KB--QCP, in contrast to $v_\fs$ and $v_\ls$ which exhibit a jump when crossing the KB--QCP. The jumps are such that $2v_\fs + 2v_\ls = n_f + n_c$ evolves smoothly across the KB--QCP, i.e.\ the jumps of $v_\fs$ and $v_\ls$ compensate each other.
In the Kondo phase, $v_\ls = 1$ and $v_\fs$ is such that together, $v_\ls$ and $v_\fs$ account for the total particle number. 
Note that $v_\ls = 1$ means that the LS volume fills the whole BZ, i.e.\ there is no LS in the Kondo phase.
In the RKKY phase on the other hand, $v_\ls = \tfrac{1}{2} \simeq \tfrac{1}{2} n_f$ while $v_\fs \simeq \tfrac{1}{2} n_c$.
$v_\ls$ takes a fractional value in the RKKY phase, which means the LS volume fills a fraction of the BZ and there is
a LS. The presence of a LS can be linked to an emergent spinon FS~\cite{Fabrizio2023}, suggesting that the RKKY phase is a
fractionalized FL~\cite{Senthil2003,Senthil2004}. We will provide a more detailed analysis of this in future work.
In the Kondo phase, $v_\fs$ is the same as in the $U=0$ limit at the same filling (the same is true for $v_\ls$). The FS in the Kondo phase
is therefore as expected in a normal FL in the PAM. $v_\fs$ ``trivially'' (in the sense that one can infer it from
the $U=0$ limit) accounts for both the $f$ and $c$ electrons which is why it is called a ``large'' FS, even though 
the value of $v_\fs$ is smaller than in the RKKY phase.
In the RKKY phase by contrast, $v_\fs \simeq \tfrac{1}{2} n_c$ takes almost the value expected 
for $V=0$ with $U\neq0$. The FS seems to account only for the $c$ electrons and hence
is called a ``small'' FS.
Based on the shape and volume of the FS in the RKKY phase, one may be 
tempted to conclude that $c$ and $f$ electrons have decoupled. However, this is not the
case, as elaborated in sections~\ref{sec:SingleParticleCluster} and~\ref{sec:Ak_vs_w} above.
We now elaborate how the jump in $v_\fs$ and $v_\ls$ is connected to the sign change of $\epsilon_{f+}^\ast$ across the KB--QCP,
which we have discussed in Sec.~\ref{sec:QP_properties}. To make this connection, we examine Eq.~\eqref{eq:FS_LS_CDMFT}
for $v_\ls$ (computed without reperiodization)
in more detail, while assuming $\im \Sigma_c(0) = -0^{\scriptscriptstyle +}$ (Eq.~\eqref{eq:Im_Sigma_cf=0}), consistent with our results.
Analogously to our discussion of $\delta_{\smallsigma\bk}$ under the aforementioned assumption (see Eq.~\eqref{eq:delta_Sigmac}), 
the corresponding phase shift of the effective 2IAM (see Eq.~\eqref{eq:LuttingerIntegral_CDMFT})
is given by $\delta_\smallsigma = \pi\theta(\re\Sigma_c(\izeroplus))$ (note that both $\delta_\smallsigma$ and $\Sigma_c(\izeroplus)$ are
$2\times2$ matrices which are in our case diagonal in the $\pm$ basis). We can identify three different cases, which lead to 
distinct values of $v_\ls$: both eigenvalues of $\re\Sigma_c(\izeroplus)$ are (i) positive, (ii) negative or (iii) the eigenvalues
of $\re\Sigma_c(\izeroplus)$ have opposite signs, i.e.\ one is positive, the other negative. 
Inserting these into $v_\ls = \tr \, \delta_\smallsigma/(2\pi)$ (see Eq.~\eqref{eq:FS_LS_CDMFT}), we find (i) $v_\ls = 1$, (ii) $v_\ls = 0$
or (iii) $v_\ls = \tfrac{1}{2}$. The value of $v_\ls$ is therefore connected to the \textit{signs of the eigenvalues} of $\re\,\Sigma_c(\izeroplus)$,
which in our case are $\re\,\Sigma_{c\pm}(\izeroplus)$. 
The signs are related to those of $\epsilon_{f\pm}^\ast$ via Eq.~\eqref{eq:signchange}, namely
$\mr{sgn}\,\re\,\Sigma_{c\pm}(\izeroplus) = 
-\mr{sgn}\, \epsilon_{f\pm}^\ast$.
Thus, we find (i) $v_\ls = 1$ if $\epsilon_{f\pm}^\ast$ are both negative,
(ii) $v_\ls = 0$ if $\epsilon_{f\pm}^\ast$ are both positive and (iii) $v_\ls = \tfrac{1}{2}$ if $\epsilon_{f\pm}^\ast$
come with opposite signs.
In the Kondo phase, both $\epsilon_{f\pm}^{\ast}$ are negative (cf. Fig.~\ref{fig:QP_para}), just as in the $U\to0$ limit for the parameters we have chosen,
which leads to $v_\ls = 1$ in the Kondo phase. 
As discussed in more detail in Sec.~\ref{sec:QP_properties}, when the KB--QCP
is crossed from the Kondo to the RKKY phase, $\epsilon_{f+}^{\ast}$ changes sign and becomes positive while
$\epsilon_{f-}^{\ast}$ remains negative, which leads to $v_\ls = \tfrac{1}{2}$ in the RKKY phase.
Since $v_\ls$ jumps due to the sign change of $\epsilon_{f+}^{\ast}$, $v_\fs$ exhibits a corresponding jump.
We note that a jump of $v_\ls$ and $v_\fs$ is not at odds with a continuous QPT.
Indeed, $\epsilon_{f+}^{\ast}$ changes smoothly across the KB--QCP.
The reason why $v_\ls$ jumps is because it is not sensitive to the absolute value
of $\epsilon_{f\pm}^{\ast}$, but only to the signs. Signs are by definition discrete 
quantities and changes can only occur via jumps, which is why both
$v_\ls$ and $v_\fs$ change via a jump, even though the QPT is continuous.
We emphasize that a prerequisite for this sign sensitivity is that $\im \Sigma_c(\izeroplus) = 0^-$,
i.e.\ that the $T\to0$ phase is a FL. If $\im \Sigma_c(\izeroplus)$
were finite, $v_\ls$ and $v_\fs$ could change continuously.
Because finite $\im \Sigma_c(\izeroplus)$ would imply that the $T\to0$ phase is not a FL,
$v_\ls$ and $v_\fs$ then would not have the interpretations of being volumes in the BZ bounded by
sharply defined Fermi or Luttinger surfaces.
Further, our analysis shows that the FS reconstruction is a priori independent of possible translation symmetry breaking like antiferromagnetic or charge density wave order. Translation symmetry breaking increases the size of the unit cell and thus reduces the size of the BZ. While this may change the FS volume 
measured in units of the smaller BZ, it does \textit{not} change the average phase shifts $v_\ls$ and $v_\fs$ as they appear in Eqs.~\eqref{eq:define-v_FS-v_SS}
and~\eqref{eq:generalizedLuttinger_CDMFT}. For instance, the onset of antiferromagnetic order without jumps in the phase shifts $v_\ls$ and $v_\fs$ 
marks a SDW--QCP while 
jumps in the phase shifts $v_\ls$ and $v_\fs$ mark a KB--QCP, regardless of whether it is or is not accompanied by 
the onset of e.g. AFM order.
The generalized Luttinger's theorem also offers a possible explanation why the QCP in the 2IAM is stabilized by the CDMFT self-consistency condition. Without self-consistency, the QCP in the 2IAM is only stable if the scattering phase shifts are constrained by symmetry~\cite{Affleck1992,Jones2007,Eickhoff2020}. 
The symmetry constraint then prevents a smooth change of the phase shifts, resulting in a QCP where the phase shifts jump between allowed values~\cite{Affleck1992,Jones2007}. However, if such symmetry constraints are 
\textit{absent}, the phase shifts will simple change without a QCP~\cite{Jones2007}. 
In the self-consistent 2IAM, the phase shifts \textit{are} constrained, not by symmetry, but by the Luttinger sum rule, as we have seen in the previous discussions in this section.
We conjecture that this Luttinger sum rule constraint is the reason why the self-consistent 2IAM does have a QCP. 
A more detailed analysis will be presented in future work.
%

%
\begin{figure}[bt!]
\includegraphics[width=\linewidth]{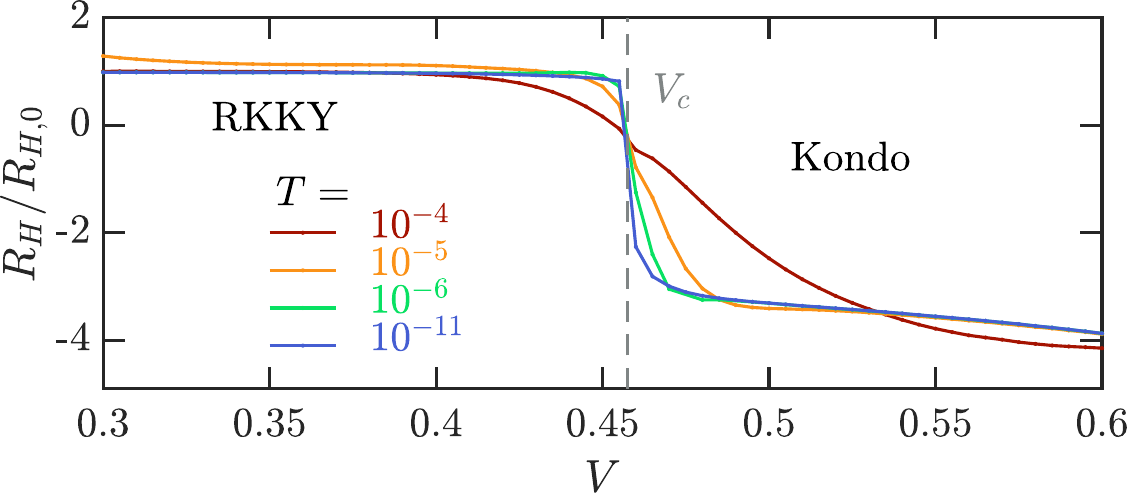}
\caption{%
Hall coefficient versus $V$ for different temperatures.}
\label{fig:Hall_PAM}
\end{figure}

%
One of the experimental hallmarks of a FS reconstruction at a KB--QCP is a sharp crossover of the Hall coefficient, $\RHall\sim 1/\nHall$, and thus of the Hall carrier density, $\nHall$~\cite{Coleman2001,Paschen2004,Maksimovic2022}.
This has been observed in experiments on \YRS~\cite{Friedemann2010,Friedemann:PhD,Paschen2004} and \CCI~\cite{Maksimovic2022}. A sign change of $\RHall$ has also been observed in \CCA when increasing $x$ from $0$ to $0.1$~\cite{Namiki2000}. To make contact to these experiments, we show the Hall coefficient calculated from our data in Fig.~\ref{fig:Hall_PAM}, as a function of $V$ at different temperatures. It is calculated using reperiodized self-energies with the formulas shown in App.~\ref{app:ssec:transport}. At all considered temperatures, we find two qualitatively distinct values of $\RHall$ deep in the Kondo regime (high $V$) and deep in the RKKY regime (low $V$). It shows a sign change across $\Vc$, reflecting a reconstruction from a particle-like FS in the Kondo regime to a hole-like FS in the RKKY regime. These are connected by a smooth crossover at high temperatures, which becomes sharper as temperature is lowered and almost step-like at the lowest temperature. This is qualitatively very similar to the experimental findings on \YRS~\cite{Friedemann2010,Friedemann:PhD,Paschen2004} and \CCI~\cite{Maksimovic2022}.
The analysis of generalized Luttinger sum rules for coupled $c$ and $f$ bands presented in this section makes no claims for generality---it focuses solely on the PAM and is based on assumptions consistent with our numerical results for this model. 
Nevertheless, similar analyses are likely possible for related models hosting QPTs with FS reconstruction, and we expect the LS to play a crucial role, there, too.
\section{Sommerfeld coefficient and entropy \label{sec:Sommerfeld}}
A further  quantity showing interesting behavior at a heavy-fermion QCP is the lattice Sommerfeld coefficient,
\begin{align}
\label{eq:defineSommerfeld}
\gamma_{\mr{latt}}=C_{\mr{latt}}/T = \partial S_{\mr{latt}}/ \partial T \, .
\end{align}
Here,  $C_{\mr{latt}}$ and $S_{\mr{latt}}$ are the lattice specific heat and entropy  per 2-site cluster (\textit{not} just the impurity contribution).
$S_{\mr{latt}}$ can be computed from the $f$-electron contribution to the entropy $S_f$ of the effective 2IAM and a correction term $S_{\mr{corr}}$~\cite{Georges1996,Gleis2023b}, as follows:
\begin{subequations}
\begin{align}
S_{\mr{latt}} &= S_f + S_{\mr{corr}} \, , \\
\label{eq:Sf}
S_{f} &= -\left.\frac{\partial \Omega_{f}}{\partial T}\right|_{\Delta_f = \mr{const.}} \, ,  \\
\Omega_f \! &= \! \Phi[G_f] \! - \! \tfrac{2}{\pi} \, \tr \! \int_{\omega} f_T(\omega) \bigl( \delta_{f} \! - \im  \left[\Sigma_{f} G_{f}\right]\bigr)  \, , \\
S_{\mr{corr}} &= \tfrac{2}{\pi} \, \tr \! \int_{\omega} \frac{\partial f_T(\omega)}{\partial T} \bigl( \delta_{c} + \delta_{\Sigma} \!- \!\delta_{f}\bigr)  \, , \\
\delta_{f}(\omegaR) &= -\im \,\ln G^{-1}_{f}(\omegaR) \, , \\
\delta_{\Sigma}(\omegaR) &= -\im \, \ln \Sigma^{-1}_{c}(\omegaR) \, , \\
\delta_{c}(\omegaR) &=  -\im \int_{\mr{cBZ}} \frac{\mr{d}\bK}{\mc{V}_{\mr{cBZ}}}  \ln G^{-1}_{c\bK}(\omegaR) \, .
\end{align}
\end{subequations}
Here, $\Phi$ is the Luttinger-Ward functional of the 2IAM, $f_T(\omega) = 1/\left[\exp(\omega/T) + 1\right]$ is the Fermi-Dirac distribution, $\tr$ is a trace over the cluster indices and $\delta_f$, $\delta_\Sigma$ and $\delta_c$ are matrix-valued phase shifts.
The derivative in Eq.~\eqref{eq:Sf}, which is evaluated while keeping the hybridization function $\Delta_f$ fixed, is accessible via NRG~\cite{Costi1994}.
The correction $S_{\mr{corr}}$ accounts for the fact that $\Delta_f(T)$ actually depends on temperature~\cite{Gleis2023b}.
To compute the Sommerfeld coefficient $\gamma_{\mr{latt}} = \partial S_{\mr{latt}}/\partial T$, we numerically differentiate $S_{\mr{latt}}(T)$.
The Sommerfeld coefficient is a measure of the
density of states. In a FL, it is proportional to the QP mass ($m^\ast$) and QP weight ($Z$), 
 $\gamma \sim m^\ast \sim Z^{-1}$~\cite{Giuliani2005} and hence is expected to be independent of temperature.
 By contrast, a $\gamma_{\mr{latt}}\sim\ln(T)$ dependence, indicating NFL behavior, has been observed almost universally for numerous compounds in strange metallic regimes at finite temperatures above QCPs~\cite{Maple1995,Coleman2002,Coleman2007,Maple2010}, e.g.\ for \YRS~\cite{Trovarelli2000,Custers2003,Gegenwart2006,Krellner2009}, \CCA~\cite{Loehneysen1996,Loehneysen1996a} and \CCI~\cite{Bianchi2003}.
Further, $\gamma_{\mr{latt}}$ has been observed to diverge when approaching 
the KB--QCP from either side~\cite{Custers2003,Gegenwart2003}; this implies a divergent effective mass $m^\ast$ at the QCP. A divergence of $m^{\ast}$ when approaching the KB--QCP from either side has been observed in 
many HF materials using different measurement techniques~\cite{Loehneysen1996,Gegenwart2002,Custers2003,Shishido2005}.
This is direct evidence for a breakdown of the FL at the QCP~\cite{Custers2003} 
%

%
\begin{figure}[bt!]
\includegraphics[width=\linewidth]{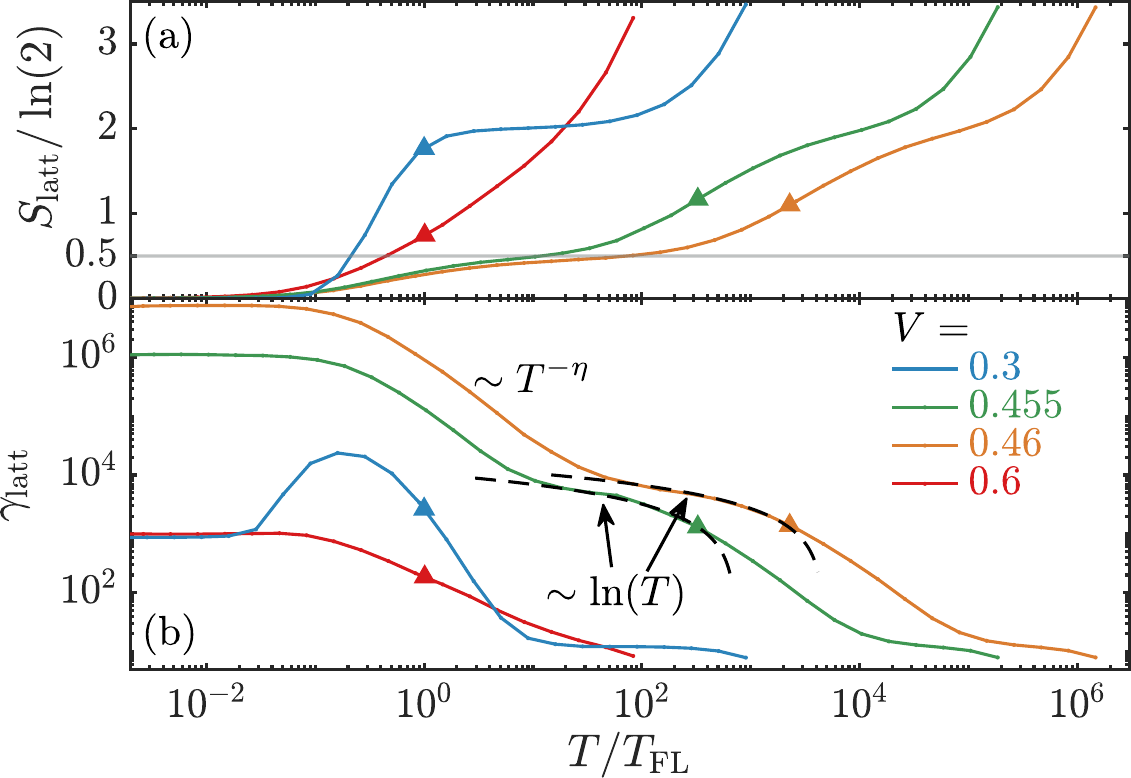}
\caption{%
Temperature dependence of (a) the lattice entropy and (b) the specific heat coefficient per 2-site cluster. Triangles mark the position of $\TNF/\TFL$.}
\label{fig:Sent_SFC_PAM}
\end{figure}

\begin{figure*}[bt!]
\includegraphics[width=\textwidth]{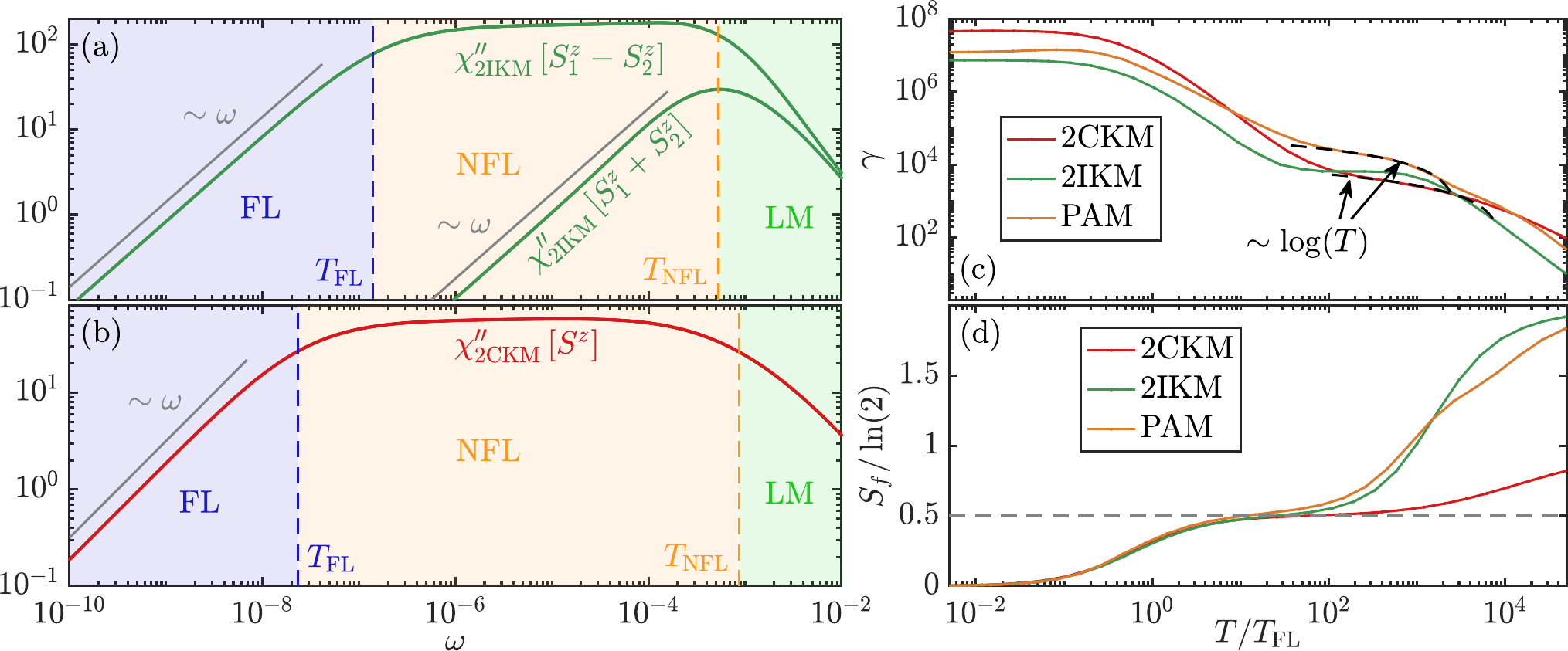}
\caption{%
(a,b) Spectral part of the spin susceptibility of (a) a 
2ICK and (b) a 2CKM close to their QCPs. 
$\chi_{\mathrm{2IKM}}''[S_1^z-S_2^z]$ and $\chi_{\mathrm{2CKM}}''[S^z]$ show similar features as $\chi''[X^{zx}]$ in the PAM close to the KB--QCP (c.f. Fig.~\ref{fig:Chi_QCP}). (c) Sommerfeld coefficient and (d) impurity contribution to the entropy of the 2IKM, 2CKM and PAM close to their QCPs. The data for the PAM is taken at $V=0.46$. Most notable, in the NFL region, $\gamma$ shows a $\ln(T)$ behavior for the 2CKM and the PAM, and $S_f \simeq \tfrac{1}{2}\cdot\ln(2)$ 
for all three models.
\label{fig:2CK_vs_PAM}
}
\end{figure*}

%
Our results for $S_{\mr{latt}}(T)$ and $\gamma_{\mr{latt}}(T)$ are shown Figs.~\ref{fig:Sent_SFC_PAM}(a,b).
At very high temperatures ($T \gg \TNF)$, $S_{\mr{latt}}$ decreases from some high-$T$ value and, for $V<0.6$, exhibits a shoulder around $2\ln(2)$ (the entropy of two local moments). This shoulder becomes more pronounced for lower $V$ (and is not visible for $V=0.6$). In the same high-temperature range, $\gamma_{\mr{latt}}$ shows a shoulder which is also more pronounced for lower $V$ (and again not visible for $V=0.6$). Thus, the entropy in this high-temperature regime has a linear $T$-dependence, $S_{\mr{latt}}(T) \simeq 2\ln(2) + a\cdot T$, with a slope $a$ which is almost independent of $V$ (since $\gamma_{\mr{latt}} = \partial_T S_{\mr{latt}}$ is roughly independent of $V$ in the high-$T$ shoulder region). This behavior can be understood in terms of thermally fluctuating $f$ moments (leading to the shoulder of $S_{\mr{latt}}$) and unrenormalized thermally excited $c$ electrons, leading to the $V$-independent linear-in-$T$ dependence and hence a $V$-independent shoulder in $\gamma_{\mr{latt}}$. Such a temperature dependence is characteristic for the LM regime, which is also expected to be more pronounced for lower $V$.
As $T$ is lowered, the free moments begin to hybridize with the $c$ electrons. This leads to screening of the moments, both by forming Kondo singlets and inter-impurity singlets, thus reducing the entropy. Far away from $\Vc$ (blue, red curves), where $\TFL \simeq \TNF$, the entropy drops to zero as $T\to0$ without notable features at intermediate temperatures. Deep in RKKY regime ($V = 0.3$), this entropy decrease reflects the formation of $f$-electron singlets and is very rapid, leading to a pronounced hump of $\gamma_{\mr{latt}}$ around $T \simeq \TFL \simeq \TNF$.
However, close to $\Vc$ (green, orange curves), $S_{\mr{latt}}(T)$ flattens considerably in the intermediate range $\TFL \!\lesssim \! T \! \lesssim \! \TNF$, resulting in a second shoulder near $\ln\sqrt{2}$.  
Concurrently, 
$\gamma_{\mr{latt}}(T)$ shows a logarithmic $T$ dependence  (black dashed lines), implying a $T \ln T$ behavior for the specific heat.  As mentioned above, this is an almost universal feature of heavy-fermion compounds with KB--QCPs.

The shoulder at  $\ln\sqrt{2}$ for the lattice entropy suggests that at the QCP, where $\TFL=0$, the zero-temperature entropy would be non-zero, $S_{\mr{latt}}(T\to0) = \ln\sqrt{2}$. 
This value is also found for the zero-temperature impurity entropy of the two-channel Kondo model and the two-impurity Kondo model, where it can be attributed to an unscreened Majorana zero mode at the QCP~\cite{Emery1992,Sengupta1994,Delft1998b} (see also the next section). 
For the PAM, this means that 2-site CDMFT predicts a non-zero, extensive entropy at $T=0$ \textit{at} the QCP. This suggests that the KB--QCP in 2-site CDMFT would be highly unstable to symmetry breaking orders tending to get rid of the non-zero entropy. It remains to be checked in future work (by studying larger cluster sizes) whether this extensive entropy is due to the finite cluster size, i.e.\ whether the entropy per lattice site at the KB--QCP scales to zero with increasing the cluster size, or whether the nonzero $T\to 0$ entropy per lattice site is robust, independent of cluster size.
When $T$ is decreased to 
about 1 to 2 orders of magnitude below $\TNF$,
 the $\ln T$ dependence of $\gamma_{\mr{latt}}$ turns to a $T^{-\eta}$ dependence, with $\eta\simeq 4/3$, before becoming constant below $\TFL$. A somewhat similar power-law $T$-dependence, with an onset temperature of less than 1 order or magnitude below $\TNF$,  has also been found in \YRS~\cite{Custers2003,Gegenwart2006,Krellner2009}, albeit with different exponent of $\eta=1/3<1$~\cite{Custers2003}. Ref.~\onlinecite{Custers2003} suggested that for \YRS, the $T^{-\eta}$ dependence is a property of the NFL. For our PAM however, this cannot be the case: since at $\Vc$ the NFL regime reaches down $T\to0$, it 
cannot support $T^{-\eta}$ behavior with $\eta > 1$, since its specific heat $C_{\mr{latt}}=T\gamma_{\mr{latt}} \sim T^{1-\eta}$ would diverge, which is thermodynamically impossible. Further, since $\gamma_{\mr{latt}} = \partial_T S_{\mr{latt}}$, the $T^{-\eta}$ behavior of $\gamma_{\mr{latt}}$ implies $S_{\mr{latt}}(T) \sim T^{1-\eta}/(1-\eta) + \mr{const}$ in this regime. If the powerlaw dependence would extend all the way down to $T=0$, $\eta > 1$ would imply $S_{\mr{latt}}(T=0) = -\infty$, which is clearly nonsense. Therefore, we view the $T^{-\eta}$ power law of $\gamma_{\mr{latt}}$ to be  a property of the NFL-FL \textit{crossover}, rather than of the 
NFL regime itself. Since $\gamma_{\mr{latt}} \sim \ln T$ in the NFL region is a much weaker singularity than the $T^{-\eta}$ crossover behavior, the latter takes over at a relatively high temperature compared to $\TFL$.
For $T < \TFL$, $\gamma_{\mr{latt}}(T)$ is constant for all $V$ values shown, as expected in a FL. It is orders of magnitude larger close to the QCP on either side 
(green, orange curves) than further away from it 
(blue, red curves), reflecting the divergence of the QP mass $m^{\ast} \sim \gamma_{\mr{latt}}(T=0) \sim Z^{-1}$ at the QCP.  It is noteworthy that far from the QCP, the value of $\gamma_\mr{latt}(T=0)$ for the RKKY phase is comparable to that in the Kondo phase. This suggests that in the RKKY phase the localized spins, though tending to lock into nearest-neighbor singlets, nevertheless contribute quite significantly to the density of states. Understanding this aspect in more detail is left as an interesting task for future work. 
%

\section{Relation to other models \label{sec:other_models}}%
The NFL regime in the PAM shares several similarities with the NFL in the two-impurity Kondo model~\cite{Affleck1992,Affleck1995,Jones2007,Mitchell2012a,Mitchell2012} and the two-channel Kondo model~\cite{Andrei1984,Tsvelick1984,Tsvelick1985,Sacramento1989,Sacramento1991,Affleck1991,Affleck1991a}:
(i)  The Sommerfeld coefficient $\gamma$ shows a region of $\ln(T)$-dependence [cf.~\Fig{fig:Sent_SFC_PAM}(b)]; (ii) the entropy takes the value of $\ln \sqrt{2}$; and (iii) plateaus in dynamical susceptibilities imply overscreening.
In this section, we compare the features of the effective 2IAM describing the PAM at $V=0.46$ close to the QCP to known features  of the two-impurity Kondo model
(2IKM) close to its  QCP.
We also include data on the two-channel Kondo model (2CKM)
 close its QCP ~\cite{Andrei1984,Tsvelick1984,Tsvelick1985,Sacramento1989,Sacramento1991,Affleck1991,Affleck1991a}, since its critical behavior  is known to be closely related to that of the 2IKM \cite{Mitchell2012,Mitchell2012a}.
The Hamiltonian describing the 2IKM is given by
\begin{equation}
H = \sum_{i=1,2} \Bigl[\sum_{k\sigma} \epsilon_k c^{\dagger}_{ik\sigma} c^\pdag_{ik\sigma} + J \vec{S}_i \cdot \vec{s}_i \Bigr] + K\vec{S}_1 \cdot \vec{S}_2 \, . 
\end{equation}
Here, $c^\pdag_{ik\sigma}$ destroys an electron in channel $i=1,2$ with energy $\epsilon_k$ and spin $\sigma$, 
$\vec{s}_i = \frac{1}{2} \sum_{kk'\sigma\sigma'} c^{\dagger}_{ik\sigma} \hat{\vec{\sigma}}_{\sigma\sigma'} c^\pdag_{ik'\sigma'}$
describes the local spin of channel $i$ at the origin, and $\vec{S}_i$ describe two impurity spin $1/2$ degrees of freedom at the origin. The impurity spins are coupled antiferromagnetically to the corresponding conduction electrons with coupling strengths $J>0$. $\vec{S}_1$ and $\vec{S}_2$ are further coupled antiferromagnetically with coupling strength $K>0$. The 2IKM can be tuned through a QCP from a Kondo regime at $K<K_c$, where $\vec{S}_i$ is screened by its corresponding bath, to an RKKY regime at $K>K_c$, where $\vec{S}_1$ and $\vec{S}_2$ form a singlet. At the QCP, $\TFL$ vanishes and an intermediate NFL region emerges, similar to the case of the PAM discussed in the main text. 
The NFL in the 2IKM is closely related to the NFL found in the 2CKM, described by the Hamiltonian
\begin{equation}
H = \sum_{i=1,2} \Bigl[\sum_{k\sigma} \epsilon_k c^{\dagger}_{ik\sigma} c^\pdag_{ik\sigma} + J_i \vec{S}\cdot\vec{s}_i \Bigr] \, .
\end{equation}
Here, there is only one impurity spin, coupled antiferromagnetically to two  channels, $i=1,2$, with coupling strengths $J_i>0$. The 2CKM has a QCP at $J_1=J_2$, where the impurity changes from being screened by channel $1$ at $J_1>J_2$ to being screened by channel $2$ at $J_1<J_2$. Close to its QCP, a NFL fixed point is found which extends to $T=0$ at the QCP and shows features similar to these of the NFL in the 2IKM and in the PAM as presented in the main part. 
In the following, we consider a 2IKM and a 2CKM tuned close to their respective QCP's. We compare their Sommerfeld coefficients, impurity contributions to the entropy and dynamical correlation functions to the $f$ electron contribution to the entropy and Sommerfeld coefficient of the self-consistent 2IAM describing the
PAM close to its QCP, shown in Fig.~\ref{fig:2CK_vs_PAM}. For both the 2IKM and the 2CKM, a box shaped, particle-hole symmetric density of states with width $2$ and height $1/2$ is used for both channels. As parameters, $J=0.256$ and $K=7.2\cdot 10^{-4}$ where used for the 2IKM and $J_1=0.29$, $J_2=0.2894$ was chosen for the 2CKM. Both models are solved with NRG using $\Lambda=3$ and keeping $N_{\mathrm{keep}}=2000$ $\mr{SU}(2)_{\mathrm{charge}}\times \mr{SU}(2)_{\mathrm{charge}}\times \mr{SU}(2)_{\mathrm{spin}}$ multiplets at every NRG iteration. 
Fig.~\ref{fig:2CK_vs_PAM}(a) shows $\chi_{\mathrm{2IKM}}''[S_1^z+S_2^z]$ and $\chi_{\mathrm{2IKM}}''[S_1^z-S_2^z]$, together with the NFL and FL scales extracted from the kinks of  $\chi_{\mathrm{2IKM}}''[S_1^z+S_2^z]$ and $\chi_{\mathrm{2IKM}}''[S_1^z-S_2^z]$ on the $\ln$-$\ln$ scale, respectively, similar as described for the PAM before. In Fig.~\ref{fig:2CK_vs_PAM}(b), we show the impurity spin susceptibility of the considered 2CKM, $\chi_{\mathrm{2CKM}}''[S^z]$, with the corresponding NFL and FL scales both extracted from  $\chi_{\mathrm{2CKM}}''[S^z]$. The similarity of $\chi_{\mathrm{2IKM}}''[S_1^z-S_2^z]$ in Fig.~\ref{fig:2CK_vs_PAM}(a), $\chi_{\mathrm{2CKM}}''[S^z]$ in Fig.~\ref{fig:2CK_vs_PAM}(b) and $\chi''[X^{zx}]$ in Fig.~\ref{fig:Chi_QCP} is evident.

The Sommerfeld coefficient $\gamma = \mathrm{d}S_f/\mathrm{d}T$, and the impurity contribution to the entropy $S_f$ are shown in Fig.~\ref{fig:2CK_vs_PAM}(c) and (d), respectively, for the 2IKM, the 2CKM and the PAM at $V=0.46$. Qualitatively similar behavior is found, most notably a plateau in $S_f$ with $S_f\simeq \tfrac{1}{2} \ln(2)$ for all three models, 
and a $\ln(T)$ dependence of $\gamma$ at intermediate temperatures for the 2CKM and the PAM (the 2IKM 
shows similar behavior, though without a clean
 $\ln(T)$ dependence). 
 A further elucidation of the nature of the NFL in the PAM will require an impurity model analysis as done in Refs.~\cite{Wang2020,Walter2019}. We leave this for future work.
 %


\section{Conclusion and Outlook}%
\label{sec:ConclusionOutlook}
%

%
We have presented an extensive 2-site CDMFT plus NRG study of the PAM, following up on and considerably extending and refining previous work done with an ED impurity solver~\cite{DeLeo2008,DeLeo2008a}. Leveraging the capabilities of NRG to resolve exponentially small energy scales on the real-frequency axis, we confirmed the existence of the KB--QCP found in Refs.~\onlinecite{DeLeo2008a,DeLeo2008}, which can be understood in terms of a continuous OSMT at $T=0$.
Beyond that, we unambiguously showed that the KB--QCP marks a second-order transition between two FL phases [Fig.~\ref{fig:FiniteSize}] which differ in their FS volumes [Sec.~\ref{sec:Luttinger}], leading to a sharp jump of the FS volume [Fig.~\ref{fig:LT_PAM}] and the Hall coefficient [Fig.~\ref{fig:Hall_PAM}].
We found that, in contrast to widespread belief~\cite{Kirchner2020}, the $f$-electron QP weight is non-zero both in the Kondo \textit{and} the RKKY phase [Sec.~\ref{sec:SingleParticleCluster}, in particular Fig.~\ref{fig:QP_para}] and only becomes zero at the QCP itself. We showed that the 
FS reconstruction across the KB--QCP can be understood in terms of a sign change of the effective level position $\epsilon^{\ast}_{f+}$ 
[Fig.~\ref{fig:QP_para} and Sec.~\ref{sec:Luttinger}], which is connected to the emergence of a dispersive pole in the RKKY phase [Fig.~\ref{fig:SECinv_vs_V_T1e-11_Cluster_log}].
An interesting consequence of the non-zero $f$-electron QP weight is that in the RKKY phase, the band structure consists of \textit{three} bands, with a very narrow band crossing the Fermi level [Sec.~\ref{sec:Ak_vs_w}]. We also showed that both in the Kondo and RKKY phases, the specific heat is linear in $T$, as expected from a FL; we find that the Sommerfeld coefficient diverges when the KB--QCP is approached from either side [Sec.~\ref{sec:Sommerfeld}].
We find that the physics at the KB--QCP and at non-zero temperatures in its vicinity is governed by a NFL fixed point [Sec.~\ref{sec:two-stage screening}], which has some resemblance to the NFL fixed points in the 2-impurity and 2-channel Kondo models [Sec.~\ref{sec:other_models}]. In this paper, we reported a strange-metal like $\sim T\ln T$ specific heat in the NFL region [Sec.~\ref{sec:Sommerfeld}]. A more detailed analysis of the NFL regime is provided in a companion paper~\cite{Gleis2023a}, where we show data regarding $\omega/T$ of the optical conductivity closely resembling experimental data~\cite{Prochaska2020} and evidence for linear-in-T resistivity. 
We should however also mention some caveats in our work, which may be addressed in future work.
For instance, while we provided some extensive formal treatment of Luttinger's theorem [Sec.~\ref{sec:Luttinger}], 
we refrained from offering any physical interpretation on how and why the FS in the RKKY phase can be small. Indeed,
Oshikawa's non-perturbative treatment of Luttinger's theorem~\cite{Oshikawa2000} implies 
that in case of a violation
of Luttinger's theorem, additional low-energy degrees of freedom must be present~\cite{Senthil2003,Bonderson2016}.
We leave their identification within the context of the PAM to future work. We expact that
this could also lead to a connection to slave particle theories~\cite{Senthil2003,Aldape2022}.
Further insights could possibly be pursued 
along the lines of recent work by Fabrizio~\cite{Fabrizio2020,Fabrizio2022,Fabrizio2023}. As we have pointed out repeatedly in this paper, the work cited above draws a connection between Luttinger surfaces
and fractionalized spinon excitations. Our work paves the way to explore this connection in detail in terms of a concrete example.

An unsatisfactory aspect of our 2-site CDMFT treatment is that it yields a non-zero entropy at the KB--QCP at $T=0$, see Sec.~\ref{sec:Sommerfeld}.
As already mentioned in that section, such a non-zero entropy would render the KB--QCP highly unstable to 
symmetry breaking. Whether this is a finite cluster size effect or continues to be the case also for larger cluster sizes needs to be checked
in future work.

Lastly, our CDMFT treatment requires reperiodization of self-energies to obtain a periodic self-energy with $\bk$-dependence. Reperiodization is an ad-hoc post-processing procedure. We have checked that our most important claims are consistent with our non-repriodized bare data. Nevertheless, in our view
it would be
important to cross-check our results in future studies, for instance with numerically exact methods. Such studies will be very useful for establishing the
range of applicability of CDMFT for describing KB physics.
Our work motivates several follow-up studies. For instance, as  mentioned before, it would be interesting to
explore the interplay between KB physics and potential symmetry breaking orders of all kind, e.g. antiferromagnetic or superconducting orders.
Similar studies could also be done for models appropriate to other classes of strongly correlated materials.
Obvious candidate material classes, which experimentally show quantum critical behavior quite similar to heavy fermions, are cuprates~\cite{Michon2019,Keimer2015,Michon2023,Badoux2016,Legros2019,Li2023}
or twisted bilayer graphene~\cite{Cao2018,Cao2020,Jaoui2022}.
\begin{acknowledgments}
We thank Assa Auerbach, Silke B\"uhler-Paschen, Andrey Chubukov, Piers Coleman, Dominic Else, Fabian Kugler, Antoine Georges, Sean Hartnoll, Andy Millis, Achim Rosch, Subir Sachdev, Qimiao Si, Katharina Stadler, Senthil Todadri, Matthias Vojta, Andreas Weichselbaum and Krzysztof W\'ojcik for helpful discussions.   This work was funded in part by the Deutsche Forschungsgemeinschaft under 
Germany's Excellence Strategy EXC-2111 (Project No.\ 390814868). It is part of the  Munich Quantum Valley, supported by the Bavarian state government with funds from the Hightech Agenda Bayern Plus. 
SSBL was supported by Samsung Electronics Co., Ltd (No. IO220817-02066-01), and also by a National Research Foundation of Korea (NRF) grant funded by the Korean government (MEST) (No. 2019R1A6A1A10073437).
G. K. was supported by the National Science Foundation Grant No. DMR-1733071.
\end{acknowledgments}
%


%
\renewcommand{\thefigure}{\thesection\arabic{figure}}
\counterwithin{figure}{section} 

\begin{appendix}
\section{CDMFT for the PAM \label{app:CDMFT}}
In the two-site CDMFT treatment of the PAM, the three-dimensional lattice is tiled into a superlattice of two-site clusters~\cite{Kotliar2001,DeLeo2008,Tanaskovic2011}. An effective action can then be constructed via the cavity method~\cite{Georges1996}. It takes the form of a 2IAM coupled to an effective bath, describing the rest of the lattice from the point of view of the two-site cluster in the CDMFT approximation. The $c$ electrons can either be treated explicitly, or they can be integrated out and merged with the effective bath, as they are non-interacting. We decided to treat the $c$ electrons explicitly, as this enables us to directly calculate dynamical susceptibilities involving $c$-electron degrees of freedom. Additional calculations, treating the $c$ electrons together with the bath, confirmed that both methods give the same results. Note that both methods have roughly the same amount of computational complexity for the NRG impurity solver. Since the $f$ electrons only couple to the bath via the $c$ electrons as they do not have any non-local hopping term, one obtains two bands of spinfull bath electrons in both cases.
For a detailed description of 2-site CDMFT for the PAM, see Ref.~\onlinecite{Tanaskovic2011}.

\subsection{Self-consistency}

In the CDMFT approximation,
the cluster Green's function 
is given by
\begin{subequations}
\begin{align}
\label{eq:app-G_loc,f}
G_{\textrm{loc}}(z) &=
\int_\bk 
\left( 
\begin{array}{cc}
z - \epsilon_f - \Sigma_f(z) & -V \\
-V & \mathcal{G}^{-1}_{0,c\bk}(z)
\end{array} \right)^{-1} \\
&= 
\left( 
\begin{array}{cc}
G_{\textrm{loc},f}(z)  & G_{\textrm{loc},fc}(z)  \\
G_{\textrm{loc},fc}(z)  & G_{\textrm{loc},c}(z) 
\end{array} \right) \, , 
\end{align}
\end{subequations}
where $\mathcal{G}_{0,c\bk}$ is the $c$-band Green's functions at $V\!=\!0$,
\begin{align}
\mathcal{G}_{0,c\bk}(z) =
\frac{1}{(z +\mu)^2-(\epsilonck^{0})^2}
\left( 
\begin{array}{cc}
z +\mu 
& e^{i k_x}\epsilonck^{0} \\
e^{-i k_x}\epsilonck^{0} & z +\mu
\end{array} \right) \, ,
\end{align} 
and $\Sigma_f(z)$ is the cluster $f$-electron self-energy (which is a proper, single-particle irreducible self-energy),
\begin{align}
\Sigma_f(z) = 
\left(
\begin{array} {cc}
\Sigma_{f11}(z) & \Sigma_{f12}(z) \\
\Sigma_{f21}(z) & \Sigma_{f22}(z) 
\end{array} \right) \, ,
\end{align}
which is $\bk$-independent in CDMFT. $G_{\textrm{loc},c}(z)$ can be computed by performing a momentum integral 
(c.f. Sec.~\ref{app:integration}),
\begin{align}
\label{Eq:Gcloc}
G_{\textrm{loc},c}(z) 
&= \int_\bk  \left[\mathcal{G}_{0,c\bk}^{-1}(z) - \Sigma_c(z)\right]^{-1} \, , 
\end{align}
while $G_{\textrm{loc},f}(z)$ and $G_{\textrm{loc},fc}(z)$ are related to $G_{\textrm{loc},c}(z)$ via Eq.~\eqref{eq:Gcluster-fc}.
$\Sigma_c(z)$ is the cluster $c$-electron self-energy (which is not single-particle irreducible), related to $\Sigma_f(z)$ via Eq.~\eqref{eq:Sigmac}.
$\Sigma_f(z)$ can be computed from an auxiliary self-consistent two-impurity Anderson model~(2IAM) [c.f. Eq.~\eqref{eq:2IAM}] with $G_\mr{loc}(z) = G_\mr{2IAM}(z)$, with the corresponding Green's functions of the 2IAM given in Eq.~\eqref{eq:Gcluster}. 

The self-consistent solution is not known a priori and has to be computed via a self-consistency cycle. For that, the hybridization function of the 2IAM  Eq.~\eqref{eq:Deltac} is initialized with some guess. Then, (i) the self-energies are computed via NRG, (ii) $G_{\mr{loc},c}(z)$ is computed via Eq.~\eqref{Eq:Gcloc} and (iii) the hybridization function is updated via
\begin{align}
\label{Eq:Deltac_CDMFT}
\Delta_{c}(z)&=z+\mu+t\cdot\tau^x -\Sigma_c(z)-G_{\textrm{loc},c}^{-1}(z) \, .
\end{align}
This cycle is repeated until convergence is reached.
\subsection{Momentum integration}
\label{app:integration}
To achieve accurate results, a method for precise momentum integration of propagators is needed in the CDMFT. For this, we employ the tetrahedron method~\cite{Bloechl1994,Kaprzyk2012}, which is applicable for integrals of the form
\begin{equation}
\int_{1.BZ} \, \textrm{d}\bk \frac{f_\bk}{g_\bk} \, ,
\end{equation}
where $f_\bk$ and $g_\bk$ are smooth functions of $\bk$. The Brillouin zone is tiled into tetrahedra and both $f$ and $g$ are interpolated linearly on this tetrahedron.  The integral can then be performed analytically, yielding
\begin{equation}
I_{\textrm{tetra}} = \sum_{i} f_{i} w_i(\{g_{i}\}) \, .
\end{equation}
Here, $f_i$ and $g_i$ are the functions $f$ and $g$ evaluated at the corners of the tetrahedron and $w_i(\{g_{i}\})$ are integration weights which depend on $g$ only. Formulas for these weights are quite lengthy and can be found in~\cite{Kaprzyk2012} for one-, two- and three-dimensional integration. We further use an adaptive momentum grid to reduce computational effort, adjusting the grid size according to the degree of difficulty of the integral in a certain region. This enables us to evaluate all our integrals with an absolute error less than $5\cdot 10^{-4}$. For this, the integration domain is tiled into a coarse and a fine grid, and the grid is iteratively refined in regions where the error bound is not fulfilled, until convergence is reached within the error bounds. Using the tetrahedron method, we then calculate $G_{\textrm{loc},c}(\omega)$ via Eq.~\eqref{Eq:Gcloc}. The computational effort for this integral can be reduced by treating the integral over $k_y$ and $k_z$ as a density of states integration, thereby mapping Eq.~\eqref{Eq:Gcloc} to a two dimensional integral:
\begin{equation}
\begin{gathered}
\epsilon(k_x,E) = -2t\left(\cos(k_x) + E\right) \\
\mathcal{G}^{-1}_{0,c}(z,k_x,E) =  
\left( \begin{array}{cc}
z+\mu & -e^{i k_x}\epsilon(k_x,E) \\
-e^{-i k_x}\epsilon(k_x,E) & z+\mu
\end{array} \right) \\
G_{\textrm{loc},c}(z) = \int_{k_x}\int_{E} \rho_{\mathrm{2D}}(E) \left[\mathcal{G}_{0,c}^{-1}(z,k_x,E) - \Sigma_c(z)\right]^{-1} \\
\rho_{\mathrm{2D}}(E) = \int_{k_y} \int_{k_z} \delta(E-\cos(k_y)-\cos(k_z)) \, ,
\end{gathered}
\end{equation}
where $\rho_{\mathrm{2D}}$ is the density of states of a square lattice.
%

\begin{figure}[bt!]
\includegraphics[width=\linewidth]{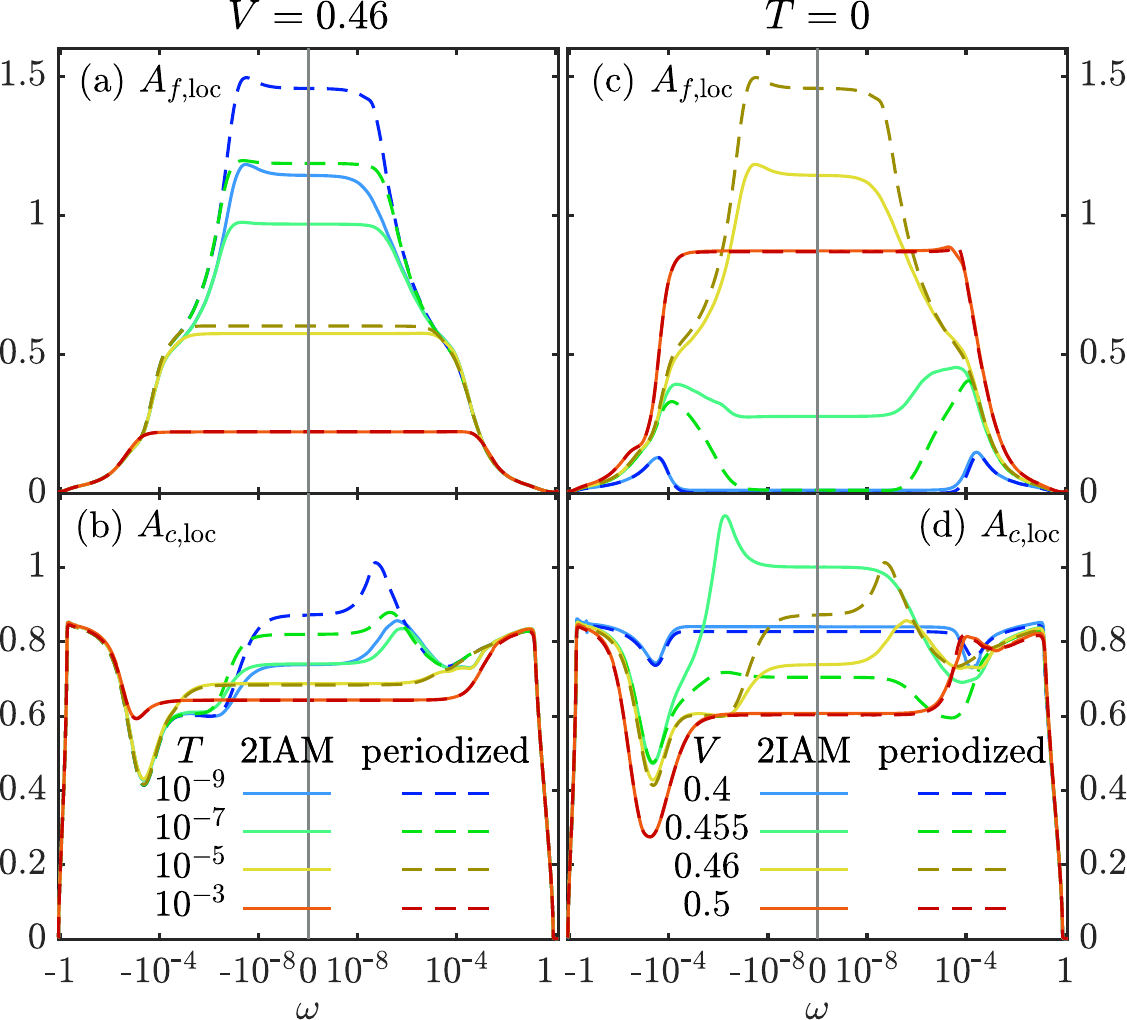}
\caption{%
Comparison of local cluster spectral functions (solid lines) and local lattice spectral function (dashed lines) after periodization of the self-energies. The upper and lower rows show the $f$- and $c$-electron spectral function, respectively. The left panels show data for four different temperatures at fixed $V=0.46$, close to $V_c$. The right panels show data for four different $V$ at $T=0$. The layout mirrors that of Fig.~\ref{fig:A_vs_V_T1e-11_Cluster_log}.
}
\label{fig:Mper_Aloc}
\end{figure}

%

\begin{figure*}[t!]
\centerline{\includegraphics[width=\textwidth]{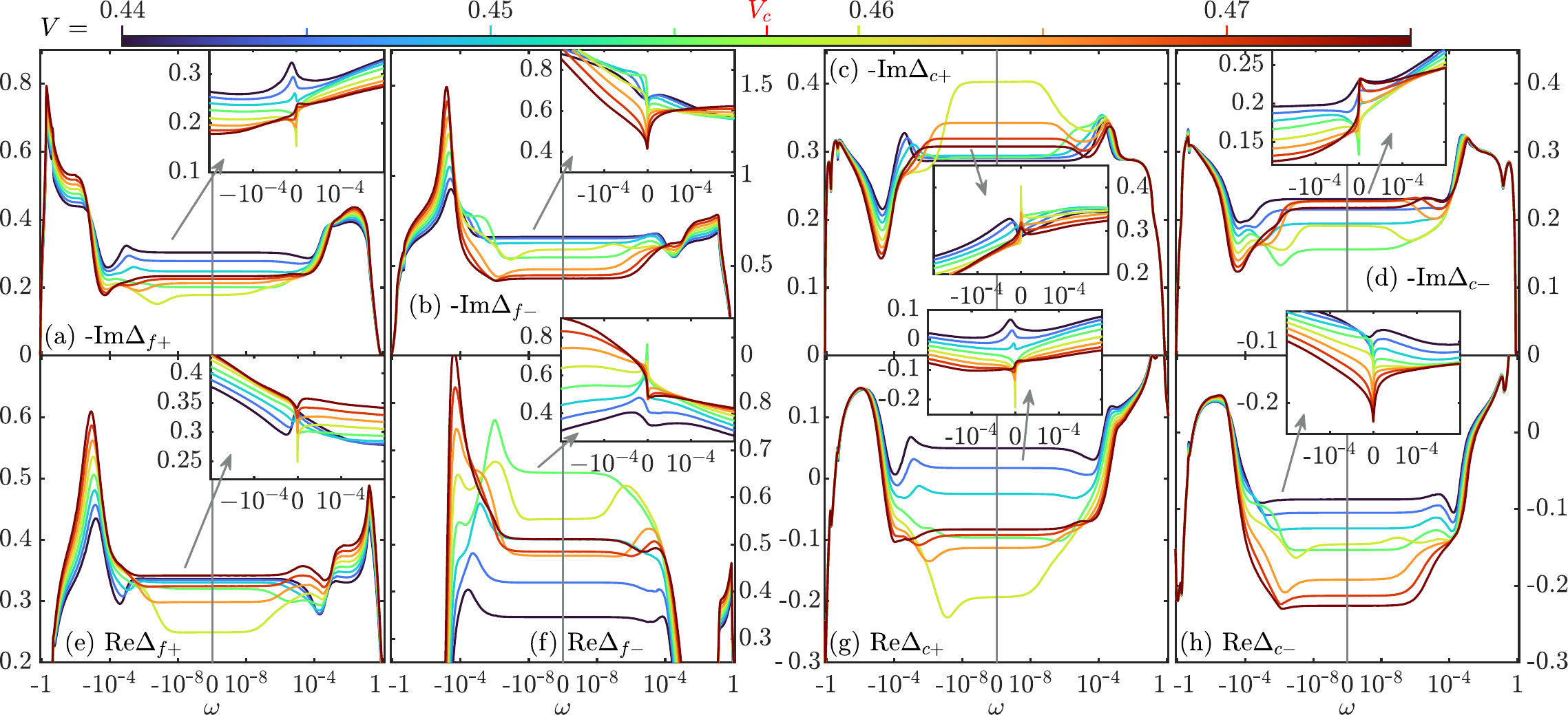}}
\caption{%
Evolution of the self-consistent hybridization functions $\Delta_{x\alpha}(\omegaR)$ at $T=0$ as $V$ is tuned across the QCP. Colored curves correspond to $V$ values marked by ticks on the color bar. The layout mirrors that of Fig.~\ref{fig:A_vs_V_T1e-11_Cluster_log} in the main text.
Top row: imaginary parts; bottom row: real parts. 
The insets show the hybridization functions on a linear frequency scale for $|\omega|<2\!\cdot\! 10^{-4}$.}
\label{fig:Hybfun_PAM_QCP} 
\end{figure*}

%
\subsection{Reperiodization of the self-energy \label{app:ssec:reperiodization}}

To determine transport properties such as the resistivity, the Hall coefficient or the FS, the lattice symmetries have to be restored by reperiodizing the cluster self-energy. We accomplish this via 
a modified
periodization of the cumulant $M(z)=[z+\mu-\Sigma_c(z)]^{-1}$ ($M$-periodization)~\cite{Stanescu2006a,Stanescu2006}.
As we discussed in Sec.~\ref{sec:Luttinger}, the Luttinger integral (at $T=0$) \textit{without reperiodization} Eq.~\eqref{eq:LuttingerIntegral_CDMFT} is zero (c.f. Fig.~\ref{fig:LT_PAM}). This property is not respected by conventional $M$-periodization; the Luttinger integral \textit{with reperiodization} 
Eq.~\eqref{eq:LuttingerIntegralc} will generically be non-zero. To ensure that the Luttinger integral vanishes also after reperiodization, we therefore
modify the cumulant by a $V$-dependent shift of the chemical potential in the denominator \textit{only for reperiodization purposes}:
\begin{subequations}
\label{Eq:Mper}
\begin{align}
\label{eq:Mshift}
\widetilde{M}(z)&=\bigl[z+\mu+\delta\mu(V)-\Sigma_c(z)\bigr]^{-1} \\
\label{eq:Mk}
\widetilde{M}_\bk(z) &= \widetilde{M}_{11}(z) + \widetilde{M}_{12}(z) 
\sum_{\alpha=1}^{3} \tfrac{1}{3}\cos(k_{\alpha}) \\
\label{eq:SEck}
&= \bigl[z+\mu+\delta\mu(V)-\Sigma_{c\bk}(z)\bigr]^{-1} \, . 
\end{align}
\end{subequations}
Here, Eq.~\eqref{eq:Mshift} defines the modified cumulant used for reperiodization, Eq.~\eqref{eq:Mk} defines the reperiodization of $\widetilde{M}(z)$ and Eq.~\eqref{eq:SEck} relates it to $\Sigma_{c\bk}$, thereby defining $\Sigma_{c\bk}$;
quantities like $\Sigma_{f\bk}$ or $G_{x\bk}$ are obtained from $\Sigma_{c\bk}$ using the relations Eqs.~\eqref{eq:Glatt}. 
Note that the shift $\delta\mu(V)$ appears both in Eq.~\eqref{eq:Mshift} and in Eq.~\eqref{eq:SEck};
it therefore does not constitute an \textit{actual} shift in the chemical potential but rather slightly redefines the quantity used for reperiodization (i.e. $\widetilde{M}$ instead of $M$ is reperiodized).
The shift $\delta\mu(V)$ is chosen such that the Luttinger integral after reperiodization (Eq.~\eqref{eq:LuttingerIntegralc}) coincides with that before reperiodization (Eq.~\eqref{eq:LuttingerIntegral_CDMFT}) at $T=0$. The same shift $\delta\mu(V)$ is then used at $T>0$ for the same $V$.
After reperiodization, we have $\widetilde{M}_\Gamma=\widetilde{M}_{11} + \widetilde{M}_{12} = \widetilde{M}_{+}$ and $\widetilde{M}_\Pi=\widetilde{M}_{11} - \widetilde{M}_{12} = \widetilde{M}_{-}$; the same relation also holds between $\Sigma_{x\bk}$ and $\Sigma_{x\alpha}$. This establishes a correspondence between $\Gamma$/$\Pi$ point in the lattice model and $+$/$-$ orbital in the effective cluster model. 
To benchmark our reperiodization scheme, we compare the local spectral functions with and without reperiodization, shown in Fig.~\ref{fig:Mper_Aloc}. For $V$ not too close to $\Vc=0.4575$ and at elevated temperatures, these two functions agree, implying that reperiodization works well here. Close to the QCP (low temperatures, $V\simeq \Vc$) however, reperiodized and cluster results show differences. 
These differences are mostly quantitative, while most of the qualitative features remain similar.
For instance, for $A_f$ at $T=0$, both the Kondo peak height at $V=0.46$ and the pseudogap at $V=0.455$ are more pronounced after periodization,
but the qualitative behavior is the same before and after periodization [Fig.~\ref{fig:Mper_Aloc}(c)].
The most severe qualitative mismatch is the $A_{c,\mr{loc}}(0)$ for $V<V_c$ [Fig.~\ref{fig:Mper_Aloc}(d)]: at $T=0$ in the RKKY phase, as $V$ is increased towards $V_c$, a Kondo-like peak develops in $A_{c,\mr{loc}}$ before periodization (see our discussion in Sec.~\ref{sec:SpectralFunctionsCluster_OSMT}), i.e. $A_{c,\mr{loc}}(0)$ increases as $V$ approaches $V_c$ (solid green curve lies above solid blue curve at $\omega=0$); after periodization, the converse happens, i.e. $A_{c,\mr{loc}}(0)$ decreases as $V$ approaches $V_c$ (dashed green curve lies below dashed blue curve at $\omega=0$). Our periodization procedure
thus misses the development of the Kondo-like peak in $A_{c,\mr{loc}}$.

We emphasize that reperiodization is an ad-hoc post-processing procedure. Features in reperiodized data should always be substantiated
by analyzing the raw data before reperiodization. We have done so repeatedly in the main text, for instance for the FS and LS volumes, the dispersive pole in $\Sigma_f$ or the two and three-band structures in the Kondo and RKKY phases, respectively.
\subsection{Self-consistent hybridization function versus $V$ \label{app:Hybfun}}
In the main text, we emphasized the importance of self-consistency to access the KB--QCP in 2-site CDMFT. 
Here, we elaborate this aspect by showing that the self-consistent hybridization functions in the vicinity of the KB--QCP 
show (i) a strong $\omega$-dependence and 
(ii) a strong $V$-dependence. This implies that self-consistency is of high importance if one wants to capture the KB--QCP. 
%

\begin{figure}[b!]
\includegraphics[width=.5\textwidth]{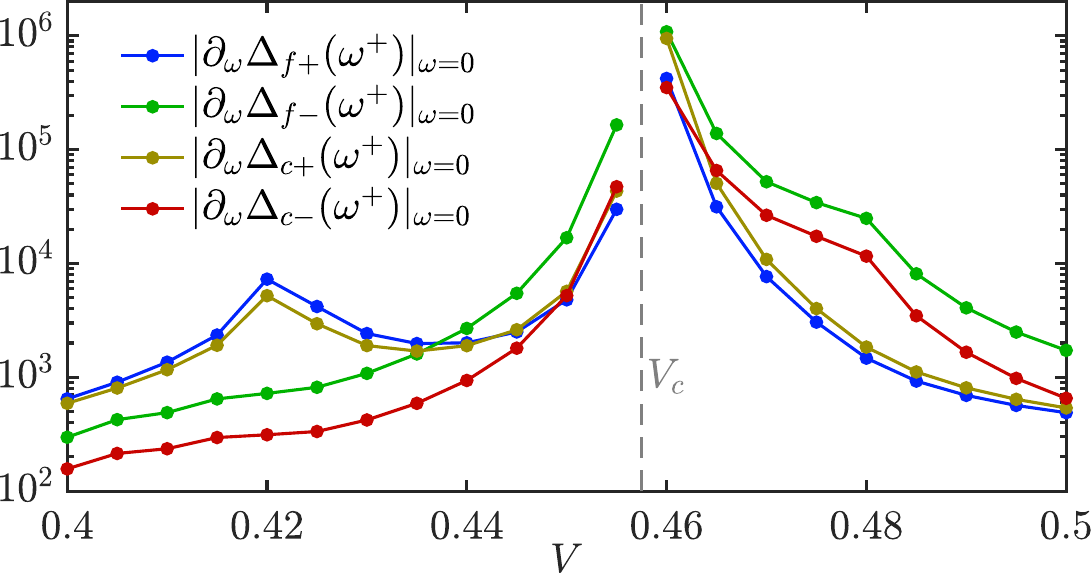}
\caption{%
Absolute value of the derivative of the hybridization function as zero frequency, $|\partial_{\omega} \Delta_{x\alpha}(\omegaR)|_{\omega=0}$ at $T=0$, plotted as a function of $V$.}
\label{fig:Hybfun_derivative}
\end{figure}

%
Figure~\ref{fig:Hybfun_PAM_QCP} shows the hybridization functions at $T=0$ on a logarithmic frequency scale, plotted for several different values of $V$ close to the KB--QCP. 
As seen from the insets, showing the same data on a linear frequency scale, the hybridization functions have sharp features at low frequencies. The sharp features appear stretched out on the log scale of the main panels, which reveal that for $V$ very close to $\Vc$ (green, yellow), they occur at frequencies as low as $|\omega| \simeq 10^{-8}$.
This shows that the hybridization functions of the effective 2IAM describing the self-consistent PAM depend strongly on frequency. For a non-self-consistent 2IKM with weakly frequency-dependent hyrbidization functions, it has been shown that the NFL fixed point and thus the QCP is unstable to the breaking of symmetries which are absent in our effective 2IAM~\cite{Sakai1990,Sakai1992,Silva1996,Jones2007}.
Our work implies that this conclusion does not generalize to the case of hybridization functions displaying a strong frequency dependence.
Figure~\ref{fig:Hybfun_PAM_QCP} shows that the hybridization functions are also strongly $V$-dependent, especially close to $\omega=0$. A change of $V$ by $5\cdot 10^{-3}$ in the vicinity of the KB--QCP induces comparably large changes in the hybridization functions, of the order of $10^{-1}$ at some frequencies. Thus, close to the KB--QCP, a tiny change in $V$ leads to a considerable readjustment of the self-consistent hybridization function by iterating the CDMFT self-consistency cycle. This shows that self-consistency is of high importance to capture the KB--QCP.
In Figure~\ref{fig:Hybfun_derivative}, we plot the absolute value of the derivative of the hybridization functions at $\omega = 0$, $|\partial_{\omega} \Delta_{x\alpha}(\omegaR)|_{\omega=0}$, at $T=0$ as functions of $V$. The zero frequency derivative of the hybridization functions has a peak at the KB--QCP at $V_c$, indicative of a divergence. This suggests that the self-consistent hybridization functions become singular at $\omega=0$ at the KB--QCP. This further emphasizes that (i) results obtained on the 2IAM with weakly frequency dependent hybridization functions are not straightforwardly applicable to the self-consistent 2IAM arising in our CDMFT solution of the
 PAM; and (ii) self-consistency is important to capture this singular behavior at the KB--QCP. A more detailed study investigating how the self-consistency equations 
manage to drive the 2IAM to a stable QCP will be subject to future work.
\subsection{Transport properties \label{app:ssec:transport}}
For the calculation of the resistivity and the Hall coefficient, the $M$-reperiodized self-energy is used. The formulas for the optical conductivity $\sigma(\omega)$ ignoring vertex corrections, resistivity $\rho_{xx}=1/\sigma_{xx}$ and the Hall coefficient $\RHall=\sigma_{xy}/(\sigma_{xx}^2H)$ are given by~\cite{Voruganti1992,Lange1999}
\begin{equation}
\label{eq:conductivity}
\begin{aligned}
\sigma(\omega) &= 2\pi e^2 \int d\epsilon \, \Phi_{xx}(\epsilon) \tilde{\sigma}(\epsilon,\omega) \, , \\
\tilde{\sigma}(\epsilon,\omega) &= \int \mathrm{d}\widetilde{\omega}\, \frac{f(\widetilde{\omega})-f(\widetilde{\omega}+\omega)}{\omega} A_{c}(\epsilon,\widetilde{\omega}) A_{c}(\epsilon,\widetilde{\omega}+\omega) \, ,\\
\sigma_{xx} &= \lim_{\omega\to 0} \sigma(\omega) \, , \\
\sigma_{xy} &= \tfrac{4}{3}\pi^2e^3H  \int \mathrm{d}\epsilon \, \Phi_{xy}(\epsilon) \int d\omega \left[-\frac{\partial f}{\partial \omega}\right] A_{c}^3(\epsilon,\omega) \, , \\
\Phi_{xx}(\epsilon) &=\int_{1.\textrm{BZ}} \frac{\mathrm{d}\bk}{(2\pi)^3} \, \left(\epsilon^x_\bk\right)^2 \delta(\epsilon-\epsilonck) \, , \\
\Phi_{xy}(\epsilon) &=\int_{1.\textrm{BZ}} \frac{\mathrm{d}\bk}{(2\pi)^3} \, 
\left|\begin{array}{cc}
\epsilon^x_\bk \epsilon^x_\bk & \epsilon^{xy}_\bk  \vspace{2mm} \\
\epsilon^y_\bk\epsilon^x_\bk & \epsilon^{yy}_\bk
\end{array} \right| \delta(\epsilon-\epsilonck) \, ,
\end{aligned}
\end{equation}
where $H$ denotes the magnetic field, $e<0$ the charge of the electrons, $f(\omega)$ the Fermi function, $\epsilon^x_\bk = \partial_{k_x} \epsilonck$ the derivative of the dispersion by $k_x$ (and correspondingly for e.g. $\epsilon^{xy}_\bk$) and $|\cdot|$ the determinant. In the above formulas, only the $c$-electron spectral function appears, as there are no terms involving the $f$ electrons which do not conserve local charge. Note also that the $\bk$-dependent spectral function depends on $\bk$ only through $\epsilonck$ after reperiodization via Eq.~\eqref{Eq:Mper}.

Eqs.~\eqref{eq:conductivity} include only the bubble contribution to the conductivities and ignore vertex corrections. In a companion paper~\cite{Gleis2023a}, we show
that vertex corrections to the conductivity are qualitatively important in order to capture the correct scaling of the optical conductivity in the NFL region.
A full treatment of vertex corrections is currently computationally unfeasible with our NRG impurity solver because the computation of 
4-point correlation functions for the 2IAM at hand is too expensive. 
To compute the Hall coefficient shown in Fig.~\ref{fig:Hall_PAM}, we have therefore \textit{not} considered vertex corrections but just used the formulas presented in
Eqs.~\eqref{eq:conductivity}.
\section{Determination of energy scales \label{app:Chi_App}}
%

\begin{figure}[bt!]
\includegraphics[width=.45\textwidth]{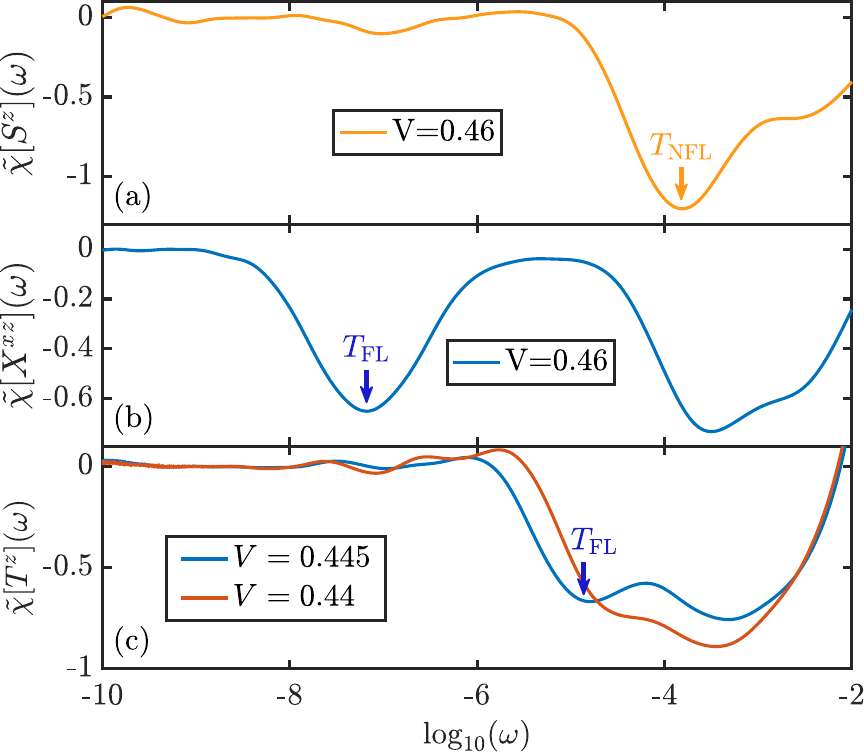}
\caption{%
(a) $\tilde{\chi}[S^{z}](\omega)$ plotted versus $\ln_{10}(\omega)$ at $V=0.46$ and $T=0$. We extract the NFL scale from the position of the minimum marked by the orange arrow.
(b) $\tilde{\chi}[X^{xz}](\omega)$ plotted versus $\ln_{10}(\omega)$ at $V=0.46$ and $T=0$. The FL scale is extracted from the position of the minimum marked by the blue arrow. Note that while the other minimum is associated with $\TNF$, it is not used to extract this scale.
(c) $\tilde{\chi}[T^{z}](\omega)$ plotted versus $\ln_{10}(\omega)$ at $V=0.445$ and $V=0.44$ at $T=0$. While the FL scale can still be extracted from $\tilde{\chi}[T^{z}](\omega)$ at $V=0.445$ because both minima in $\tilde{\chi}[T^{z}](\omega)$ are still clearly distinguishable, it is not possible any more for $V=0.44$ as the minima have merged.
}
\label{fig:fit_TFL}
\end{figure}
%

\begin{figure}[htb!]
\includegraphics[width=.45\textwidth]{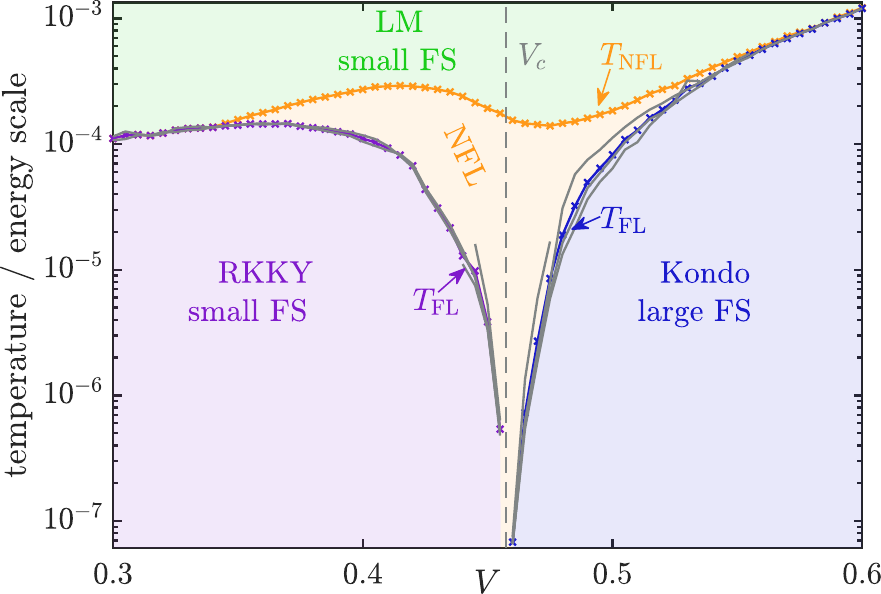}
\caption{%
FL scales, extracted from different correlators (grey lines). For more information on the energy scales, see Fig.~\ref{fig:Escale} in the main text and the corresponding discussion there.
}
\label{fig:Escale_PAM_all}
\end{figure}
To determine the crossover scales $\TFL$ and $\TNF$, we exploit the fact that the spectral functions of particular susceptibilities show a well-defined power law dependence in the fixed point regions, which changes when traversing the crossover regions. On a log-log scale this leads to straight lines with kinks in the crossover regions, as can be seen in Fig.~\ref{fig:Chi_QCP} in the main text. The second derivative of $\chi''[O](\omega)$ on the log-log scale,
\begin{equation}
\tilde{\chi}[O](\omega) = \frac{\partial^2 \ln_{10}(\chi''[O](\omega))}{\partial \ln_{10}(\omega)^2} \, ,
\end{equation}
tracks the change in slope at the crossover, enabling us to determine the corresponding scale. We use
$\chi''[S^z]$
to determine the NFL scale and
$\chi''[X^{xz}]$, $\chi''[T^{z}]$, $\chi''[T^{y}]$ and $\chi''[P^{z}]$
to determine the FL scale. The corresponding operators are defined in Eq.~\eqref{eq:operators} in the main text. $\chi''[X^{xz}]$, $\chi''[T^z]$, $\chi''[T^{y}]$ and $\chi''[P^{z}]$ are shown in Fig.~\ref{fig:Escale}.
Examples for the determination of the NFL scale via $\chi''[S^{z}]$ and of the FL scale via $\chi''[X^{xz}]$ are shown in Figs.~\ref{fig:fit_TFL}(a) and (b), respectively. While $\chi''[T^{y}]$ and $\chi''[P^z]$ are much smaller than $\chi''[T^z]$ and $\chi''[X^{xz}]$, they are more suitable than the latter two far away from the QCP, especially for $V<\Vc$.  The reason for this is shown in Fig.~\ref{fig:fit_TFL}(c) for $\chi''[T^z]$: the second derivative on the $\ln$-$\ln$ scale shows two minima, the first related to $\TFL$ and the second related to $\TNF$. The same is true for  $\chi''[X^{xz}]$ (not shown). These two minima merge away from the QCP, preventing the determination of $\TFL$. The determination via $\chi''[T^y]$ and $\chi''[P^z]$ is not faced with these difficulties, as they have only one minimum, associated with $\TFL$. 
We note that the determination of $\TFL$ via $\chi''[X^{xz}]$ works well in the Kondo regime for all $V$ but in the RKKY regime only close to the QCP, while $\chi''[T^z]$ works only close to the QCP both in the Kondo and the RKKY regime. As the FL scale extracted from different correlators in are generally not exactly equal, we define $\TFL$ as their geometric mean as shown in Fig.~\ref{fig:Escale_PAM_all}. Each grey line there show a FL scale extracted from a different correlator, while the blue and purple lines show their geometric mean. The grey lines do not exactly lie on top of each other, but they are sufficiently similar to justify the averaging described above.
\section{Determination of $T_{\mr{Hall}}$}\label{app:THall_PAM}

\begin{figure}[b!]
\includegraphics[width=.5\textwidth]{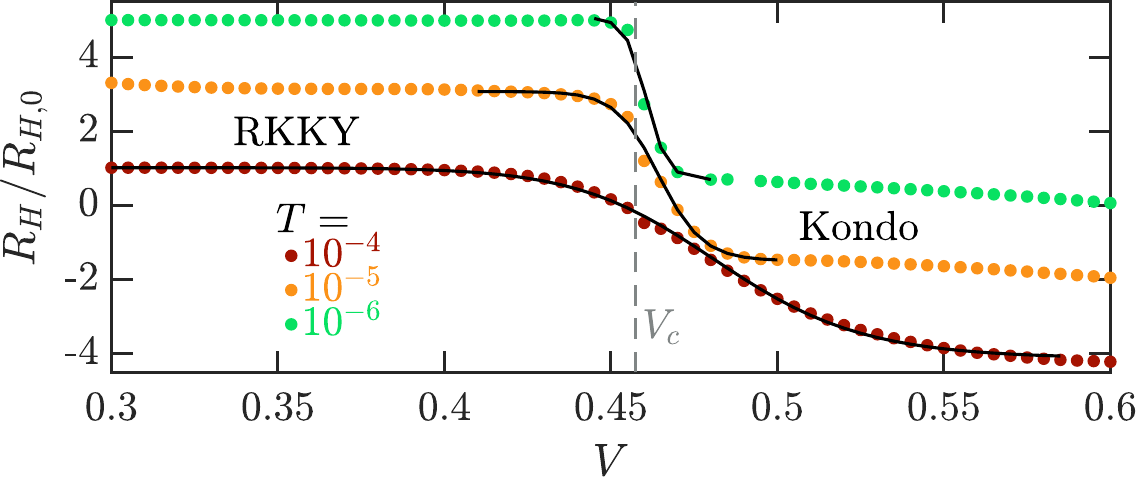}
\caption{%
Hall coefficient data from Fig.~\ref{fig:Hall_PAM} (symbols) and corresponding fits via Eq.~\eqref{eq:THall_fit} (black lines). To improve visibility, the data for $T=10^{-5}$ and $T=10^{-6}$ is offset by $2$ and $4$, respectively.
}
\label{fig:THall_fit}
\end{figure}

Figure~\ref{fig:Escale} includes data points (red dots) marked  $T_\mr{Hall}(V)$, showing how the crossover from a large FS in the Kondo phase to a small FS in the RKKY phase evolve with temperature. Here, we
describe how $T_{\mr{Hall}}$ was determined.

We closely follow the procedure used in Ref.~\onlinecite{Paschen2004}. We fit our numerical $V$-dependent Hall coefficient data (see Fig.~\ref{fig:Hall_PAM}) at $T=10^{-4}$, $10^{-5}$ and $10^{-6}$ to the form
\begin{align}
\label{eq:THall_fit}
\frac{R_{H}(V)}{R_{H,0}} \overset{\mr{fit}}{\simeq} a + \frac{b}{1+(V/V_{\mr{Hall}})^{p}} \, ,
\end{align}
with fit parameters $a$, $b$, $p$ and $V_{\mr{Hall}}$,
i.e. we impose the same functional form as used in Ref.~\onlinecite{Paschen2004} (except for the different tuning parameter, $V$ in our case and $B$-field in Ref.~\onlinecite{Paschen2004}). To closely mirror the procedure used in Ref.~\onlinecite{Paschen2004}, we constrain our fit 
 for a given, fixed $T$ to the crossover region from Kondo to RKKY regime  by only fitting in a $V$-region determined by $T>0.1\! \cdot\! \TFL(V)$ (i.e. we omit datap oints deep in the FL regions). This yields $V_{\mr{Hall}}(T)$, and inverting this function
 yields $T_{\mr{Hall}}(V)$. Figure~\ref{fig:THall_fit} shows our Hall effect data with the corresponding fits.

\end{appendix}

\bibliography{references}

\end{document}